\documentclass[twocolumn,showpacs,aps,floatfix,prd,nofootinbib]{revtex4}
\usepackage{graphicx}
\usepackage{dcolumn}
\usepackage{array, longtable}
\usepackage{epsfig}
\usepackage{amsmath}
\usepackage{colordvi}
\usepackage{hhline}
\newcommand{\BaBarYear}       {07}

\newcommand{\BaBarPubNumber}  {066}
\newcommand{\SLACPubNumber} {13023}
\newcommand{\BaBarType}      {PUB}  
\input pubboard/babarsym
\newcommand{\bei}{\begin{itemize}}
\newcommand{\eei}{\end{itemize}}
\newcommand{\beq}{\begin{equation}}
\newcommand{\eeq}{\end{equation}}
\newcommand{\beqn}{\begin{eqnarray}}
\newcommand{\eeqn}{\end{eqnarray}}
\newcommand{\beqns}{\begin{eqnarray*}}
\newcommand{\eeqns}{\end{eqnarray*}}

\newcommand{\tabtworef}[2]{Tables~\ref{tab:#1} and \ref{tab:#2}}
\newcommand{\tabsref}[2]{Tables~\ref{tab:#1}--\ref{tab:#2}}
\newcommand{\equaref}[1]{Eq.~(\ref{eq:#1})}
\newcommand{\equtworef}[2]{Eq.~(\ref{eq:#1}) and (\ref{eq:#2})}

\renewcommand{\figref}[1]{Fig.~\ref{fig:#1}}

\def\ea{{\em et al.}}
\def\min{{\rm min}}

\def\rPTbarkappa{\kern 0.18em\overline{\kern -0.18em r}{}^{\kappa}{}}
\def\rPTbarsigma{\kern 0.18em\overline{\kern -0.18em r}{}^{\sigma}{}}

\def\deltabarkappa{\kern 0.18em\overline{\kern -0.18em \delta}{}_r^{\kappa}}
\def\deltabarsigma{\kern 0.18em\overline{\kern -0.18em \delta}{}_r^{\sigma}}
\def\deltaTbarkappa{\kern 0.18em\overline{\kern -0.18em \delta}{}_T^{\kappa}}
\def\deltaTbarsigma{\kern 0.18em\overline{\kern -0.18em \delta}{}_T^{\sigma}}

\def\OC{X}
\def\OCbar{{\kern 0.18em\overline{\kern -0.18em \OC}}}

\def\mprime{\ensuremath{m^\prime}}
\def\thetaprime{\ensuremath{\theta^\prime}}
\def\deprime{\ensuremath{{\de^\prime}{}}}

\def\de{\DeltaE}
\def\demin{\de_{\rm min}}
\def\demax{\de_{\rm max}}

\def\a{\kappa}

\def\Amptpbar{\kern 0.18em\overline{\kern -0.18em {\cal A}}_{{\overline B^0} \rightarrow K^-\pi^+\pi^0}}

\def\Amptpbarkappa{\kern 0.18em\overline{\kern -0.18em A}{}^{\kappa}{}}
\def\Amptpbarsigma{\kern 0.18em\overline{\kern -0.18em A}{}^{\sigma}{}}

\def\Tbarkappa{\kern 0.18em\overline{\kern -0.18em T}{}^{\kappa}{}}
\def\Tbarsigma{\kern 0.18em\overline{\kern -0.18em T}{}^{\sigma}{}}
\def\Pbarkappa{\kern 0.18em\overline{\kern -0.18em P}{}^{\kappa}{}}
\def\Pbarsigma{\kern 0.18em\overline{\kern -0.18em P}{}^{\sigma}{}}

\def\Nbpm{{\kern 0.18em\overline{\kern -0.18em N}}^{+-}}
\def\Nbmp{{\kern 0.18em\overline{\kern -0.18em N}}^{-+}}

\def\Mu{\mu}
\def\Chi2MinaMu{\chi^2_{\min ;\a,\Mu}}
\def\Chi2MinMu{\chi^2_{\min ;\Mu}(a)}

\def\TM{{\rm TM}}

\def\fscfave{\kern 0.18em\overline{\kern -0.18em f}_{\rm SCF}}

\def\abar{\bar{a}}

\def\Bbar{\kern 0.18em\overline{\kern -0.18em B}{}\xspace}

\def\BRpmb{{\cal \kern 0.18em\overline{\kern -0.18em  B}}{}_{\rho\pi}^{+-}}
\def\BRmpb{{\cal \kern 0.18em\overline{\kern -0.18em  B}}{}_{\rho\pi}^{-+}}

\def\BRipmb{{\cal \kern 0.18em\overline{\kern -0.18em  B}}{}_{\rho^+\pi^-}}
\def\BRimpb{{\cal \kern 0.18em\overline{\kern -0.18em  B}}{}_{\rho^-\pi^+}}

\def\Abar{\kern 0.18em\overline{\kern -0.18em A}{}}
\def\abar{\kern 0.18em\overline{\kern -0.18em a}{}}

\long\def\inst#1{\par\nobreak\kern 4pt\nobreak
    {\it #1}\par\vskip 10pt plus 3pt minus 3pt}

\begin{document}
\begin{flushleft}
\babar-\BaBarType-\BaBarYear/\BaBarPubNumber \\
SLAC-PUB-\SLACPubNumber\\
arXiv:0711.4417 [hep-ex]\\
\end{flushleft}

\title{\large \bf
\boldmath
Dalitz Plot Analysis of the Decay~$\Bz(\Bzb) \to \Kpm\pimp\piz$
} 

%% author list as of 04-Sep-2007 (556 authors)
%
\author{B.~Aubert}
\author{M.~Bona}
\author{Y.~Karyotakis}
\author{J.~P.~Lees}
\author{V.~Poireau}
\author{X.~Prudent}
\author{V.~Tisserand}
\author{A.~Zghiche}
\affiliation{Laboratoire de Physique des Particules, IN2P3/CNRS et Universit\'e de Savoie, F-74941 Annecy-Le-Vieux, France }
\author{J.~Garra~Tico}
\author{E.~Grauges}
\affiliation{Universitat de Barcelona, Facultat de Fisica, Departament ECM, E-08028 Barcelona, Spain }
% ...
\collaboration{The \babar\ Collaboration}
\noaffiliation
 
\date{\today}

\begin{abstract}
We report a Dalitz-plot analysis of the charmless hadronic decays of
neutral \B mesons to $\Kpm\pimp\piz$. With a sample of
$(231.8\pm2.6)\times 10^6 \FourS \to \B\Bbar$ decays collected by the \babar\ detector 
at the \pep2\ asymmetric-energy \B~Factory at SLAC, we measure the
magnitudes and phases of the intermediate resonant and nonresonant
amplitudes for $\Bz$ and $\Bzb$ decays and determine the corresponding
$CP$-averaged branching fractions and charge asymmetries. The inclusive branching fraction and $CP$-violating charge asymmetry are measured to be 
$\mathcal{B}(\Bz\to\Kp\pim\piz)=(35.7_{-1.5}^{+2.6}\pm2.2)\times10^{-6}$, and $\mathcal{A}_{CP}=-0.030^{+0.045}_{-0.051}\pm0.055$
where the first errors are statistical and the second systematic. We observe the decay $\Bz\to\Kstarz(892)\piz$
with the branching fraction $\mathcal{B}(\Bz\to\Kstarz(892)\piz)=(3.6_{-0.8}^{+0.7}\pm0.4)\times10^{-6}$. 
This measurement differs from zero by 5.6 standard deviations (including the systematic uncertainties).
The selected sample also contains $\Bz \to \Dzb \piz$ decays where $\Dzb \to \Kp\pim$, and we measure
$\mathcal{B}(\Bz\to\Dzb\piz)=(2.93\pm0.17\pm0.18)\times 10^{-4}$.
\end{abstract}


%\vfill
%\centerline{To be submitted to \jprBase}

\maketitle

\setcounter{footnote}{0}

\section{INTRODUCTION}
\label{sec:Introduction}
Amplitude analyses of three-body decays of \B mesons with no charm
particle in the final state are well suited to study the Cabibbo-Kobayashi-Maskawa (CKM)
framework~\cite{CKM} for charged current weak interactions. In the analysis of a Dalitz plot the strong phase motions along
the lineshapes of interfering resonances are measured and can be used to constrain the 
weak phases related to the CKM parameters which, in the Standard Model govern $CP$-violation.
Following the path~\cite{SnyderQuinn,rhopibabar,rhopibelle} of the 3-pion \B meson decays which give
constraints on the CKM angle $\alpha_{\rm CKM}\equiv \arg(-V_{td}V_{tb}^*/V_{ud}V_{ub}^*)$, it has been shown in~\cite{Ciuchini:2006kv,Gronau:2006qn} 
that \B decays into a kaon and two pions are sensitive to the angle $\gamma_{\rm CKM}\equiv \arg(-V_{ud}V_{ub}^*/V_{cd}V_{cb}^*)$.  
\par
In this paper we present a time independent amplitude analysis of the flavor-specific $\Bz\to\Kp\pim\piz$ decay~\cite{Cmodeimplied} which finalizes 
preliminary studies~\cite{josezhitang,Zhitang}. The analysis compares the Dalitz plots of the \Bz and \Bzb
decays where low-mass $K\pi$ and $\pi\pi$ resonances
interfere. Previous measurements of the three-body final
state~\cite{belle,cleo} and subdecays~\cite{cleorhok,Feng} to a vector and a
pseudoscalar meson have been published. Other $\B\to K\pi\pi$ decays
have been studied in~\cite{latham, bellek+pi+pi-, babar-kspipi, belle-kspipi}. A phenomenological study of three-body \B meson
decays without charm in the final state is presented in~\cite{Cheng:2007si}. \par
This paper is organized as follows. We first present in~\secref{DecayAmplitudes} the decay model based on an isobar
expansion of the three-body decay amplitude. The complex coefficients of the expansion
are the unknowns we seek to determine by means of an unbinned extended maximum likelihood fit. 
We describe the detector and dataset in~\secref{DetectorAndData}, the procedure used to select the data sample in~\secref{EventSelection},
and the fit method in~\secref{TheFit}. The results are then described
in~\secref{Results} together with the accounting of the systematic
uncertainties (in~\secref{Systematics}). Finally
in~\secref{Interpretation}, we summarize our findings and present an interpretation of our results. 

\section{DECAY AMPLITUDES}
\label{sec:DecayAmplitudes}
The $\Bz\to\Kp\pim\piz$ decay amplitude is a function of two independent kinematic variables commonly chosen to be
the invariant masses squared\footnote{We use natural units where $\hbar=c=1$ in our algebraic equations.}, 
$x=m_{\Kpm\pimp}^2$ and $y=m_{\Kpm\piz}^2$. The Dalitz plot (DP) is the $x,\ y$ bidimensional distribution.  
It is customary to express the decay amplitude as a sum over intermediate (isobar) states:
\begin{equation}
\label{eq:isobarB}
\mathcal{A}(x,y)             = \sum_j t_j e^{i\phi_j} f_j(x,y), 
\end{equation}
and similarly for the $\Bzb\to\Km\pip\piz$ Dalitz plot,
\begin{equation}
\label{eq:isobarBbar}
\overline{\mathcal{A}}(x,y)  = \sum_j \overline{t}_j e^{i\overline{\phi}_j} f_j(x,y).
\end{equation}
The isobar coefficients $t_j e^{i\phi_j}$ are constant over the Dalitz plot. For each decay channel, the isobar phase $\phi_j$ 
is the sum of a strong phase, the same for \Bz and \Bzb decays, and a weak phase which changes sign.
The decay dynamics of an intermediate state are described by the $f_j(x,y)$ function which structures the Dalitz plot.
For instance a resonance formed in the $\Kp\pim$ system gives a contribution which factorizes as:
\begin{equation}
\label{eq:fequalRT}
f_j(x,y) = R_j(x) \times T_j(x,y) \times WB_j(x),
\end{equation}
where $R_j(x)$ is the resonance mass distribution or lineshape and  $T_j(x,y)$ models the angular dependence. 
$WB_j(x)=\sqrt{B_B(Rp^*(x))\;B_j(Rq(x))}$, the product of Blatt-Weisskopf damping factors~\cite{BlattWeissk}, slightly deviates from unity as
a function of $x$ through the breakup momenta\footnote{$p^*$, the momentum of the bachelor particle in the \B meson rest frame, is equal
to the breakup momentum of the studied \B meson decay.} of the (quasi) two body \B and resonance decays multiplied by a range parameter $R$.    
The $f_j$ are normalized,
\begin{equation}
\label{eq:isobarnorm}
\int_{DP}|f_j(x,y)|^2dx\ dy=1.
\end{equation}
\par
We use the Zemach tensor formalism~\cite{Asner,Zemach} for the angular distribution $T_j^{(J)}(x,y)$ of a process by which a pseudoscalar 
$\B$ meson produces a spin-$J$ resonance in association with a bachelor pseudoscalar meson. For $J=0,\ 1,\ 2$, we have:
\begin{eqnarray}
\label{zemach}
T_j^{(0)} &=& 1, \nonumber \\
T_j^{(1)} &=&  -2 \vec{p}\cdot\vec{q}, \nonumber \\
T_j^{(2)} &=&  \frac{4}{3} [3(\vec{p}\cdot\vec{q})^2 - (|\vec{p}||\vec{q}|)^2],
\end{eqnarray}
where\footnote{For simplicity, we have dropped the $j$ index in  $\vec{p}$ and $\vec{q}$.} $\vec{p}(x,y)$ [$\vec{q}(x)$] is the momentum vector of the bachelor particle 
(the resonance decay product $Q$ defined below) 
measured in the resonance rest frame. For a neutral (charged) $K\pi$ resonance, $Q$ is the pion (kaon), and for a dipion 
resonance, $Q$ is the $\piz$. Notice that these choices define for each two-body system the helicity angle  $\theta_j=(\vec{p_j},\ \vec{q_j})$ 
between 0 and $\pi$.\par
Our nominal model (\tabref{nominal}) for the decay $\Bz\to\Kp\pim\piz$ includes a nonresonant contribution which is uniformly distributed over the Dalitz plot, 
and five resonant intermediate states: $\rho^-(770)$, $K^*(892)^{+,0}$
and $(K\pi)^{*+,0}_0$. The latter notation introduced by the \babar\ experiment~\cite{latham} denotes phenomenological
amplitudes describing the neutral and charged $(K\pi)$ S-waves each by a coherent superposition of an elastic effective range term and a
term for the $\Kstar_0(1430)$ scalar resonance. It describes current knowledge on low energy $K\pi$ systems with a small number of parameters. 
Variations to the nominal model are used to estimate the model-dependent systematic uncertainty in the results.
\begin{table}
\begin{center}
\caption{\label{tab:nominal}The {\it nominal model} for the decay $\Bz\to\Kp\pim\piz$ comprises a nonresonant part and five intermediate states. The three types of 
lineshape are described in the text. The resonances masses and widths are from~\cite{PDG2004}, except for the LASS shape~\cite{LASS}. We use the same LASS 
parameters for both neutral and charged $K\pi$ systems. {\it Additional resonances} that may contribute are included in extended models which
we study to estimate the systematic uncertainties.}
\begin{tabular}{ccc} \hline\hline
Intermediate state & Lineshape & Parameters  \\ \hline
\multicolumn{3}{c}{{\it Nominal model}}\\ 
&& \\
Nonresonant & Constant & \\ 
&& \\
$\rho^-(770)$              & GS  & \\ 
                           &     & \\ 
&& \\
$\Kstarp(892)$             & RBW & \\
$\Kstarz(892)$             & RBW  &\\ 
&& \\
$(K\pi)^{*+}_0$            & LASS & $m^0 = 1415 \pm 3\ \ \ \mevcc$ \\
$(K\pi)^{*0}_0$            &      & $\Gamma^0 = \ 300 \pm 6\ \ \ \mev\ \ $ \\
                           &      & cutoff $m_j^{max}=2000 \mevcc$  \\
                           &      & $a = 2.07 \pm 0.10 ~(\gevc)^{-1}$ \\ 
                           &      & $r = 3.32 \pm 0.34 ~(\gevc)^{-1}$\\ \hline
\multicolumn{3}{c}{{\it Additional resonances}} \\ 
&& \\
$\rho(1450)$                & GS  & $m=1439\ \ \ \mevcc$ \\
                            &     & $\Gamma^0=\ 550\ \ \ \mev\ \ $ \\
$\rho(1700)$                & GS  & $m=1795\ \ \ \mevcc$ \\ 
                            &     & $\Gamma^0=\ 278\ \ \ \mev\ \ $ \\
$K_2^*(1430)^{+,0}$        & RBW &\\
$K^*(1680)^{+,0}$          & RBW & \\ \hline \hline
\end{tabular}
\end{center}
\end{table}
Three lineshape parameterizations $R_j(x)$ are used. Parameters are
taken from~\cite{PDG2004}  unless stated otherwise. 
\subsection{LINESHAPES}
\label{subsec:lineshapes}
\subsubsection{The relativistic Breit-Wigner distribution}
The relativistic Breit-Wigner~(RBW) parameterization 
is used for $K^*(892)^{+,0}$, $K_2^*(1430)^{+,0}$, and $K^*(1680)^{+,0}$:
\beq
\label{eq:nominalBW}
        R^{(J)}_j(x;m_j,\Gamma_j^0) \;=\; 
                \frac{1}{m_j^2 - x - i m_j\Gamma^{(J)}_j(x)}~.
\eeq
The mass-dependence of the total width $\Gamma^{(J)}_j$ can be ignored for high-mass states. For the low-mass states which decay only elastically, it is defined by
\beq
\label{eq:s-dependentWidth}
        \Gamma^{(J)}_j(x) \;=\; 
                \Gamma_j^0
                \frac{m_j}{\sqrt{x}}
                \left(\frac{q(x)}{q(m_j^2)}\right)^{\!2J+1}
                \frac{B^{(J)}(Rq(x))}{B^{(J)}(Rq(m_j^2))}~,
\eeq
where $m_j$ is the mass of the resonance $j$, $\Gamma_j^0=\Gamma_j(m_j^2)$ 
its width, and the barrier factors (squares of the Blatt-Weisskopf damping factors~\cite{BlattWeissk}) are:
\beqn
\label{eq:barrier}
        B^{(0)}      &=& 1, \\
        B^{(1)}      &=& \frac{1}{1 + R^2 q^2}, \nonumber \\ 
        B^{(2)}      &=& \frac{1}{9 + 3 R^2 q^2 + R^4 q^4}. \nonumber
\eeqn
All range parameters were set to $R=0$ in the analysis but we checked that the
systematic deviations associated with
more realistic values taken or interpolated from~\cite{PDG2006} are
below the numerical accuracy we use to quote our results.
\subsubsection{The Gounaris-Sakurai distribution}
The Gounaris-Sakurai~(GS) parameterization~\cite{rhoGS} [\equaref{GS}] 
is used for $\rho^-(770)$, $\rho^-(1450)$ and $\rho^-(1700)$:
\beq
\label{eq:GS}
R^{GS}_j(x;m_j,\Gamma_j^0) \;=\; \frac{1+d_j \;\Gamma^0_j/m_j}{m_j^2+f_j(x)-x -i m_j\Gamma_j(x)},
\eeq
with the same $x$-dependence of the width as for the RBW. The expressions of the constant $d_j$ and the function $f_j(x)$ in terms of 
$m_j$ and $\Gamma^0_j$ are given in Reference~\cite{rhoGS}.
\subsubsection{The LASS distribution}
For the $K\pi\, S$-wave resonances, $(K\pi)^{*+,0}_0$, which dominate for $m_{K\pi}$ below $m_j^{max}=2~\gevcc$, an effective-range 
parameterization was suggested~\cite{Estabrooks} to describe the slowly increasing phase as a function of the $K\pi$ mass.
We use the parameterization as in the LASS experiment~\cite{LASS}, tuned 
for \B decays:
\begin{eqnarray}
\label{eq:LASS}
R_j^{LASS} (x; m_j^0,\Gamma_j^0,a,r) =\frac{\sqrt{x}}{q \cot{\delta_B} - iq} \ \ \ \ \ \ \ \ \ \\ 
                    \ \ \ \ \ \ \ \ \ \ \ \ \ +e^{2i \delta_B} \frac{m_j^0 \Gamma_j^0 \frac{m_j^0}{q_0}}{[(m_j^0)^2 - x] - i m_j^0 \Gamma_j^0 \frac{q}{\sqrt{x}} \frac{m_j^0}{q_0}},\nonumber
\end{eqnarray}
\noindent where
         \begin{equation}
          \label{eq:LASSphase}
           \cot{\delta_B} = \frac{1}{a q(x)} + \frac{1}{2}\,r\,q(x)\,,
         \end{equation}
        $a$ is the scattering length, and $r$ the effective range (\tabref{nominal}). 
A LASS isobar coherently comprises the actual $\Kstar_0(1430)$ (82\%), the effective range term (57\%) and the (destructive) interference term (39\%). \par
In this analysis we use a maximum likelihood fit to measure the inclusive branching fraction and $CP$-violation asymmetry, 
\beqn
\mathcal{B}^{incl}   & \equiv & \mathcal{B}(\Bz\to\Kp\pim\piz) \\
\mathcal{A}_{\rm CP} & \equiv & \frac{\int_{DP}[|\mathcal{\overline{A}}(x,y)|^2-|\mathcal{A}(x,y)|^2]dx\ dy}{\int_{DP}[|\mathcal{\overline{A}}(x,y)|^2+|\mathcal{A}(x,y)|^2]dx\ dy}, 
\label{eq:global-bf-acp}
\eeqn 
as well as the isobar fractions $FF_j$, ($CP$-averaged over $\Bz$ and $\Bzb$) and the $CP$-violation charge asymmetries:
\small
\begin{eqnarray} \label{eq:PartialFractions}
  FF_k &=& \frac{\int_{DP}[|t_ke^{i\phi_k}f_k(x,y)|^2+|\overline{t}_ke^{i\overline{\phi}_k}\overline{f}_k(x,y)|^2]dx\ dy} 
                {\int_{DP}[|\sum_jt_je^{i\phi_j}f_j(x,y)|^2+|\sum_j\overline{t}_je^{i\overline{\phi}_j}\overline{f}_j(x,y)|^2]dx\ dy}, \nonumber \\
  A_{\rm CP}^k &=& \frac{\overline{t_k}^2 - t_k^2}
                         {\overline{t_k}^2 + t_k^2}~, 
\end{eqnarray}
\normalsize
and the isobar phases relative to those of the $\Kstarpm\pimp$ channel.
In those expressions,  $t_j$ and $\overline{t_j}$ are the fitted magnitudes for the intermediate state $j$; 
$\phi_j$ and $\overline{\phi}_j$ are the fitted relative phases. 
Note that, due to interference, the fractions $FF_k$ in general do not add up to unity.
Nevertheless, we define the quasi-two-body branching fraction for an intermediate state $j$ as follows:
\begin{equation}
\label{eq:partialbf}
   ~~{\cal B}_j \equiv FF_j \cdot \mathcal{B}^{incl}.
\end{equation}

\subsection{THE SQUARE DALITZ PLOT}
The accessible phase space of charmless three-body \B decays is unusually large. Most contributing resonances have masses much lower than the \B mass. 
Hence signal events cluster along the Dalitz plot boundaries. This is also true for background events. 
Past experience has shown that another set of variables, defining the {\it Square Dalitz Plot} (SDP)~\cite{Dunwoodie-SDP} is well suited to such configurations. 
It is defined by the mapping:
\begin{eqnarray}
\label{eq:SDPvariables}
dx\ dy &\longrightarrow &d\mprime\ d\thetaprime \\
\ \ \mprime&\equiv&\frac{1}{\pi}\arccos(2\frac{m-m_{min}}{m_{max}-m_{min}}-1)\;,\;\thetaprime\equiv\frac{1}{\pi}\theta, \nonumber
\end{eqnarray}
where $m=\sqrt{x}$ and $\theta$ are respectively the invariant mass and helicity angle of the $K^{\pm}\pi^{\mp}$ system. 
$m_{max}=m_B-m_{\piz}$ and $\m_{min}=m_{\Kp}+m_{\pim}$ are the kinematic limits of $m$. The new variables both range between 0 and 1.

\section{THE \babar\ DETECTOR AND DATASET}
\label{sec:DetectorAndData}

The data used in this analysis were collected with the \babar\ detector
at the \pep2\ asymmetric \epem\ storage rings between October 1999 and July 2004.
The sample consists of about 210.6 \invfb\ corresponding to $N_{\BB}=231.8\pm2.6$ million $\BB$ pairs taken on the peak of the 
$\FourS$ resonance (on resonance) and 21.6 \invfb\ recorded at a center-of-mass (CM) energy 40~$\mev$ below (off resonance).\par
A detailed description of the \babar\ detector is given in~\cite{babarNim}. 
Charged-particle trajectories are measured by a five-layer, double-sided silicon vertex tracker (SVT) and a 40-layer
drift chamber (DCH) immersed in a 1.5~T magnetic field. Charged-particle identification is achieved by combining
the information from a ring-imaging Cherenkov device 
with the ionization energy loss (\dedx ) measurements 
from the DCH and SVT. Photons are detected in a CsI(Tl) electromagnetic calorimeter (EMC) inside the coil. 
Muon candidates are identified in the instrumented flux return 
of the solenoid.We use GEANT4-based~\cite{geant4} software to simulate the detector response and account for the varying beam and
environmental conditions. 

\section{EVENT SELECTION}
\label{sec:EventSelection}
\subsection{SIGNAL SELECTION AND BACKGROUND REJECTION}
\label{subsec:Cuts}
\begin{figure}
  \centerline{\epsfxsize8cm\epsffile{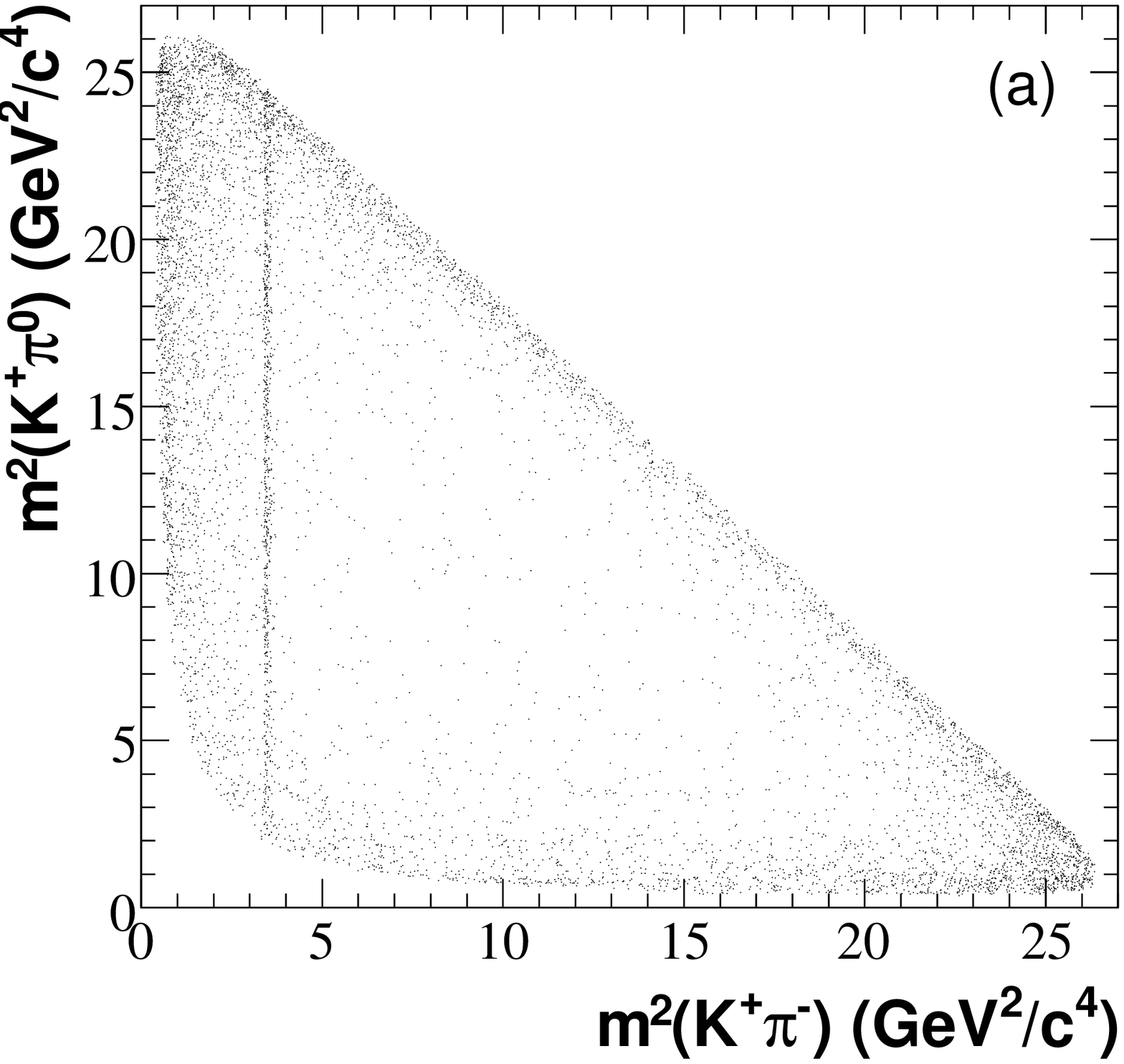}}
  \centerline{\epsfxsize8cm\epsffile{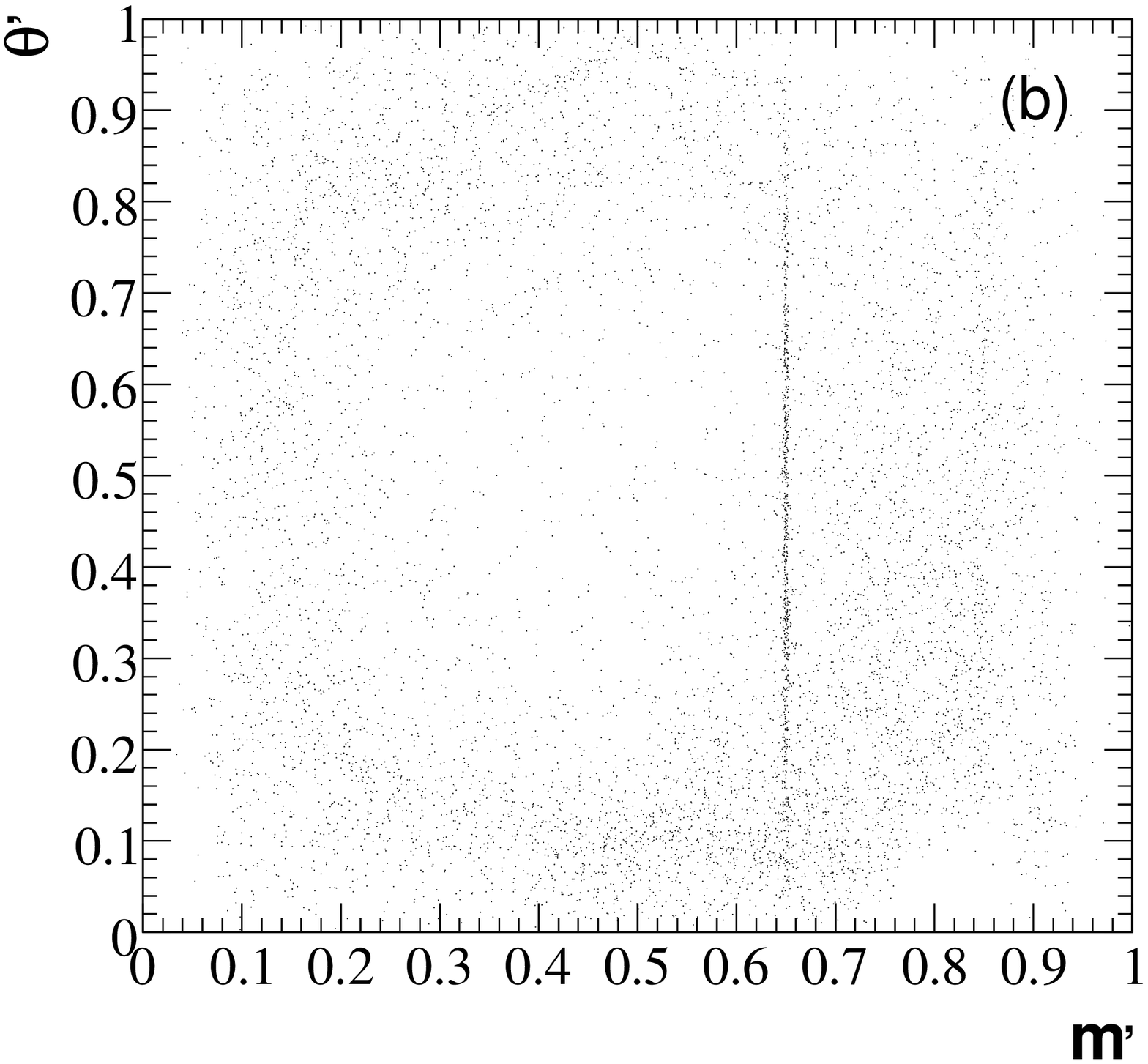}}
  \vspace{-0.3cm}
  \caption
       {\label{fig:DalitzPlots}
		The standard (a) and square (b) Dalitz plots of the
		selected data sample of 8014 events. The structures
		are more spread out in the square Dalitz plot. The
		$\Dzb\to\Kp\pim$ narrow band is preserved with the
		choice made for the \mprime\ variable.}
\end{figure}
To reconstruct $\Bz\to\Kp\pim\piz$ decays, we select two charged particles and two photons. The charged particle candidates are required to have 
transverse momenta above $100\ \mevc$ and at least 12 hits in the DCH. They must not be identified as electrons or muons or protons. 
A $\Kp$ candidate must be identified as a kaon (with a
misidentification probability smaller than 4\%) and a $\pim$
candidate must not be identified as a kaon (with a misidentification
probability smaller than 7\%). The misidentification probabilities are
momentum dependent and therefore vary across the Dalitz plot.  
A $\piz$ candidate is built from a pair of photon candidates, each with an energy greater
than $50\mev$ in the laboratory frame (LAB) and a lateral energy deposition profile in the EMC consistent with an electromagnetic shower. 
The invariant mass of a $\piz$ candidate must satisfy $0.11
<m_{\gamma\gamma}<0.16 \gevcc$, a wide enough window to
accommodate the variation of the resolution across the Dalitz Plot
from 4.5 to 8 \mevcc.\par
At the $\FourS$ resonance, $\B$ mesons are characterized by two nearly independent kinematic variables, the beam energy substituted mass and the energy 
difference:
\beqn
\label{eq:mesde}
\mes&=&\sqrt{(s/2+\vec{p_0} \cdot \vec{p_B})^2/E_0^2-p_B^2}, \\
\Delta E&=&E_B^*-\sqrt{s}/2,
\eeqn
where $E$ and $p$ are energy and momentum, the
subscripts 0 and $B$ refer to the \epem-beam system and the \B candidate respectively; $s$ is the square of the center-of-mass energy and the asterisk 
labels the CM frame. We require that $\mes>5.27\gevcc$. To avoid a bias in the Dalitz plot from the dependence on the $\piz$ energy of the
resolution in \de,  we introduce the dimensionless quantity: 
\beq
\label{eq:deprime}
\deprime \equiv \frac{2\de-(\demax+\demin)}{\demax-\demin}
\eeq
where the bounds obtained from simulation, $\demax = 0.08 - 0.0014 \cdot m_{K^+\pi^-}$, $\demin = -0.14 + 0.005 \cdot m_{K^+\pi^-}$, all in units of $\gev$
follow the variation of the \de\ resolution. We require $|\deprime| \leq 1$.\par
Continuum $e^+e^-\to q\bar{q}$ ($q = u,d,s,c$) events are the dominant  background. To enhance discrimination between signal and continuum, 
we select events by using a neural network~\cite{NN} with an output $NN$ which combines three discriminating variables: the angles of the \B momentum and the \B thrust axis 
with respect to the $e^+$ beam direction in the CM frame and the difference $2L_2-L_0$ between two variables characterizing the energy flow about 
the $\B$ thrust axis. We have $L_n\equiv\sum_i p_i\cdot|\cos\theta_i|^n$, where the sum runs over all charged and neutral particles in the event 
(except for those in the \B candidate) whose momenta $\vec{p_i}$ make angles $\theta_i$ with the $\B$ thrust axis. 
The neural network was trained on {\it off  resonance} data and correctly reconstructed signal
Monte Carlo events. A data sample with about 4000 fully reconstructed $\Bz\to\Dstarm\pip$ decays with $\Dstarm\to\Dzb\pim$ and $\Dzb\to\Kp\pim$ 
is used to validate the shapes of the distributions on which the selection procedure is based.\par
Between 3\% and 17\% of the signal events have multiple reconstructed $\B$ candidates (usually two) depending on their location in the Dalitz plot.  
When distinct $\pi^0$ candidates make acceptable \B combinations, we
choose that with the reconstructed $\pi^0$ mass closest to the nominal
value of 0.1349~$\gevcc$.  When several acceptable \B combinations can
be made of distinct charged particle sets, we arbitrarily choose one 
(in a reproducible fashion by using the date and time at which the event was recorded as a random number). \par
There are 8014 events in the data sample after the selection.  The \B\ meson candidate in each event is mass constrained to ensure that the measurement
falls within the Dalitz plot boundary. The resulting standard and
square Dalitz plots are shown in \figref{DalitzPlots}. We now describe
the composition of the selected data sample. 

\subsection{TRUTH-MATCHED AND SELF-CROSS-FEED SIGNAL EVENTS}
\begin{figure}[h]
 \epsfig{file=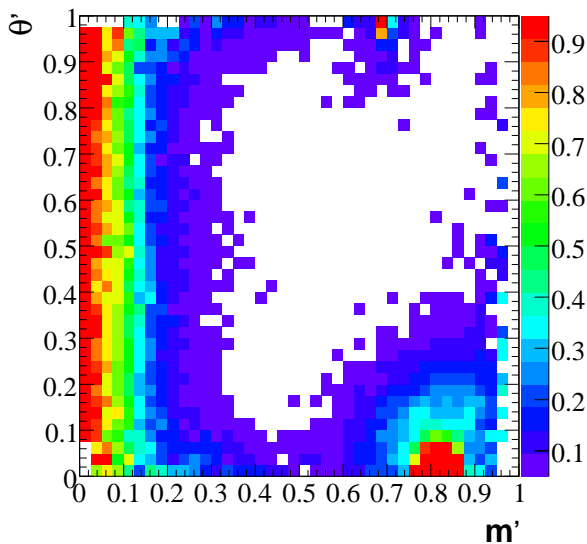, width=8cm}
 \caption{\label{fig:fractionscfDP}
	Fraction of misreconstructed events ($f_{SCF}$), in the square
	Dalitz plot. This figure includes $4.9\times 10^6$ signal MC
	events. In white is the SCF-free region. 
}
\end{figure}

Using the Monte Carlo simulation as in~\cite{rhopibabar}, we distinguish between the correctly reconstructed and the misreconstructed signal events.  
A correctly reconstructed event where the three particles of the \B candidate match the generated ones, is called a {\it Truth-Matched} (TM) event. 
A misreconstructed signal event contains a \B meson  which decays to the signal mode, but one or more reconstructed particles in the \B candidate are not 
actually from the decay of that \B. Misreconstructed signal is called  {\it Self-Cross-Feed} (SCF). 
Misreconstruction is primarily due to the presence of low momentum pions. Consequently the SCF fraction varies across the Dalitz plot as 
shown in~\figref{fractionscfDP}. For each point in the Dalitz plot there is an efficiency $\varepsilon(\mprime,\thetaprime)$ to reconstruct an event 
either correctly or incorrectly. The quality of the reconstruction is poor where the SCF fraction~$f_{\rm SCF}(\mprime,\thetaprime)$, is high. 
This occurs in the corners of the Dalitz plot where one of the final-state particles has a low momentum in the LAB frame. These variations can be seen 
in~\figref{fractionscfDP} and in~\tabref{finalEff} which compares efficiencies and $f_{\rm SCF}$ values averaged over the 
Dalitz plot for different $\Bz\to\Kp\pim\piz$ subdecays\footnote{We choose to classify the $\Bz\to\Dzb\piz, \Dm\Kp$ subdecays as background
since they proceed via a $b \to c$ quark-level transition while we are studying charmless processes.} 
computed using high statistics Monte Carlo samples ($4.9\times10^6$ 
$\Bz\to\Kp\pim\piz$ events). It is important to keep a high efficiency in the Dalitz plot corners where the low-mass vector resonances interfere. 
Overall the total efficiency is close to 15\%. The main sources of inefficiency are the \deprime\ selection ($\varepsilon \approx 70-80\%$), 
the kaon identification ($\varepsilon \approx 69\%$) and the neural network selection ($\varepsilon \approx 60\%$). 
\label{subsec:tmscf}
\begin{table}[h]
\begin{center}
\caption{\label{tab:finalEff}
Signal selection efficiencies (overall and truth-matched) and fraction of misreconstructed events (SCF) for different signal modes after full selection.
As explained in the text, $\varepsilon$ is the overall efficiency for (TM and SCF) signal events. Hence $\varepsilon_{\TM}=\varepsilon(1-f_{\rm SCF})$. 
The relative statistical precision on these numbers is a few tenths of
a percent. 
} 
\begin{tabular}{lcccc} \hline\hline
 Decay mode & $\varepsilon(\%)$ & $\varepsilon_{\rm TM}(\%)$ &$f_{\rm SCF}(\%)$ \\\hline
&&& \\
$K^{*+}(892)\pi^-$ & 13.9 & 10.0 & 28.1 \\
$K^+\rho^-$        & 14.3 & 10.3 & 28.3 \\
$K^{*0}(892)\pi^0$ & 15.7 & 13.6 & 13.3 \\
Nonresonant       & 16.2 & 15.2 &  6.2 \\ \hline
&&& \\
$\Dzb\pi^0$        & 16.4 & 15.7 &  4.0 \\
$\Dm\Kp$           & 15.8 & 14.6 &  7.5 \\
\hline\hline
\end{tabular}
\end{center}
\end{table}
\subsection{BACKGROUND}
\subsubsection{Continuum background}
\label{subsubsec:continuum}
Although the neural network selection rejects 96\% of the continuum events, this background is the dominant class of events in
the data sample, representing about two thirds of its size. 
\subsubsection{Background from other \B decays}
\label{subsubsec:bbackground}
Since there is no restriction on any two-body invariant mass of the
final state particles, large backgrounds from other \B decays
occur. We use high statistics Monte Carlo samples to study these
backgrounds. Conservative assumptions about unknown branching
fractions are made. For instance when only an upper limit $U$ is known for
the branching fraction of a decay channel, we use $U/2\pm U/2$ as an
input to the simulation.

Inclusive and exclusive \B decays with or without
charm are grouped into ten classes to be used in the fit. Rates, and
topological and kinematical similarities are studied to define 
the classes listed in~\tabref{listBb}. There are two
classes with inclusive $b\to c$ decays which are distinguished
according to whether a genuine $\Dz$ is part of the \B candidate.
We keep the exclusive decays $\Bz\to\Dzb\piz$ with $\Dzb\to\Kp\pim$ in the data sample because the
copious yield of approximately 400 well identified events helps to control the fit algorithm. 
\begin{table}[h]
\begin{center}
\caption{\label{tab:listBb} The list of \B-backgrounds retained for the fit (\secref{TheFit}). For each channel, we
give (anticipating~\secref{Results}) either the fitted number of events in the data sample if 
its yield is allowed to vary in the fit procedure or the expected number otherwise.}
\begin{tabular}{clrr}\hline\hline
Class   & Mode  &Events&  \\ \hline
0	& $\Bz\to\Kp\pim$                                           &$10.4\pm 0.5$  & fixed   \\
1	& $\Bz\to\Kstarz(892)\g,\Kstarz1430)\g$, $\rho^+\pim$       &$95.6\pm 5.0$  & fixed   \\
2	& $\Bz\to\rho^+\rho^-$, $\Kstarp(892)\rho^-$, $\Kstar(1680)\rho$ &$10.7\pm 3.8$  & fixed   \\ \hline
3	& $\Bz\to\Dzb(\Kp\pim)\piz$                                 &$424\pm25$ & varied \\ 
4	& $\Bp\to\Kp\piz$                                           &$17.4\pm 1.5$  & fixed   \\
5	& $\Bp\to$ 3-body (mainly $\Kp\pim\pip$)                    &$119\pm34$ & fixed \\
6	& $\B\to$ 4-body                                            &$30.3\pm 3.4$  & fixed   \\
7	& Generic $\B\to$ charm w/o true $\Dz$                      &$382\pm49$ & varied \\  
8	& Generic $\B\to$ charm with true $\Dz$                     &$147\pm17$ & varied \\ \hline
9	& $\Bz\to\Dm(\to\pim\piz)\Kp$                               &$20.4\pm 7.8$  & fixed   \\
\hline\hline
\end{tabular}
\end{center}
\end{table}

\section{THE MAXIMUM LIKELIHOOD FIT}
\label{sec:TheFit}
We perform an unbinned extended maximum likelihood fit to determine the total $\Bz\to\Kp\pim\piz$ event yield, the magnitudes 
$t_j, \overline{t}_j$ and the phases $\phi_j, \overline{\phi}_j$ of the complex isobar coefficients of the decay amplitude defined in~\equtworef{isobarB}{isobarBbar}.
The fit uses the variables \mprime, \thetaprime, \mes, \deprime\  and NN, to discriminate signal from background.
\subsection{THE LIKELIHOOD FUNCTION}
\label{subsec:thelikeli}
The selected on resonance data sample is assumed to consist of signal, continuum-background and background from other \B decays. 
Accordingly the likelihood function of event $i$ is written:
\beq
\label{eq:eventlik}
\mathcal{L}_i= \mathcal{L}_{sig,i}+\mathcal{L}_{\qqbar,i}+\sum_c\mathcal{L}_{Bbg,c,i},
\eeq
where the sum runs over all the \B-background classes listed in ~\tabref{listBb}. All background likelihood functions have the same expression:
\beq
\label{eq:bkgLik}
\mathcal{L}_{back,i}=N_{back}\frac{1}{2}(1-q^K_iA_{back})\mathcal{P}_{back,i},
\eeq
where $q^K_i$ is the kaon charge in event $i$, and $A_{back}$ is the charge asymmetry.

We consider \Bz and \Bzb separately to build the signal likelihood
function.
\beqn 
\label{eq:signalLik}
N_{sig}=N^+_{sig} + N^-_{sig}. \\
\mathcal{L}_{sig,i}=\mathcal{L}^+_{sig,i}+ \mathcal{L}^-_{sig,i}.
\eeqn
Each part
has two terms, one for the TM and one for the SCF events:
\beqn
\label{eq:signalLikBz}
\left.
\begin{array}{lll}
\mathcal{L}^+_{sig,i}&\multicolumn{2}{l}{= \mathcal{L}^+_{TM,i}+\mathcal{L}^+_{SCF,i}} \\
                   &=N^+_{sig}[\ &(1-\overline{f}_{SCF})\mathcal{P}^+_{TM,i}  \\
                   &        &+\overline{f}_{SCF}\mathcal{P}^+_{SCF,i}\ ], \\
\end{array}
\right.
\eeqn
and similarly for $\mathcal{L}^-_{sig,i}$.  
$\overline{f}_{SCF}$, the fraction of SCF-events averaged over the
Dalitz plot, assumed to be the same for both flavors, is discussed below.
We denote by $N_{comp}$ the expected number of events for species {\it comp}.
The fit maximizes the extended likelihood
function\footnote{The canceling factors $1/N_{tot}$ in \equtworef{bkgLik}{signalLik}, and $N_{tot}^N$ in \equaref{extendedLik} required for the 
likelihood functions to be properly normalized have been omitted for simplicity.}:
\beq
\label{eq:extendedLik}
\mathcal{L}= e^{-N_{tot}}\prod_{i=1}^{N} \mathcal{L}_i,
\eeq
where $N=8014$ events is the size of the data sample and $N_{tot}=N_{sig}+N_{\qqbar}+\sum_cN_{Bbg,c}$, is the expected number of events.\par
The five-dimensional probability density functions (PDF) $\mathcal{P}$ are the products of the four PDFs of the measured discriminating variables 
$v=\{\mes, \deprime, NN, (\mprime,\thetaprime)\}$,
\beq
\label{eq:PDFproduct}
\mathcal{P}=\prod_{k=1}^4 \mathcal{P}(v_k).
\eeq
The correlations among the measurements are handled by building
conditional PDFs where appropriate (between NN and the Dalitz
variables for the continuum, and between \deprime\ and the Dalitz variables
for TM signal events). Systematic uncertainties account for the
correlations we neglect.\par
A total of 37 parameters are varied in the fit (see Section V.D). 
A summary of the PDF parametizations is given in ~\tabref{PDF}.
\begin{table*}
\begin{center}
\caption{\label{tab:PDF} Summary of the PDF parameterizations. G=Gaussian, P1=1st order polynomial, NP=non-parametric, and CB=Crystal Ball. 
The notation GG(DP) designates a double Gaussian PDF with parameters which vary over the Dalitz plot. The Dalitz plot signal model is described 
in~\secref{DecayAmplitudes}. The numbers associated with the \B background in the last row are the class indices in~\tabref{listBb}.}
\setlength{\tabcolsep}{1.2pc}
\begin{tabular}{ccccc}\hline\hline
Component    & $\mes$   & $\deprime$ & NN       & Dalitz \\ \hline
signal (TM)  &CB      & GG(DP)       & NP       & see text \\
signal (SCF) &NP      & G            & NP       & see text \\
Continuum    &Argus   & P1           & see text & NP in patches \\ 
\B background (non-$D$)&\multicolumn{2}{c}{two-dim. NP}  &NP&  NP       \\
\B background (3 and 9)&signal-like  & signal-like&   one NP for both  & see text \\ 
\hline\hline
\end{tabular}
\end{center}
\end{table*}  
\subsection{THE DALITZ PROBABILITY DENSITY FUNCTIONS}
\label{subsec:DalitzPDF}
Since the decay $\Bz\to\Kp\pim\piz$ is flavor-specific (the charge of the kaon identifies the $b$ flavor), the \Bz and \Bzb Dalitz plots are independent.
However, because the backgrounds are essentially flavor blind, we get a more robust procedure by fitting them simultaneously. 
It is enough to describe only the \Bz\ Dalitz plot PDF\footnote{We
  drop the superscript in $\mathcal{P}^+$ in the following.}. A change from $\mathcal{A}$ to $\mathcal{\overline{A}}$ 
(\equtworef{isobarB}{isobarBbar}) gives the \Bzb\ PDF.
\subsubsection{Signal}
The signal Dalitz model has been described in \secref{DecayAmplitudes}. The free parameters are the magnitudes and phases defined
in~\equtworef{isobarB}{isobarBbar} for all the intermediate states of the signal model given in~\tabref{nominal}. 
Since the measurement is done relative to 
the $\Kstarp(892)\pim$ final state, the phases of this and the charge conjugate channels are fixed to zero. The amplitude of $\Bz\to\Kstarp(892)\pim$ 
is also fixed 
but not that of $\Bzb\to\Kstarm(892)\pip$ in order to be sensitive to direct $CP$-violation. \par

The normalization of the component signal PDFs:
\beq
\label{eq:unnormalizedTMDalitzPDFs}
\mathcal{P}_{TM,i}  \propto \varepsilon_i (1-f_{SCF,i})|det\mathcal{J}_i||\mathcal{A}_i|^2,
\eeq
\beq
\label{eq:unnormalizedSCFDalitzPDFs}
\mathcal{P}_{SCF,i} \propto \varepsilon_i   f_{SCF,i}  [|det\mathcal{J}||\mathcal{A}|^2 \otimes R_{SCF}]_i,
\eeq
is model dependent. $\mathcal{J}$ is the Jacobian matrix of the mapping to the square Dalitz plot. 
The symbol $\otimes$ stands for a convolution and the $R$ matrix is described below [\equaref{theMatrix}]. 
The normalization requires the computation of the integrals 
\beqn
\label{eq:partialDPnorm}
\int_0^1d\mprime\int_0^1d\thetaprime\  \varepsilon (1-f_{SCF})|det\mathcal{J}|f_kf_l^*,\\
\int_0^1d\mprime\int_0^1d\thetaprime\  \varepsilon f_{SCF}|det\mathcal{J}|f_kf_l^*,
\eeqn
and
\beq
\label{eq:fullDPnorm}
\int_0^1d\mprime\int_0^1d\thetaprime\  \varepsilon|det\mathcal{J}|f_kf_l^*,
\eeq
where the notations of ~\equaref{isobarB} are used. 
The integrations over the square Dalitz plot are performed numerically. 
The weight
\beq
\label{eq:fscfbar}
\overline{f}_{SCF}= 
\frac{\int_0^1d\mprime\int_0^1d\thetaprime\  \varepsilon f_{SCF}|det\mathcal{J}||\mathcal{A}|^2}
     {\int_0^1d\mprime\int_0^1d\thetaprime\  \varepsilon|det\mathcal{J}||\mathcal{A}|^2}
\eeq
in ~\equaref{signalLik} ensures that the total signal PDF is normalized. The PDF normalization depends on the decay dynamics and is computed iteratively.
In practice the computation of $\overline{f}_{SCF}$ rapidly converges to a value which we fix after a few exploratory fits.  \par
Studies in simulation have shown that the experimental resolutions of \mprime\ and \thetaprime\ need not be introduced in the TM signal PDF. On the contrary, 
misreconstructed events often incur large migrations, when the reconstructed $\mprime_r, \thetaprime_r$ are far from the true values $\mprime_t, \thetaprime_t$. 
We use the Monte Carlo simulation to compute a normalized
two-dimensional resolution function
$R_{SCF}(\mprime_r,\thetaprime_r;\mprime_t,\thetaprime_t)$, with
\beq
\label{eq:theMatrix}
\int_0^1d\mprime_r\int_0^1d\thetaprime_r R_{SCF}(\mprime_r,\thetaprime_r;\mprime_t,\thetaprime_t)=1. 
\eeq 
$R_{SCF}$ is convolved with the signal model in the expression of $\mathcal{P}_{SCF}$ [\equaref{unnormalizedSCFDalitzPDFs}].
\subsubsection{Background}
Except for events coming from exclusive $\B\to D$ decays, all background Dalitz PDF are non-parametric smoothed histograms. 
The continuum distributions are extracted from a combination of off resonance data and a sideband ($5.20<\mes<5.25\gevcc$) of the on resonance 
data from which the \B-background has been subtracted. 
The square Dalitz plot is divided into three regions where different smoothing parameters are applied in order to optimally reproduce the observed wide and 
narrow structures by using a two-dimensional kernel estimation technique~\cite{cranmer}. 
For $\mprime>0.8$ and all $\thetaprime$, a fine grained model is needed to  follow the peak from $\Dz$ continuum production. 
For $\mprime<0.8$ there are two different wide structures corresponding to the continuum production of the $\rho$'s for $\thetaprime$ below 0.8 and the
charged $\Kstar$'s above.\par
The \B-background (\tabref{listBb}) Dalitz PDFs are obtained from the Monte Carlo simulation. For the components which model $b\to c$ decays with 
real $\Dz$ mesons, a fine grained binning around the $D$ mass is used to construct the smoothed histograms.
The exclusive signal-like components $\Bz\to\Dzb\piz$, $\Dzb\to\Kp\pim$ and $\Bz\to\Dm\Kp$, $\Dm\to\pim\piz$ are modeled
with TM-like and SCF-like PDFs. The former are products of a Gaussian distribution in $m_{K\pi}$ (then transformed into \mprime) and a fifth order polynomial 
in \thetaprime. 
The \Dz\ Gaussian mass and widths are free parameters in the fit, the TM-like \Dz\ helicity polynomial
coefficients were obtained via an ancillary fit to data, all other parameters are fixed to their value in the Monte Carlo.
For the $\Bz\to\Dm\Kp$ where $\Dm\to\pim\piz$, the Dalitz Plot PDF is a two-dimensional 
smoothed histogram obtained from the Monte Carlo. 
\subsection{THE OTHER PDFS}
\label{subsec:nondalitzPDFs}
\subsubsection{Signal}
The \mes PDF for TM-signal events is a Crystal Ball function~\cite{XtalBall}, the mean and the width of which are free parameters in the fit. 
For SCF-signal events we use a non-parametric shape taken from the Monte Carlo simulation. \par
\deprime\ is correlated with the Dalitz plot variables for TM-signal events. To account for the correlation, we choose a double Gaussian PDF 
the mean and standard deviation of which vary linearly with $m_{\Kp\pim}^2$. These parameters (intercept and slope) are taken from the Monte Carlo except 
for a global mean of the double Gaussian distribution. A wide single Gaussian distribution with mean and standard deviation taken from the Monte Carlo is used for the 
SCF-signal \deprime\ PDF. \par
The NN PDFs for TM and SCF events are non-parametric distributions taken from the Monte Carlo. 
\subsubsection{Background}
We use the Argus function~\cite{Argus}
\beq
\label{eq:argus}
f(z=\frac{\mes}{m_{\rm ES}^{\rm max}})\propto z\sqrt{1-z^2} e^{-\xi(1-z^2)}
\eeq 
as the continuum \mes\ PDF. The endpoint $m_{\rm ES}^{\rm max}$ and the shape parameter $\xi$ are fixed to $5.2897\gevcc$ and $-22$ 
respectively ($\xi=-22\pm7$ from a fit to off resonance data). 
The \deprime\ PDF is a linear polynomial whose slope is free to vary in the fit. The shape of the NN distribution for 
continuum is correlated with the event location in the Dalitz plot. To account for that effect we use for the NN PDF a function that varies
with the closest distance $\Delta_{\mathrm{dalitz}}$ between the point representing the event and the boundary of the standard Dalitz plot,
\beqn
    \label{eq:NNDP}
    \mathcal {P} (NN ;\Delta_{\mathrm{dalitz}}) = &&(1-NN)^A \\
                                                  &&\times (B_0 + B_1 NN + B_2 NN^2). \nonumber
\eeqn
The $A$ and $B$ coefficients are linear functions of
$\Delta_{\mathrm{dalitz}}$. Their expressions in terms of the
$a_i$ parameters that are varied in the likelihood fit are:
\begin{eqnarray}
\label{eq:nndeltadalitz}
  A&=&a_1+a_4 \Delta_{\mathrm{dalitz}}, \\
  B_0&=&c_0+c_1\Delta_{\mathrm{dalitz}}, \nonumber \\
  B_1&=&a_3+c_2\Delta_{\mathrm{dalitz}}, \nonumber \\
  B_2&=&a_2+ c_3\Delta_{\mathrm{dalitz}}. \nonumber 
\end{eqnarray} 
We use two-dimensional non-parametric distributions to describe the \B-background classes in the (\mes, \deprime) plane, except for the exclusive 
\B decays to $D$ mesons (classes 3 and 9) in~\tabref{listBb} for which we use the same shapes as for the signal.
For each \B-background class the NN PDF is a non-parametric distribution taken from the Monte Carlo. Classes 3 and 9 share the same NN PDF.
\subsection{THE FIT PARAMETERS}
\label{subsec:fitparam}
The following parameters are varied in the fit:
\begin{itemize}
\item Five yields for signal ($N_{sig})$, continuum ($N_{\qqbar}$) and  three \B\ background classes (c=3, 7 and 8 defined in~\tabref{listBb}).
\item One $CP$-asymmetry for the continuum events.
\item Four parameters related to narrow particle masses: the mass and mass resolution for the \Bz and the \Dz. 
\item Two parameters, the global mean (slope), of the \deprime\ distribution for the TM-signal (continuum) events.  
\item Four parameters which account for the correlation between the NN output and the event location in the Dalitz plot [\equaref{nndeltadalitz}]. 
\item Twenty-one isobar amplitudes and phases. There are 6 intermediate states (5 resonances and a nonresonant term) and two Dalitz plots. We fix
      one reference amplitude, that of $\Bz\to\Kstarp(892)\pim$ and two phases for the latter and its conjugate. 
      Therefore we end up with 11 magnitudes and 10 phases to be determined by the fit. 
\end{itemize}
\begin{figure}[!h]
\epsfig{file=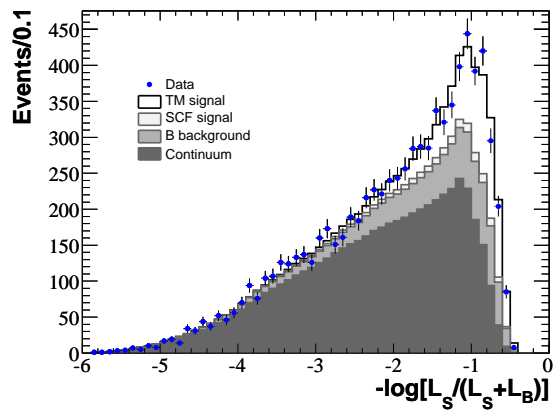,width=8.5cm}
\caption{\label{fig:rlik}
Ratio of the TM term $\mathcal{L}_{TM,i}$ [first relevant term in~\equaref{signalLik}] over the full likelihood function $\mathcal{L}_i$ 
[\equaref{eventlik}] for all events. 
The data are shown as points with error bars. The solid histogram
shows the projection of the fit result. The dark and grey shaded areas
represent the continuum and \B background, respectively. The light grey region shows the SCF contribution.}
\end{figure} 

\section{RESULTS}
\label{sec:Results}
\begin{figure}[!t]
\epsfig{file=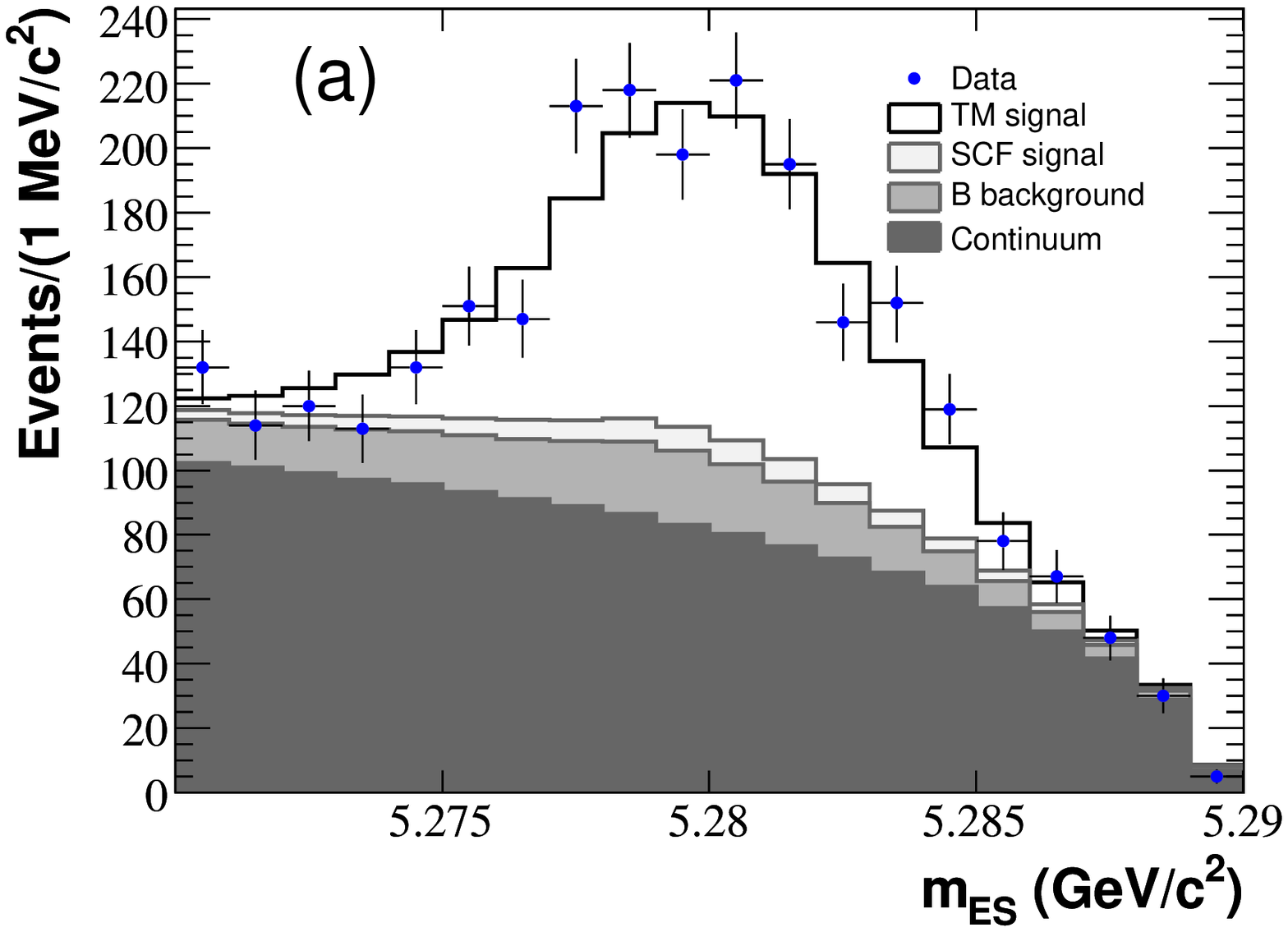,width=8.5cm}
\epsfig{file=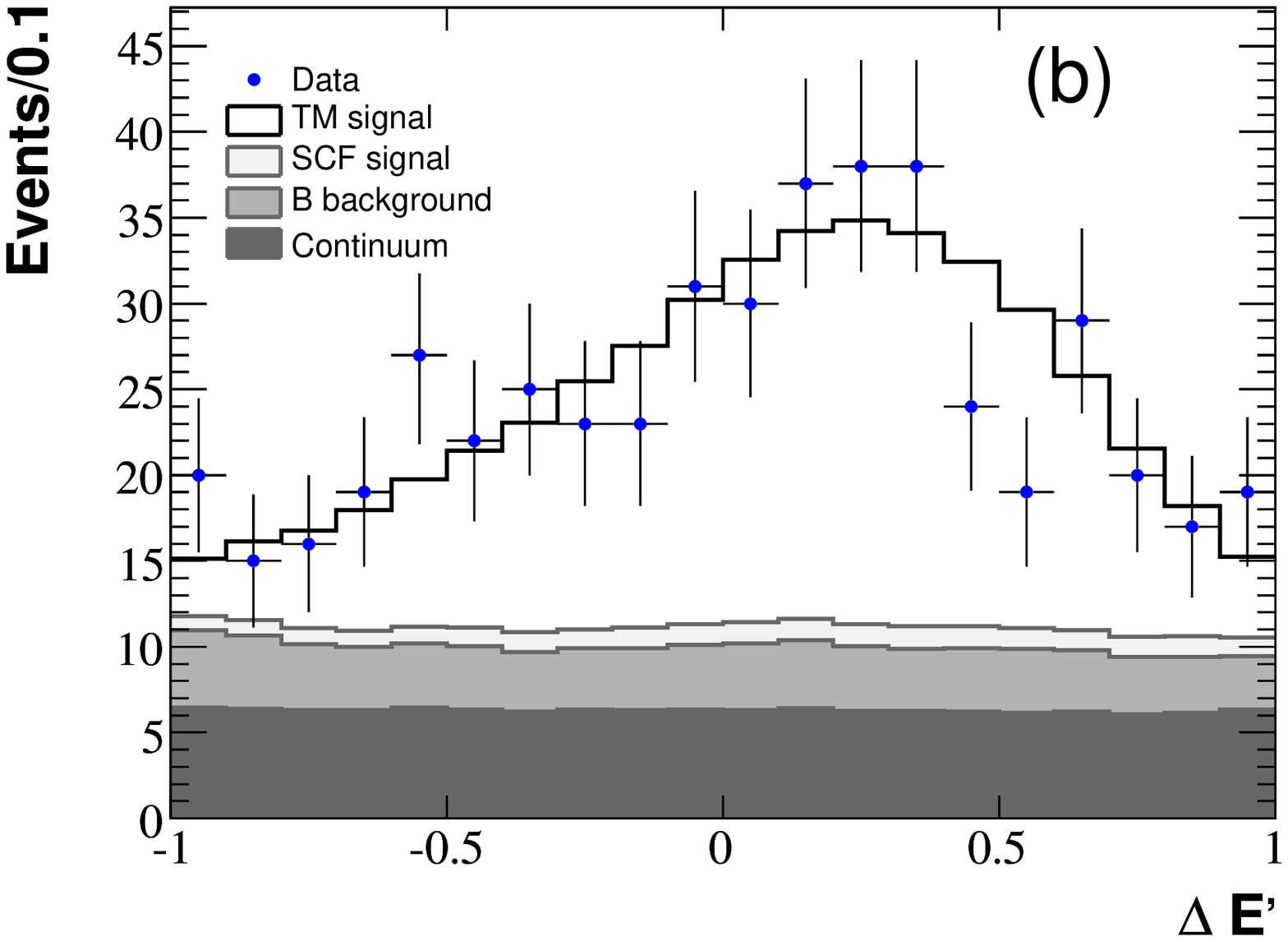,width=8.5cm}
\epsfig{file=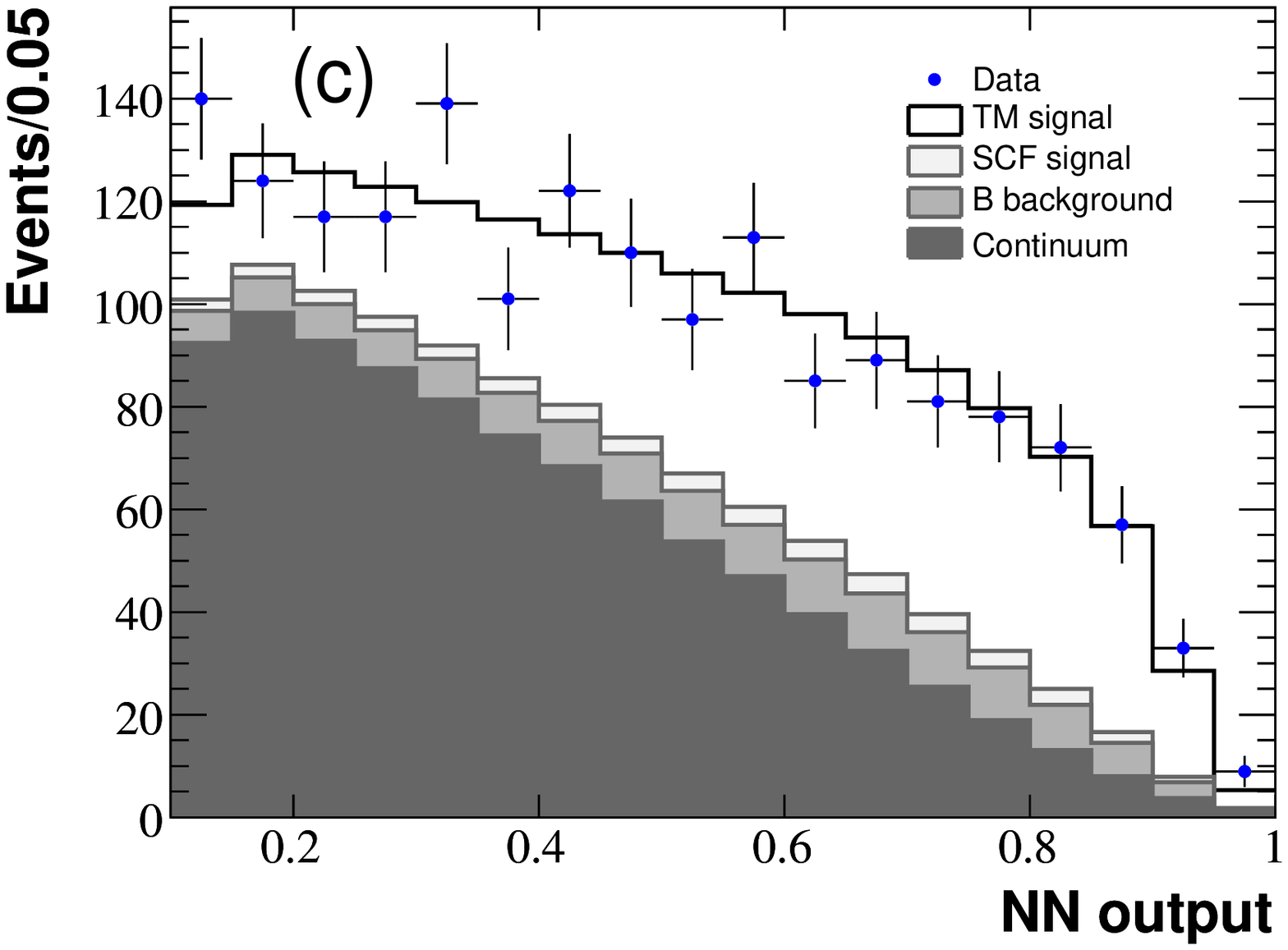,width=8.5cm} 
\caption{\label{fig:mesdeprimeNN}
(a) \mes, (b) \deprime\ and (c) NN distributions after restricting
  $r_{lik}$ to enhance signal. The fractions of accepted signal 
  and rejected continuum events are respectively 
  65\% and 71\% (a), 21\% and 97\% (b), and
  49\% and 83\% (c). The data are shown as points with error bars.  
The solid histogram shows the projection of the fit result. The dark
and grey shaded areas represent the continuum and \B background, respectively. The light grey region shows the SCF contribution.}
\end{figure} 
The maximum likelihood fit results in a $\Bz(\Bzb) \to \Kpm\pimp\piz$ event yield of $N_{sig}=1377\pm70$ events, where the uncertainty is statistical only. 
There are $5395\pm104$ continuum events and $424\pm25$ exclusive
$\B\to\Dzb\piz$ decays where $\Dzb\to\Kp\pim$, (not included in~$N_{sig}$). The remaining 833~\B background events are
detailed in \tabref{listBb}. \par
When the fit is repeated starting from input parameter values randomly chosen within wide ranges of one order of magnitude above and below the nominal values for the amplitudes 
and within the [$0,\ 2\pi$] interval for the phases, we observe convergence toward four
solutions with minimum values of the negative loglikelihood function (NLL$\equiv-\log(\mathcal{L})$) that are equal within 0.2 units. There are four degenerate solutions.
The event yields we quote are the averages of the four solutions for which the relative spreads are less than 1\%. 
Monte Carlo ($\mprime,\ \thetaprime$) scatter plots generated according to the fitted signal model in the four solutions are barely distinguishable. 
We checked with simulated datasets that multiple quasi degenerate solutions are to be expected when applying our fitting procedure 
to samples containing as many events as the experiment.\par 
In the Appendix, the fitted parameters are given for the four solutions in~\tabref{fourfitsolutions}, 
together with their correlation matrix (for solution-I) in~\tabtworef{correlationB-I}{correlationBbar-I}. 
The isobar fractions, which are not required to sum to one in order to accomodate interference, are not significantly different from unity. 
We observe that the fit fractions and the $CP$ asymmetries are
consistent within less than three standard deviations among the
solutions, and indeed 
agree much better for the subdecays to a pseudoscalar meson and a vector meson. The phases 
differ substantially. The four solutions are actually two solutions for the $\Bz$ phases and two solutions for the $\Bzb$ phases. The uncertainties and 
correlation coefficients given in these tables are statistical only. They are underestimated because the fitting program assumes a parabolic shape 
for the NLL in the vicinity of its minimum. This assumption overlooks the fit degeneracy. 
Before we explain how we derive consistent statistical and systematic uncertainties, 
we proceed to describe the goodness of the fit. We postpone the
discussion of the physical meaning of the fitted signal model
to~\secref{Interpretation}. \par
To check the validity of the fits and to study the results, we have generated 1000 pseudo experiments with as many events as in the data sample using
PDFs with the fitted parameter values. We check that the NLL of the experimental fit falls well within the NLL distribution in the pseudo experiments. 
The goodness of the fits and their ability to reproduce the real data are similar for all four solutions. We show the 
results of solution-I in the following. We compare the likelihood ratio $r_{lik}\equiv\frac{\mathcal{L}_{TM}}{\mathcal{L}}$ 
(see \equtworef{eventlik}{signalLikBz}) in the data and in the pseudo experiments and see good agreement (\figref{rlik}). 
The distributions of the discriminating variables (\mes, \deprime\ and NN) are shown in~\figref{mesdeprimeNN} for samples that have been enriched 
in signal events by restricting $r_{lik}$ 
(computed without the plotted variable) to large enough values in order to maximize the signal significance. \par
\begin{figure}[p]
\epsfig{file=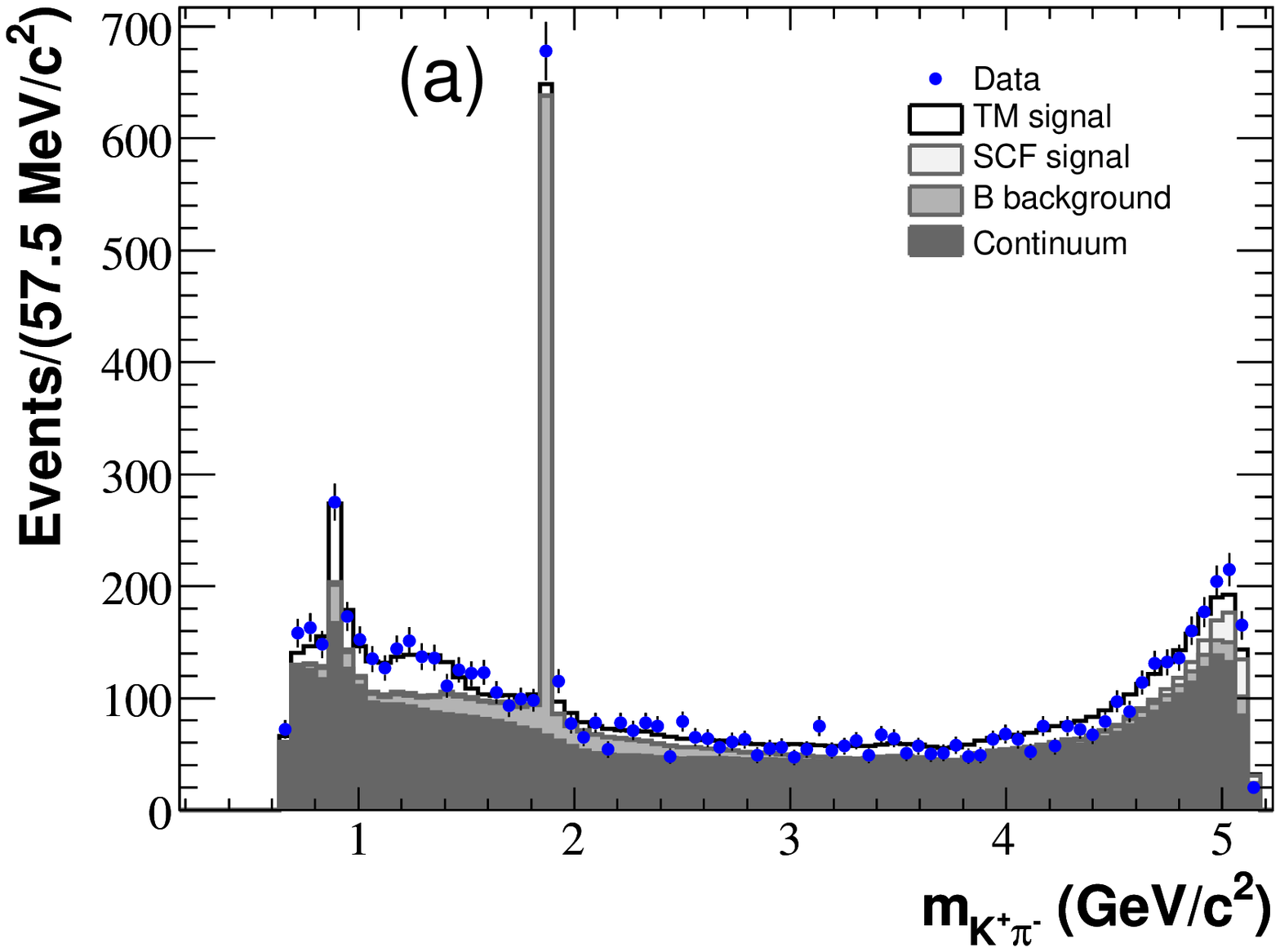,width=8.5cm}  
\epsfig{file=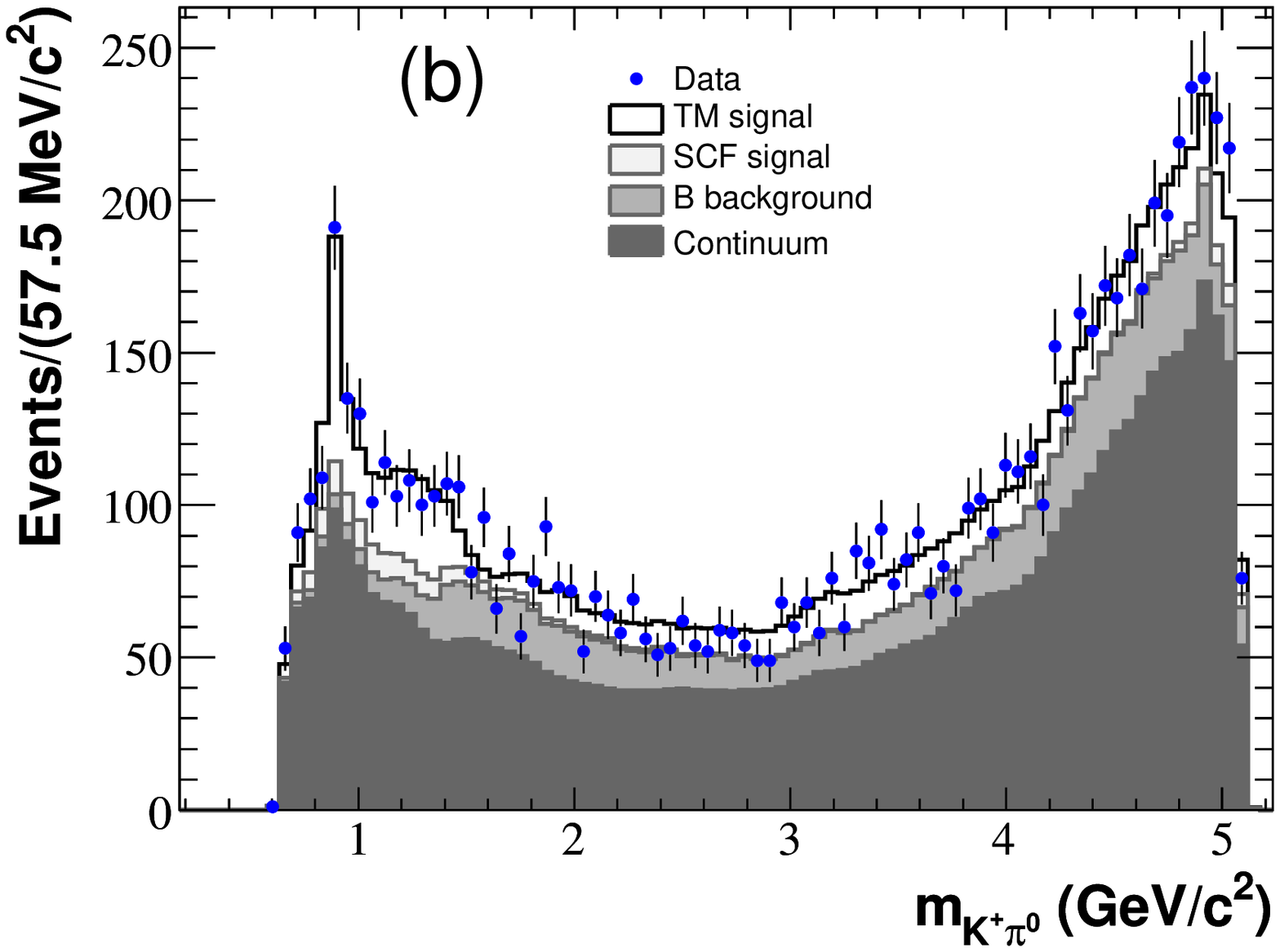,width=8.5cm}  
\epsfig{file=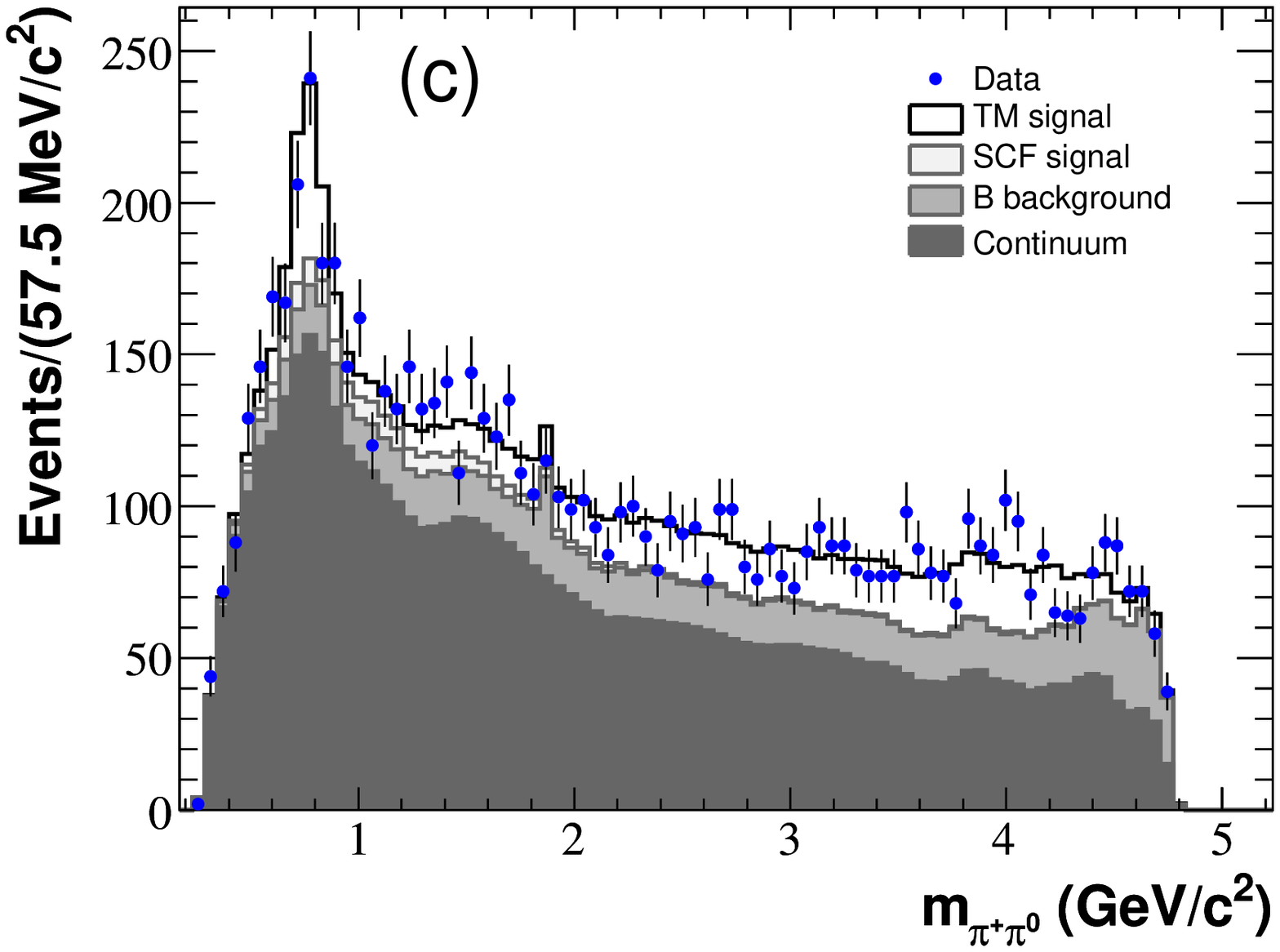,width=8.5cm}
\caption{\label{fig:DalitzMasses}
The invariant mass spectra for all events: $m_{\Kp\pim}$ (a), $m_{\Kp\piz}$ (b) and $m_{\pim\piz}$ (c).
The data are shown as points with error bars. 
The solid histogram shows the projection of the fit result. The dark
and grey shaded areas represent the continuum and \B background, respectively. 
The light grey region shows the SCF contribution.
The $\Dzb$ mass peak is prominent in the top plot.}
\end{figure} 
\begin{figure}[htb]
\epsfig{file=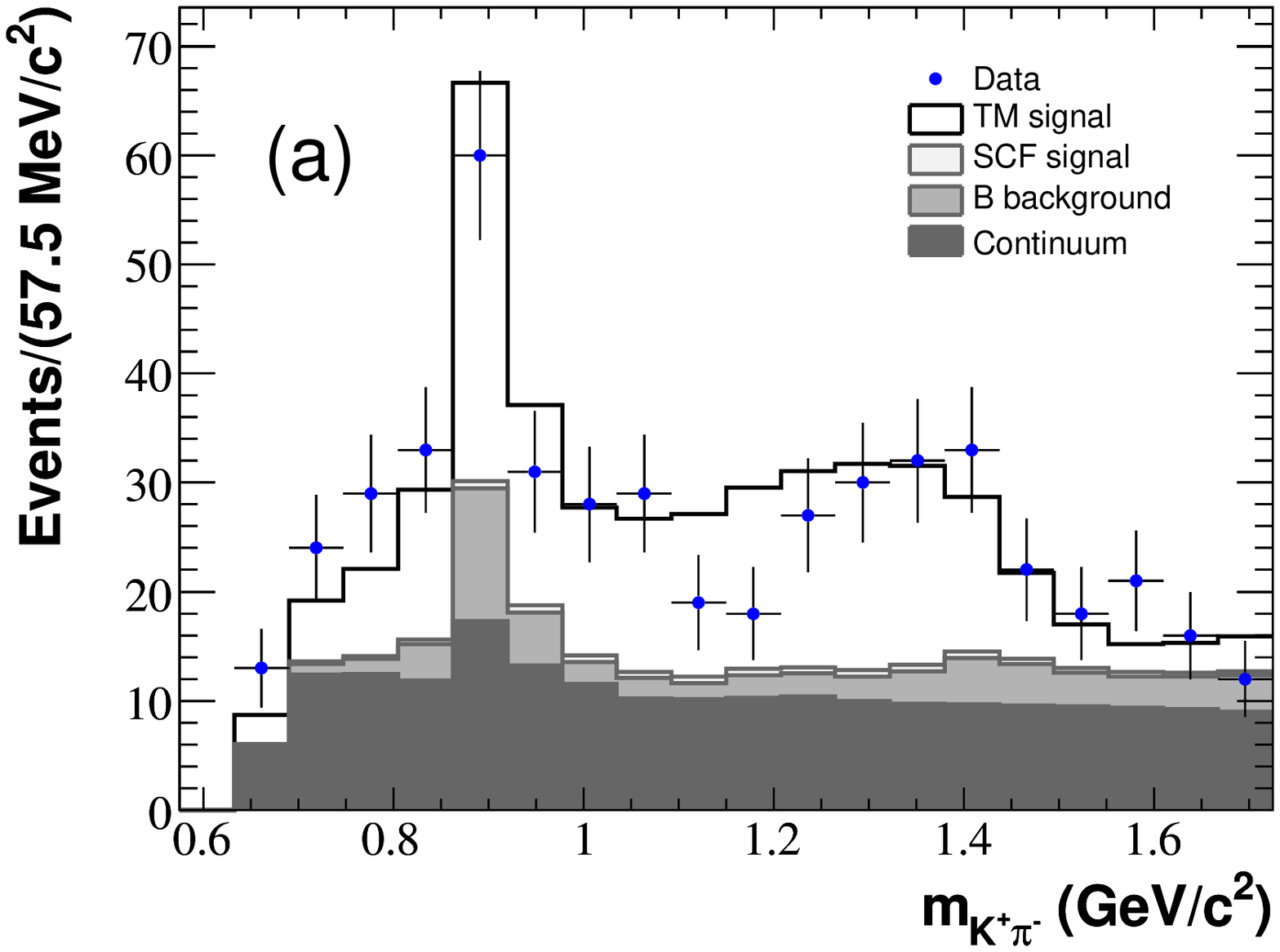,width=8.5cm}  
\epsfig{file=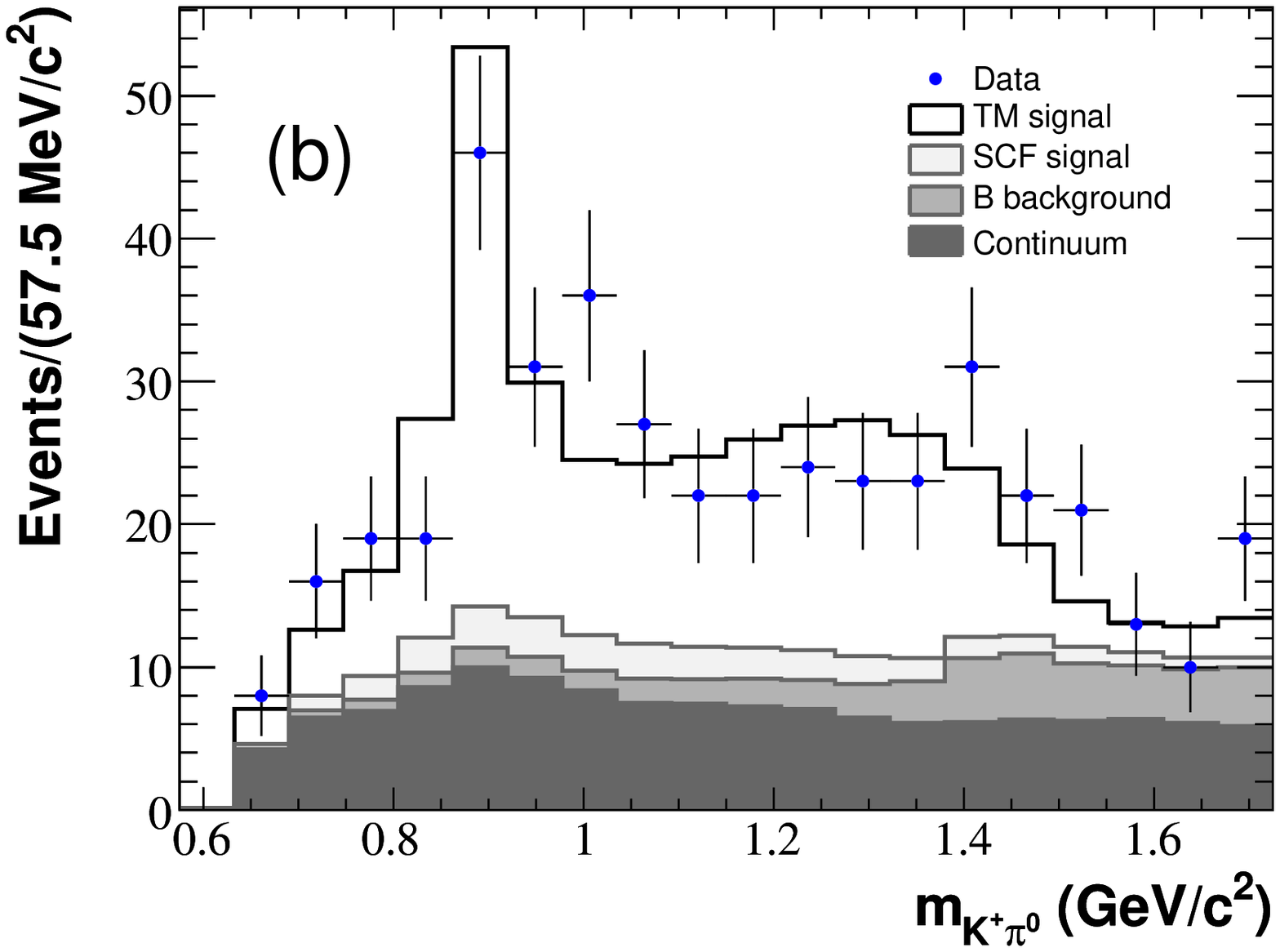,width=8.5cm}  
\epsfig{file=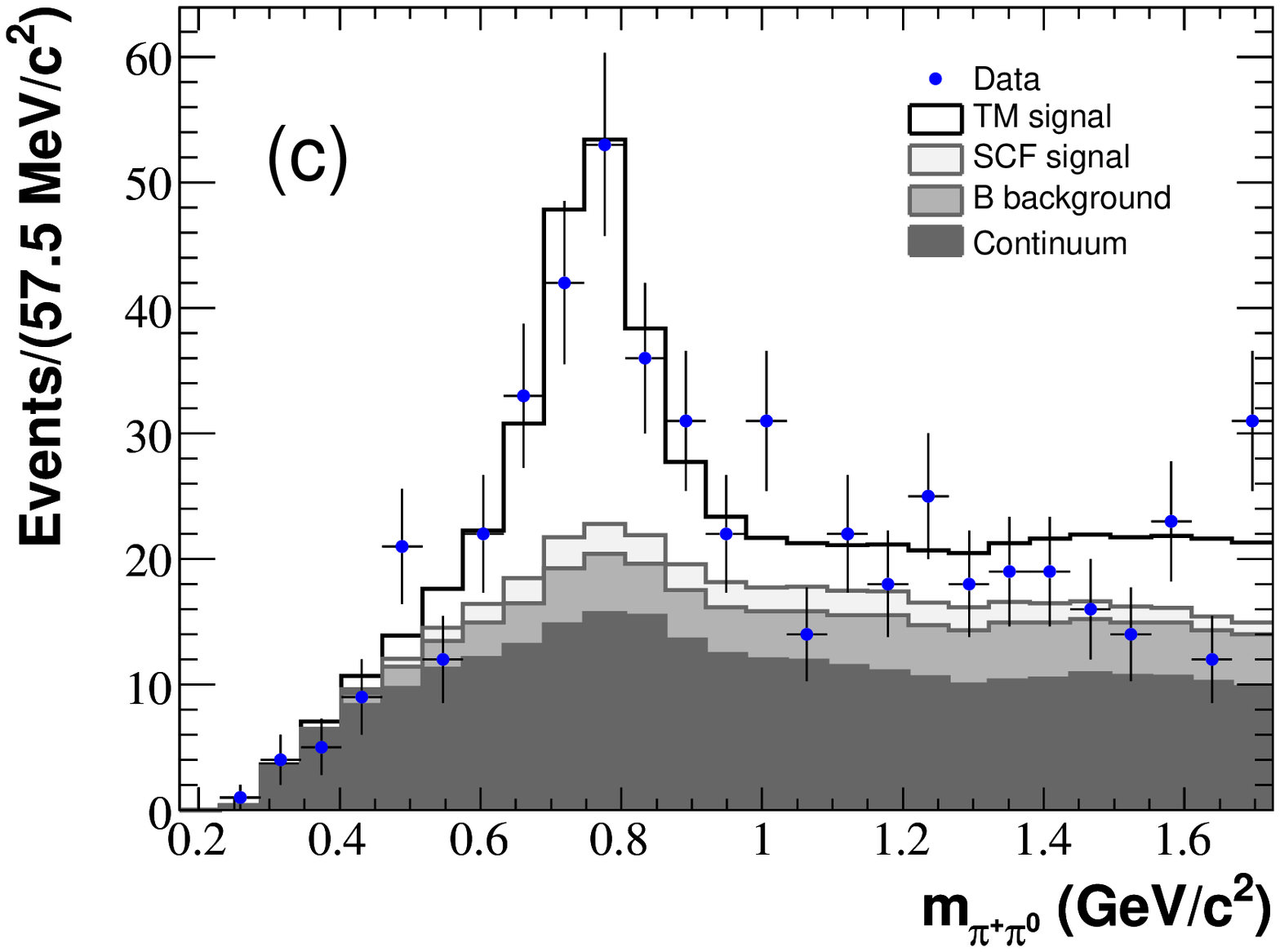,width=8.5cm}
\caption{\label{fig:DalitzMasseslowmass}
The signal-enriched spectra for masses between threshold and $1.75\gevcc$: $m_{\Kp\pim}$ (a), $m_{\Kp\piz}$  (b) and 
$m_{\pim\piz}$ (c). The fractions of accepted signal and rejected
continuum events are respectively 51\% and 89\%.
The data are shown as points with error bars. 
The solid histogram shows the projection of the fit result. The dark
and grey shaded areas represent the continuum and \B background, respectively. The 
light grey region shows the SCF contribution.}
\end{figure} 
\begin{figure}[p]
\epsfig{file=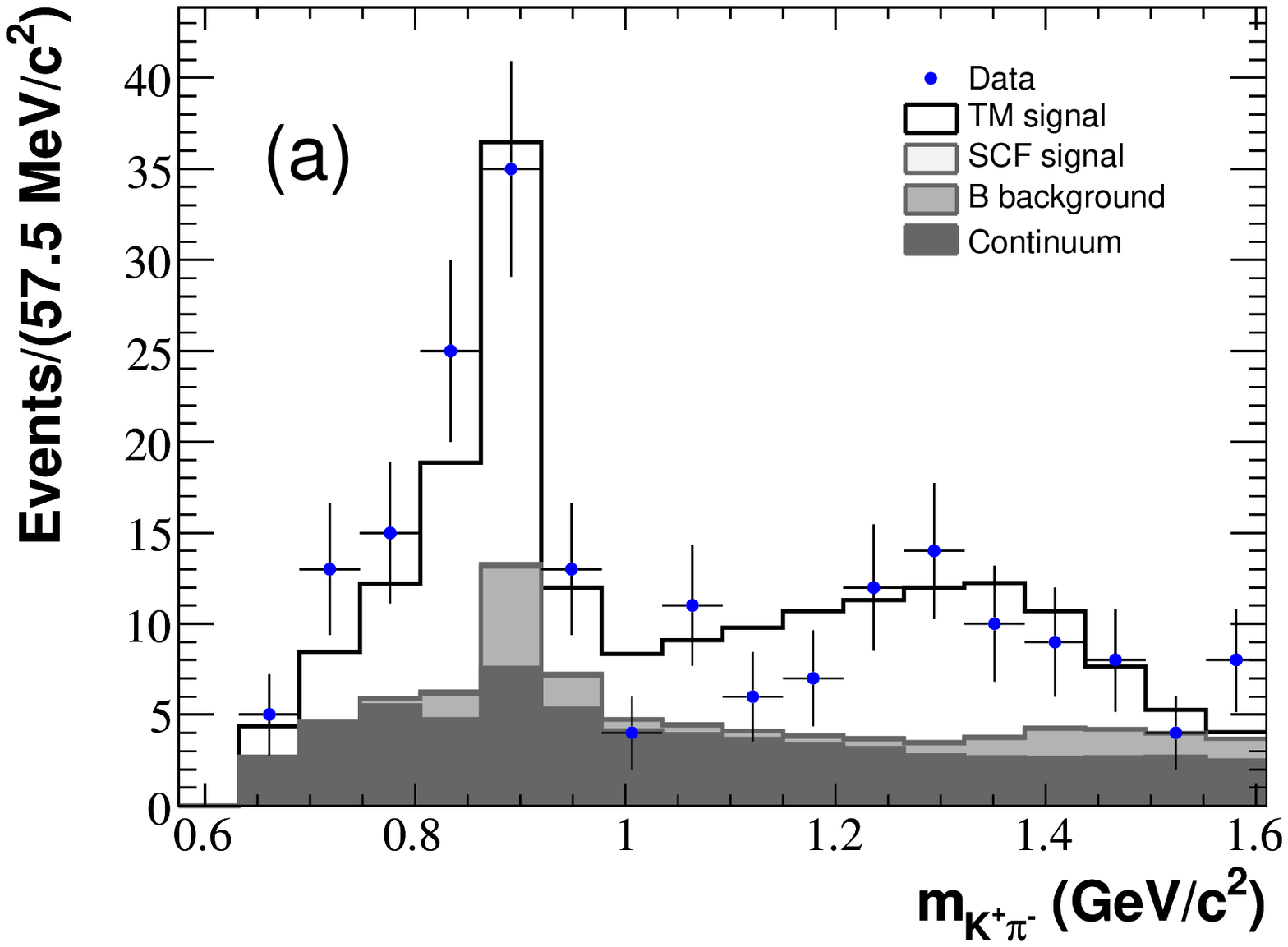,width=8.5cm}
\epsfig{file=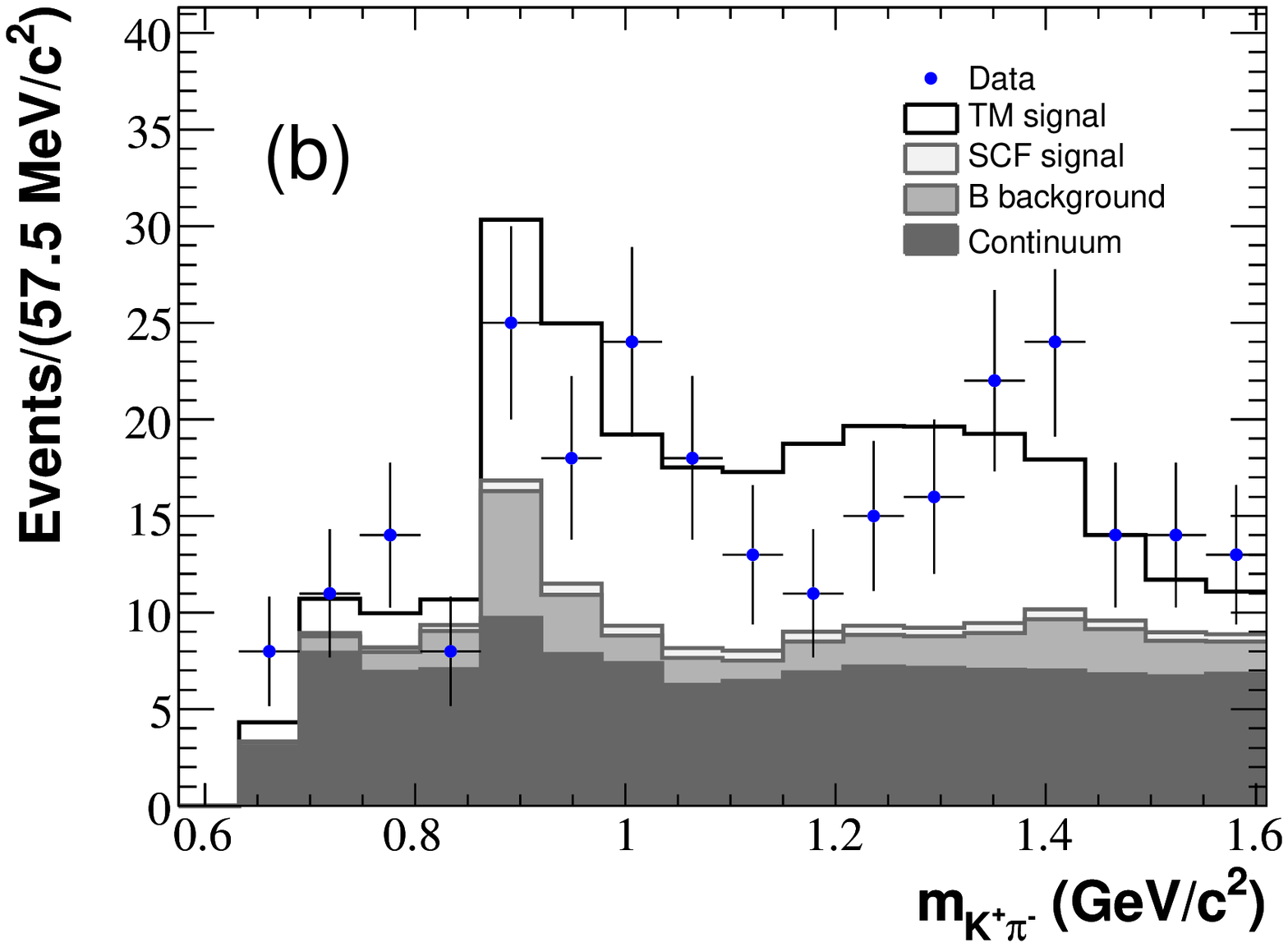,width=8.5cm}
\caption{\label{fig:KaPizoomhelicity} Signal-enriched spectra of $m_{\Kp\pim}$ in the low-mass resonance region and different ranges of the helicity angle, $\theta_{\Kp\pim}$.
(a) $0 <\theta_{\Kp\pim}<90$ degrees, 
(b) $90 <\theta_{\Kp\pim}<180$ degrees. The data sample is enriched in
  signal events as in \figref{DalitzMasseslowmass}.
An interference between the vector and scalar $\Kstarz$ is apparent through a positive forward-backward asymmetry below the $\Kstar(892)$ 
and a negative one above. 
The data are shown as points with error bars. 
The solid histogram shows the projection of the fit result. The dark
and grey shaded areas represent the continuum and \B background, respectively. 
The light grey region shows the SCF contribution.}
\end{figure} 
\begin{figure}[p]
\epsfig{file=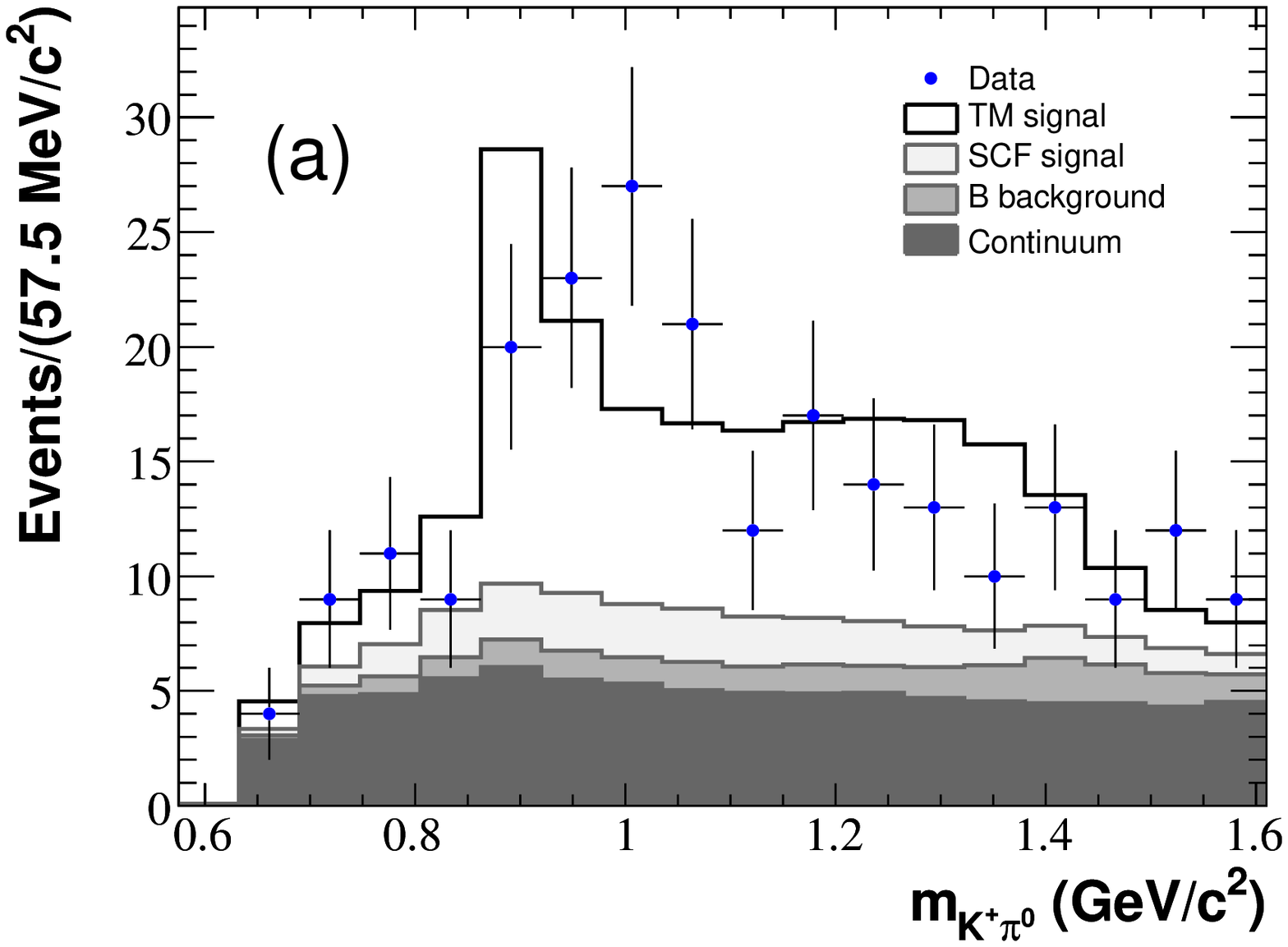,width=8.5cm}
\epsfig{file=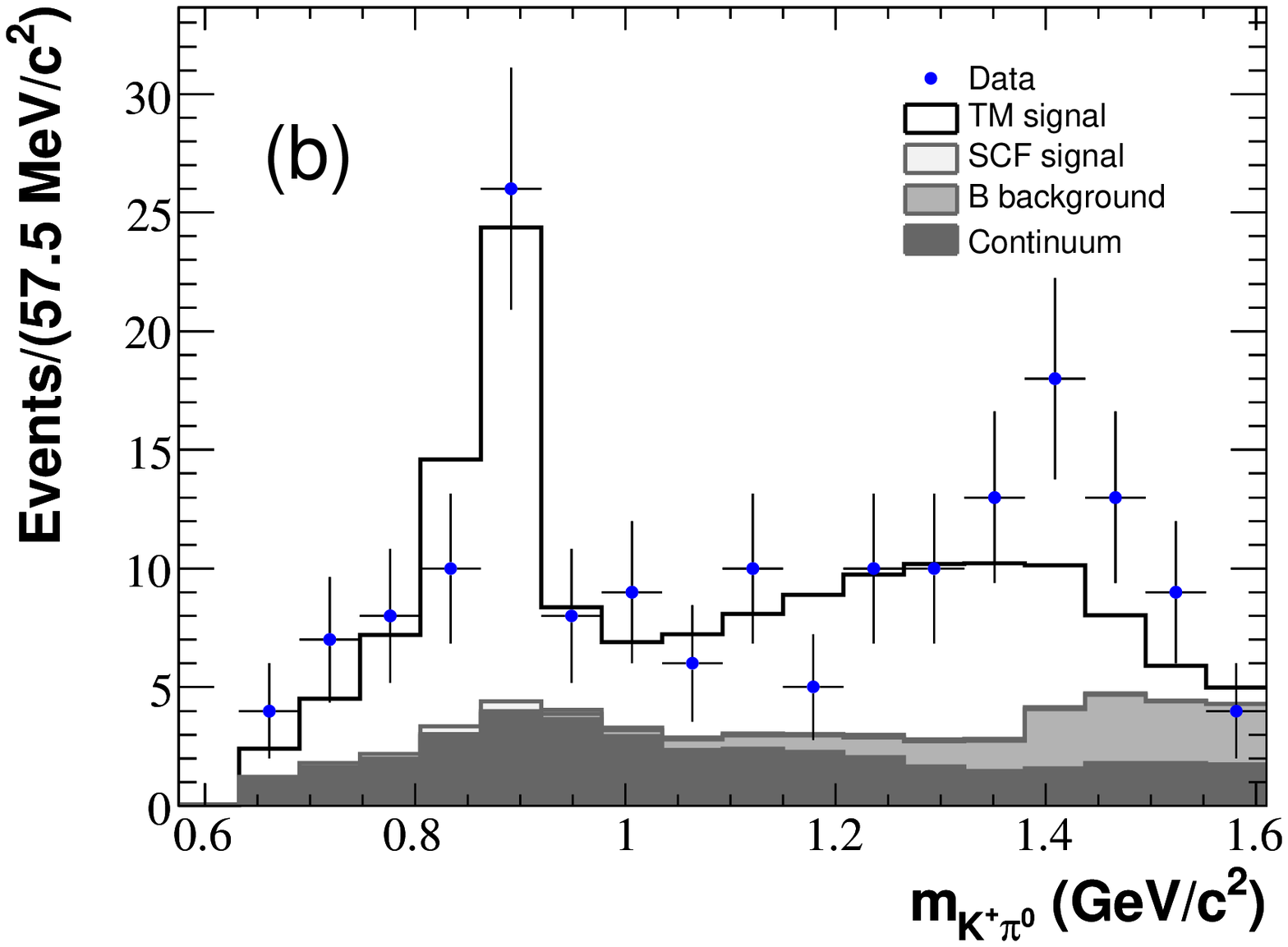,width=8.5cm}
\caption{\label{fig:KaP0zoomhelicity} Signal-enriched spectra of $m_{\Kp\piz}$ in the low-mass resonance region and different ranges of the helicity angle, $\theta_{\Kp\piz}$.
(a) $0 <\theta_{\Kp\piz}<90$ degrees, 
(b) $90 <\theta_{\Kp\piz}<180$ degrees. The data sample is enriched in
  signal events as in \figref{DalitzMasseslowmass}.
An interference between the vector and scalar $\Kstar$ is apparent through a negative forward-backward asymmetry below the $\Kstar(892)$ 
and a positive one above. 
The data are shown as points with error bars. 
The solid histogram shows the projection of the fit result. The dark
and grey shaded areas represent the continuum and \B background, respectively. 
The light grey region shows the SCF contribution.}
\end{figure} 
\figref{DalitzMasses} shows the Dalitz plot mass spectra over their full range with no restriction on $r_{lik}$.
A zoom of the low-mass resonance region (below $1.75\ \gevcc$) is shown for signal-enriched events in~\figref{DalitzMasseslowmass}. More details are shown 
in~\figref{KaPizoomhelicity} (\figref{KaP0zoomhelicity}) 
which display $m_{\Kp\pim}$, ($m_{\Kp\piz}$) for different intervals of the helicity angles $\theta_{\Kp\pim}$, ($\theta_{\Kp\piz}$). 
The interference between the scalar and vector $\Kstar$ is evident from the opposite sign of 
the forward-backward helicity asymmetries below and above the $\Kstar(892)$. This effect is seen with sufficient statistics in these plots which include
both $\Bz$ and $\Bzb$ decays, because the measured phase differences are similar in both cases (\tabref{fourfitsolutions}). \clearpage %\par 
We take into account all four solutions of the fit to quote the final results of the analysis. 
To determine the statistical uncertainty on a physical parameter $p$,
we fix it at successive values spanning its range of interest and
repeat the fit, maximizing the projected likelihood function
(\equaref{extendedLik} where $p$ is frozen). 
NLL($p$), the minimum where the fit converges given $p$, is not always associated with the same solution. Therefore the NLL($p$) envelope curve is far from a 
parabola. The flatness of its shape around the overall minimum reflects the fit degeneracy.  
The parameter values at which $\Delta\chi^2\equiv2\,(NLL-NLL_{bestfit})$ reaches unity bound the one-standard deviation confidence interval. 
The scan of the $\Kstarz(892)\piz$ isobar fraction is shown in~\figref{BR_Kstar0Pi0}. The graph also shows the envelope curve obtained when  
the systematic uncertainty is included. The null value of the branching 
fraction is excluded with a statistical and systematic significance of 5.6 standard deviations. Thus, this is the first observation of the 
decay $\Bz\to\Kstarz(892)\piz$. Other scans are presented in the~Appendix. 
%\clearpage
\section{SYSTEMATIC UNCERTAINTIES}
\label{sec:Systematics}
\begin{figure}[t]
\epsfig{file=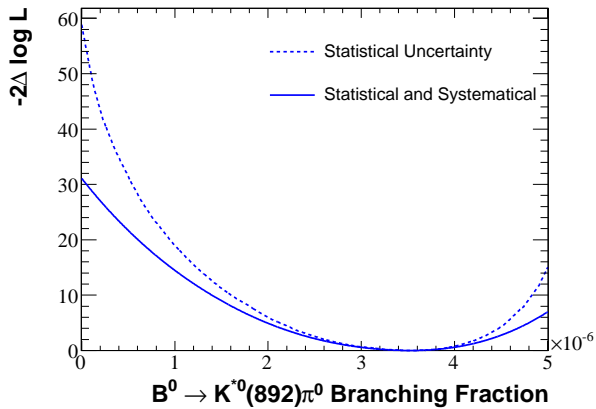,width=8.5cm}
\caption{\label{fig:BR_Kstar0Pi0} The NLL close to its minimum as a function of the $\Kstarz(892)\piz$ branching fraction. The shapes of the curves are 
non-parabolic and shallow because the plotted $\Delta\chi^2$ is the smallest of those from all solutions. The dashed line scan corresponds to the statistical 
uncertainty. The full line is the scan after smearing by the systematic uncertainties described in~\secref{Systematics}. 
}
\end{figure} 
\begin{table*}[p]
\begin{center}
\caption{\label{tab:Systematics}
Summary of the systematic uncertainties. 
The systematic uncertainties are equal for the phases $\phi$ and $\overline{\phi}$. All phases $\phi$ are referenced to $\phi_{\Kstarp(892)\pim}$ and 
$\overline{\phi}$ to $\phi_{\Kstarm(892)\pip}$. The systematic uncertainties associated with the Dalitz plot model and the lineshapes
are negligible for the phases of the $K\pi$ S-waves which are the only phase angles that are measured with some statistical accuracy. For the others which are essentially 
undetermined by the fit we do not quote these {\it undefined} systematic uncertainties. 
}
\setlength{\tabcolsep}{1.2pc}
\begin{tabular}{ccccc}
\hline\hline
                          &                      & Fraction (\%)        & $A_{CP}$          & $\phi$ (deg.)       \\\hline
$\Kstarp(892)\pim$        &  Dalitz Plot Model   &$\pm0.32$             & $\pm0.03$         &$-$                  \\
                          &  Shape Parameters    &$^{+0.31}_{-0.34}$    & $^{+0.01}_{-0.02}$&$-$                  \\ 
                          &  $\B$ Background     &$^{+0.06}_{-0.14}$    & $\pm0.02$         &$-$                  \\
                          &  Lineshapes          &$^{+0.06}_{-0.04}$    & $\pm0.001$        &$-$                  \\
                          &  Continuum DP PDF     &$\pm0.31$             &                   &$-$                 \\
                          &  Fit Bias            &$\pm0.23$             & $\pm0.01$         &$-$                  \\ 
                          &  {\bf Total}         &$^{+0.59}_{-0.62}$    & $\pm0.04$         &$-$                  \\ \hline

$\Kstarz(892)\piz$        &  Dalitz Plot Model   &$\pm0.11$             & $\pm0.04$         &undefined                 \\
                          &  Shape Parameters    &$^{+0.30}_{-0.20}$    & $^{+0.03}_{-0.02}$&                     \\
                      &  $\B$ Background         &$^{+0.19}_{-0.15}$    & $^{+0.07}_{-0.08}$&                     \\ 
                          &  Lineshapes          &$^{+0.01}_{-0.02}$    & $\pm0.001$        &undefined                 \\
                          &  Continuum DP PDF     &$\pm0.49$             &                   &$\pm\ 6.3$          \\
                          &  Fit Bias            &$\pm0.24$             & $\pm0.01$         &$\pm16.5$            \\ 
                          &  Other               &                      &                   &$\pm\ 4.1$           \\ 
                          &  {\bf Total}         &$^{+0.66}_{-0.61}$    & $\pm0.09$         &$\pm18.2$            \\ \hline

$(K\pi)^{*+}_0\pim$       &  Dalitz Plot Model   &$\pm1.56$             & $\pm0.07$      &                 \\
                          &  Shape Parameters    &$^{+1.14}_{-0.91}$    & $\pm0.01$         &                     \\ 
                      &  $\B$ Background         &$^{+0.32}_{-0.28}$    & $\pm0.05$         &                     \\ 
                          &  Lineshapes          &$^{+0.02}_{-0.04}$    & $\pm0.000$        &                 \\
                          &  Continuum DP PDF     &$\pm0.82$             &                   &$\pm\ 2.6$           \\
                          &  Fit Bias            &$\pm0.10$             & $\pm0.01$         &$\pm\ 9.6$           \\ 
                          &  Other               &                      &                   &$\pm\ 1.5$           \\
                          &  {\bf Total}         &$^{+2.12}_{-2.00}$    & $\pm0.09$         &$\pm10.1$            \\ \hline

$(K\pi)^{*0}_0\piz$       &  Dalitz Plot Model   &$\pm2.81$             & $\pm0.09$      &                  \\
                          &  Shape Parameters    &$^{+2.30}_{-0.57}$    & $^{+0.08}_{-0.03}$&                     \\ 
                      &  $\B$ Background         &$^{+0.40}_{-0.46}$    & $^{+0.04}_{-0.05}$&                     \\ 
                          &  Lineshapes          &$^{+0.05}_{-0.06}$    & $\pm0.002$        &                     \\
                          &  Continuum DP PDF     &$\pm0.73$             &                   &$\pm\ 5.2$           \\
                          &  Fit Bias            &$\pm0.19$             & $\pm0.01$         &$\pm14.5$            \\ 
                          &  Other               &                      &                   &$\pm\ 4.4$           \\ 
                          &  {\bf total}         &$^{+3.73}_{-3.00}$     & $^{+0.13}_{-0.11}$&$\pm16.0$            \\ \hline

$\rho(770)^-\Kp$          &  Dalitz Plot Model   &$\pm0.98$             & $\pm0.04$         &undefined                 \\
                          &  Shape Parameters    &$^{+0.34}_{-0.43}$    & $\pm0.01$         &                     \\ 
                      &  $\B$ Background         &$^{+0.17}_{-0.25}$    & $\pm0.05$         &                     \\ 
                          &  Lineshapes          &$^{+0.04}_{-0.03}$    & $\pm0.001$        &undefined                 \\
                          &  Continuum DP PDF     &$\pm0.45$             &                  &$\pm\ 3.6$           \\
                          &  Fit Bias            &$\pm0.15$             & $\pm0.01$         &$\pm13.7$            \\ 
                          &  Other               &                      &                   &$\pm\ 2.6$           \\ 
                          &  {\bf Total}         &$^{+1.15}_{-1.19}$     & $\pm0.07$         &$\pm14.4$            \\ \hline

$NR$                      &  Dalitz Plot Model   &$\pm0.41$             & $\pm0.04$         &undefined                 \\
                          &  Shape Parameters    &$^{+0.46}_{-0.51}$    & $\pm0.04$         &                     \\ 
                      &  $\B$ Background         &$^{+0.64}_{-0.24}$    & $^{+0.10}_{-0.08}$&                     \\ 
                          &  Lineshapes          &$^{+0.04}_{-0.03}$    & $\pm0.001$        &undefined                 \\
                          &  Continuum DP PDF     &$\pm0.91$             &                  &$\pm\ 5.8$           \\
                          &  Fit Bias            &$\pm0.22$             & $\pm0.01$         &$\pm6.8$             \\ 
                          &  Other               &                      &                   &$\pm\ 2.6$           \\ 
                          &  {\bf Total}         &$^{+1.29}_{-1.17}$    & $^{+0.11}_{-0.10}$&$\pm\ 9.3$           \\ \hline \hline
\end{tabular}
\end{center}
\end{table*}
Variations around the nominal fit are tried to study the dominant systematic effects which are summarized in~\tabref{Systematics}.
For each parameter of interest $p$ ($FF$, $A_{CP}$, $\phi$), the positive
(negative) deviations from each effect are summed in quadrature to
obtain total upward (downward) systematic errors $\delta p_+$ ($\delta p_-$). The total projected likelihood function $\mathcal{L}_{tot}(p)$ is computed
as the convolution of the fit projected likelihood function defined at
the end of the previous section, by a bifurcated Gaussian distribution with a lower (upper) standard deviation $\delta p_-$ ($\delta p_+$).
The scan of $\mathcal{L}_{tot}$ drawn as a solid line on~\figref{BR_Kstar0Pi0} (and subsequent ones) is used to determine the total confidence interval which 
accounts for both statistical and systematic uncertainties. 
Finally, the upward (downward) systematic uncertainty on $p$ (\tabref{finalresults}) is the quadratic difference between the upward (downward) limits of the total and 
statistical confidence intervals. Note that when $p$ central values significantly differ between the four fit solutions, the lowest $p_{min}$ and highest $p_{max}$ 
are used in the determination of the downward (upward) uncertainties and an additional systematic uncertainty $\pm (p_{max}-p_{min})/2$ is quoted in \tabref{finalresults}.\par
While simple methods can be used to estimate the systematic effects on the isobar fractions and $CP$-asymmetries, it is often necessary to
estimate upper limits for phases. In such cases (the rows labeled {\it Other} in~\tabref{Systematics}) we conservatively use the
upper limits at the 90\% confidence level as the systematic uncertainties.  \par  
To estimate the contribution of other resonances, we fit the on resonance data with extended signal models including one {\it extra}-resonance on
top of those in the nominal signal model. The $\rho^-(1450)K^+,\ \rho^-(1700)K^+,\ \Kstarz_2(1430)\piz,\ \Kstarp_2(1430)\pim$, $\Kstarz(1680)\piz$ and $\Kstarp(1680)\pim$
final states have been studied. These fits are not significantly better than the nominal fit. The isobar fractions of all extra resonances are below 5\%. None is significant. 
The isobar fractions of the nominal resonances change by non
significant amounts which we record as ({\it Dalitz Plot Model}) systematic uncertainties 
in~\tabref{Systematics}. Changes in $A_{CP}$ are also recorded. As we explain in~\secref{Interpretation} our fitting procedure is sensitive to some
phase differences but not others. When there is sensitivity the systematic {\it Dalitz Plot Model} effects are insignificant. When there is no sensitivity
dramatic deviations occur. However since these phase differences are
essentially undetermined by the fit we do not record these changes as
systematic effects. Note that we have not included the low
mass $\kappa$ resonance which is not established as it is an
alternative to the $(K\pi)^{*0}_0$ isobar from the model. It has
been verified however that adding it to the model results into a
destructive interference with the latter state, an insignificant
$\kappa$ amplitude and no change in the $\Kstarz(892)\piz$ numeric results.\par
There are fixed parameters in the nominal fit model. We estimate the associated ({\it Shape parameters, and $\B$ Background}) systematic uncertainties by 
repeating the fit giving the studied parameter values at $\pm1$ standard deviation from its fixed value or at the limits of a conservative range. 
The fixed parameters to which the isobar fractions are sensitive are
$\overline{f_{SCF}}$ [\equaref{fscfbar}] and the \deprime\ signal PDF
parameters. $\overline{f_{SCF}}$ was varied between 10 and 16\%, a
range that is inferred from the comparison of the data and simulation of $\B\to
D\rho$ events as explained in References~\cite{rhopibabar,rhopiq2bbabar}.
The $A_{CP}$ mainly depend on the fixed $CP$-asymmetries of \B\ background classes~1 and~7 (see~\tabref{listBb}). We vary them by $
\pm0.5$ and $\pm0.2$ to determine the ({\it \B background }) systematic uncertainties. These intervals were chosen after inspection of the latest 
available measurements~\cite{hfag}.\par
The variations of the physical parameters of the resonances in the nominal signal model are recorded as ({\it Lineshape}) systematic uncertainties. 
In particular, variations of the barrier coefficients $R$ in [\equaref{barrier}] cause no significant effects.\par
The method used to determine the continuum square Dalitz plot PDF~(Section V.B)
has been extensively tested in the Monte Carlo. The spread of the
yields across those tests is used to estimate the associated ({\it continuum DP PDF}) systematic errors. Among the dominant effects is the variation of the distortion 
of the Dalitz plot distribution as a function of \mes\ due to the \B\ meson mass constraint.   \par  
To estimate the {\it Fit Bias} uncertainties stemming from the
imperfection of the fit model (most importantly the too simple description of the SCF
but also the neglected correlations between \mes and \deprime\ for TM
signal and \B-background events, and the neglected resolution effect
smearing the \Kstarpm(892) peak), 
we record the fitted biases and spreads in fits performed on large Monte Carlo samples with GEANT4 signal events and the actual amount of 
background generated along with their PDFs. Since the SCF prevails in the Dalitz plot corners where the resonances interfere, 
this kind of systematic uncertainty is the dominant one for the phase measurements. \par
For the \B meson branching fraction measurements, we have to include additionally the effect of
imperfections in the event reconstruction on the efficiency. Adding in 
quadrature the uncertainties associated with tracking (1.6\%), charged particle identification (2\%), $\piz$ reconstruction (3\%), the efficiency of 
the selection requirements (0.3\% for \mes, 1.2\% for \de, 2\% for the NN), and the integrated luminosity (1.1\%), we obtain a global
systematic percent error of $4.7\%$. \par

\section{INTERPRETATION}
\label{sec:Interpretation}
\begin{table*}[h]
\begin{center}
\caption{\label{tab:finalresults}
Final results for rates and $CP$-asymmetries. 
The quasi-two-body branching fractions $\mathcal{B}_j$ have been computed from the isobar fractions $FF_j$ using ~\equaref{partialbf}. The statistical uncertainties are given first.
The statistical and systematic uncertainties are calculated by scanning the NLL close to its minimum taking into account the four fit solutions 
and recording the values where the NLL increases by one unit above 
its minimum. For the final states with $K\pi$ S-waves, a second systematic uncertainty covers the spread between the best fit values from the 
four solutions. For resonances that are considered in extended fit models, we quote upper limits at the 90\% confidence level (based on
statistical uncertainties only) for the isobar fractions and the quasi-two-body branching fractions.} 
\setlength{\tabcolsep}{1.2pc}
\begin{tabular}{llll} 
\hline\hline
\multicolumn{1}{c}{isobar $j$}                & \multicolumn{1}{c}{$FF_j$ ($\%$)}                             & \multicolumn{1}{c}{$\mathcal{B}_j$ ($10^{-6}$)}                     
                & \multicolumn{1}{c}{$A^j_{CP}$} \\ \hline \\
$K^{*+}(892)\pi^{-}$    & $\ \ 11.8_{-1.5}^{+2.5}\pm0.6  $               & $ \ 4.2_{-0.5}^{+0.9}\pm0.3  $  & $\ -0.19_{-0.15}^{+0.20}  \pm0.04  $  \\
$K^{*0}(892)\pi^{0}$    & $\ \ \ 6.7_{-1.5}^{+1.3}$$_{-0.6}^{+0.7} $     & $ \ 2.4\pm0.5\pm0.3  $& $\ -0.09_{-0.24}^{+0.21}  \pm0.09  $  \\
$(K\pi)^{*+}_0\pi^{-}$&$\ 26.3_{-3.8}^{+3.1}$$_{-2.0}^{+2.1}\pm4.9$      & $ \ 9.4_{-1.3}^{+1.1}$$_{-1.1}^{+1.4}\pm1.8$&$\ +0.17_{-0.16}^{+0.11}\pm0.09\pm0.20$\\
$(K\pi)^{*0}_0\pi^{0}$&$\ 24.3_{-2.6}^{+3.0}$$_{-3.0}^{+3.7}\pm6.7$      & $ \ 8.7_{-0.9}^{+1.1}$$_{-1.3}^{+1.8}\pm2.2$&$\ -0.22\pm0.12 _{-0.11}^{+0.13}\pm0.27$\\
$\rho^{-}(770)K^{+}$        &$\ 22.5_{-3.7}^{+2.2}\pm1.2  $              & $ \ 8.0_{-1.3}^{+0.8}\pm0.6   $             & $\ +0.11_{-0.15}^{+0.14}  \pm0.07  $  \\
N.R.                        &$\ 12.4\pm2.6_{-1.2}^{+1.3}  $              & $ \ 4.4\pm0.9\pm0.5           $ & $ +0.23 _{-0.27}^{+0.19}$$_{-0.10}^{+0.11}$ \\
&&&\\
{\bf Total}     & $102.3_{-4.0}^{+7.1}\pm4.1  $                          & $ 35.7_{-1.5}^{+2.6}\pm2.2$                & $-0.030_{-0.051}^{+0.045}\pm 0.055$     \\
&&&\\  \hline
               & \multicolumn{1}{c}{$FF_j$, [Upper Limits] (\%)} &   \multicolumn{1}{c}{Upper Limits ($10^{-6}$)  } &                            \\
 \hline \\
$\rho^-(1450)K^+$        &$\ \ 2.2_{-1.5}^{+2.2}$, [$5.9$] & $2.1$                                             & \\
$\rho^-(1700)K^+$        &$\ \ 0.7_{-0.6}^{+1.2}$, [$3.1$] & $1.1$                                             & \\
$K^{*0}_2(1430)\piz$     &$\ \ 1.2_{-1.0}^{+1.5}$, [$3.6$] & $1.3$                                             & \\
$K^{*+}_2(1430)\pim$     &$\ \ 5.2_{-1.4}^{+1.6}$, [$7.5$] & $2.7$                                             & \\
$\Kstarz(1680)\piz$      &$\ \ 3.0_{-1.3}^{+1.6}$, [$5.5$] & $2.0$                                             & \\
$\Kstarp(1680)\pim$      &$\ \ 5.7_{-1.7}^{+2.0}$, [$8.9$] & $3.2$                                             & \\ 
                         &      &                                                   & \\
\hline\hline
\end{tabular}
\end{center}
\end{table*}
\begin{table*}
\begin{center}
\caption{ \label{tab:phasedifferences} 
Final results for phases.
When there is sensitivity, the results are the one standard deviation confidence interval ($1\;\sigma$~c.i.) for the phase angle measurements 
(in degrees). The statistical and systematic uncertainties, determined by the NLL scan method described in~\figref{BR_Kstar0Pi0} are included. 
The interval bounds can be seen on the graphs referenced in the second column, as the intersections of the solid scan curves with the 
($\Delta\chi^2=1$) horizontal dashed lines. The $\Delta\chi^2$ evaluated for zero $\delta\phi$ measures the significance (squared) for direct $CP$-violation. 
When there is no sensitivity we give the maximum $\Delta\chi^2$ reached over the scanned region.}
\setlength{\tabcolsep}{0.5pc}
\begin{tabular}{ccccccc}\hline\hline
Interference pattern                  & Graph &   Results   & $\Delta\phi$ for $\Bz$ & $\Delta\overline{\phi}$ for $\Bzb$ & $\delta\phi \equiv \Delta\overline{\phi} - \Delta\phi$ & $\Delta\chi^2(\delta\phi=0)$\\\hline
                                      & &             &                        &                         &                &          \\
$K\ \pi$ neutral$-$charged P-waves    & \figref{PzerominusPplusphasediff}&$\Delta\chi^2_{\rm MAX}$    & $2.2$         & $5.4$                   & $0.88$&          \\
                                      & &             &                        &                         &                &          \\
$K\ \pi$ neutral$-$charged S-waves    & \figref{SzerominusSplusphasediff}&$\Delta\chi^2_{\rm MAX}$    & $2.2$         & $7.6$                   & $0.84$&          \\
                                      & &             &                        &                         &                &          \\
$\rho^{\mp}\ \Kpm - \Kstarpm\pimp $   & \figref{RhoKPwave}&$\Delta\chi^2_{\rm MAX}$    & $7.6$         & $1.9$                   & $1.0$&          \\
                                      & &             &                        &                         &                &          \\
                                      & &             &                        &                         &                &          \\
Charged $K\ \pi$ P-wave - S-wave      &
                                      \figref{PplusminusSplus}&$1\;\sigma$~c.i.& $[\ 177,\ 209]$& $[232,\ 305]$  & $[\ 44,116]$&3.1    \\
                                      & &             &                        &                         &                 &         \\
Neutral $K\ \pi$ P-wave - S-wave      & \figref{PzerominusSzero}&$1\;\sigma$~c.i. & $[\ -6,\ 41]$& $[\ -12,\ 46]$ & $[\ -32,  \ 38]$&0  \\
                                      & &             &                        &                         &                 &         \\
NR $-$ charged S-waves                & \figref{NRminusSplus}&$1\;\sigma$~c.i. & $[-87,-41]$ & $[\ -84,\ 38]$ & $[-151,107]$&0      \\
                                      &               &          & $[\ \ 20,\ \ 81]$ &  & &      \\
                                      & &             &                        &                         &                 &         \\
NR $-$ neutral S-waves                & \figref{NRminusSzero}&$1\;\sigma$~c.i. & $[\ -96,\ -41]$ & $[\ -84,\ 67]$& $[-145,145]$&0      \\
                                      &               &          & $[\ -3,\ \ \ 75]$ &         & &      \\
\hline\hline
\end{tabular}
\end{center}
\end{table*} 

The final results are given in~\tabsref{finalresults}{isospin-corrected-bf}. 
The total branching fraction $\mathcal{B}(\Bz\to\Kp\pim\piz)=(35.7_{-1.5}^{+2.6}\pm2.2)\times 10^{-6}$ 
and the global $CP$-asymmetry $\mathcal{A}_{CP}=-0.030^{+0.045}_{-0.051}\pm0.055$ 
are compatible with the published measurements from Belle~\cite{belle}
$(36.6\ ^{+4.2}_{-4.3}\pm3.0)\times 10^{-6}$ and $0.07\pm0.11\pm0.01$ respectively. 
The decay\footnote{Isospin conservation is assumed for the strong decays of a I=1/2 K meson resonance $k^*$ to $K\pi$ when we compute the
branching fraction of the quasi-two-body process $\Bz \to k^*\pi$, namely
$\frac{\Gamma(k^{*0}\to\Kp\pim)}{\Gamma(k^{*0}\to K\pi)}=2/3$, and $\frac{\Gamma(k^{*+}\to\Kp\piz)}{\Gamma(k^{*+}\to K\pi)}=1/3$.} 
$\Bz\to\Kstarz(892)\piz$ is observed with a significance of
5.6~standard deviations (including systematics). We measure
$\mathcal{B}(\Bz\to\Kstarz(892)\piz)=(3.6\pm 0.7\pm 0.4)\times 10^{-6}$, just at the
edge of the 90\% upper limits of previous experiments ($3.5 \times 10^{-6}$ by Belle~\cite{belle} and $3.6\times 10^{-6}$ by CLEO~\cite{cleorhok})
combined to $3.5\times 10^{-6}$ in Reference~\cite{PDG2006}.
The subdecay branching fraction for $\Bz\to\Kstarp(892)\pim$ agrees with previous measurements from Belle~\cite{belle,belle-kspipi} and
\babar~\cite{babar-kspipi} in both $\Kstarp \to \Kp\piz$ and $\KS\pip$ decay channels.
Averaging the \babar~measurements one obtains $\mathcal{B}(\Bz\to\Kstarp(892)\pim)=(11.7_{-1.2}^{+1.3})\times 10^{-6}$ 
and $A_{CP}(\Bz\to\Kstarp(892)\pim)=-0.14\pm0.12$.
The branching fraction for $\Bz\to\rho^-K^+$ is lower than those measured by Belle~\cite{belle} and CLEO~\cite{cleorhok} but in agreement within errors.
If we assume that the $(K\pi)^{*+,0}_0$ isobars are pure isospin-1/2 and
neglect possible non $K\pi$ final states, we determine the following
effective branching fractions:
$\mathcal{B}(\Bz\to (K\pi)^{*+}_0\pi^{-})=(28.2_{-4.1}^{+3.3}$$_{-3.3}^{+4.3}\pm5.2)\times 10^{-6}$, and
$\mathcal{B}(\Bz\to (K\pi)^{*0}_0\pi^{0})=(13.1_{-1.5}^{+1.6}$$_{-1.9}^{+2.7}\pm3.6)\times 10^{-6}$.
A consistency check of our analysis is provided by our measurement of the branching ratio, 
$\mathcal{B}(\Bz\to\Dzb\piz)=(2.93\pm0.17\pm0.18)\times 10^{-4}$ in agreement with its world average and that of the branching fraction of
the decay $\Dz\to\Km\pip$~\cite{PDG2006}. \par
There is no evidence of direct $CP$-violation in any resonant subdecay. In~\tabtworef{finalresults}{isospin-corrected-bf} 
we give upper limits at the 90\% statistical confidence level on the branching fractions of resonances that might contribute to $\Kp\pim\piz$ but are not part of the nominal signal model. \par
\begin{figure}[p]
\epsfig{file=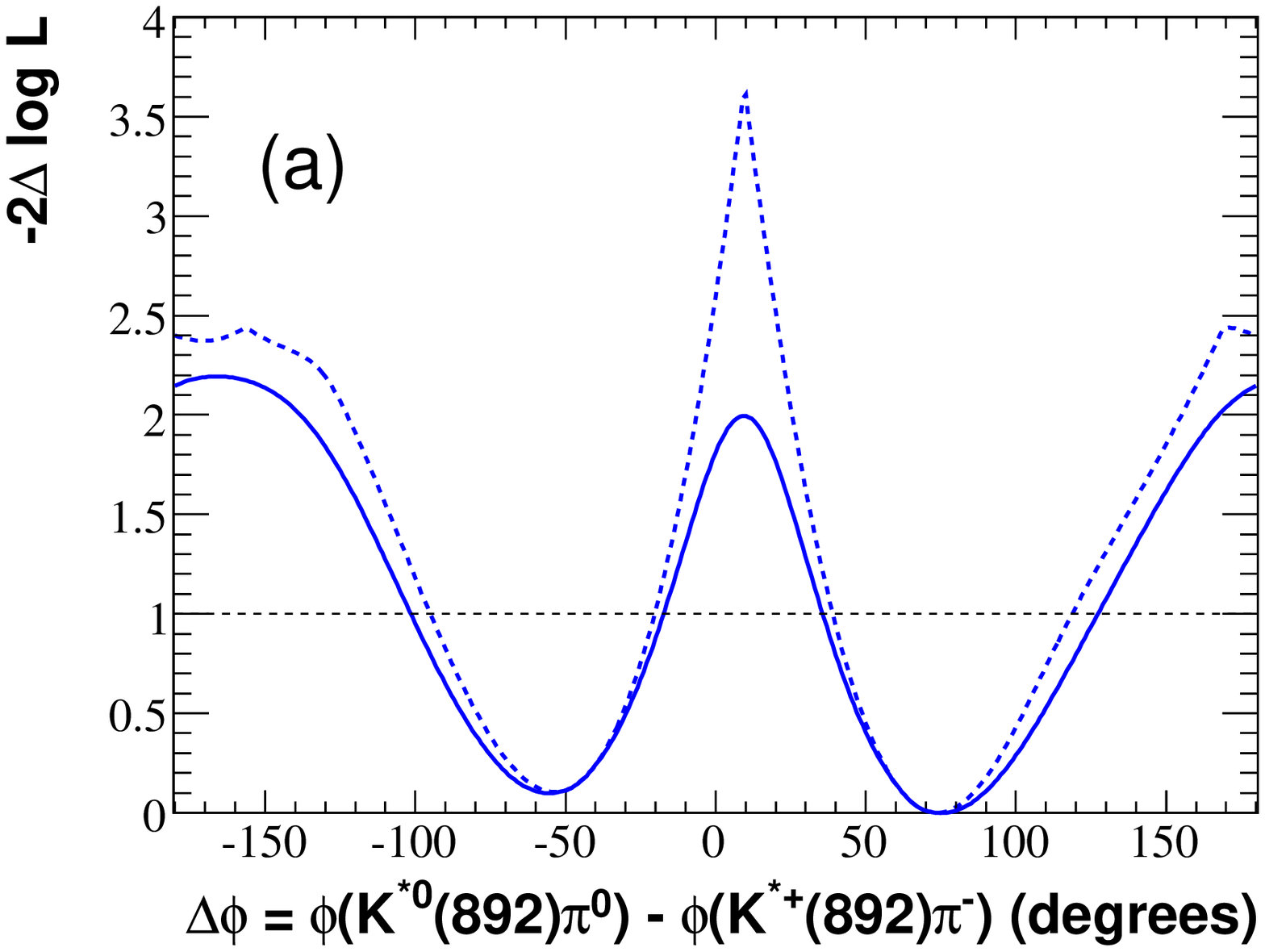,width=8.5cm} 
\epsfig{file=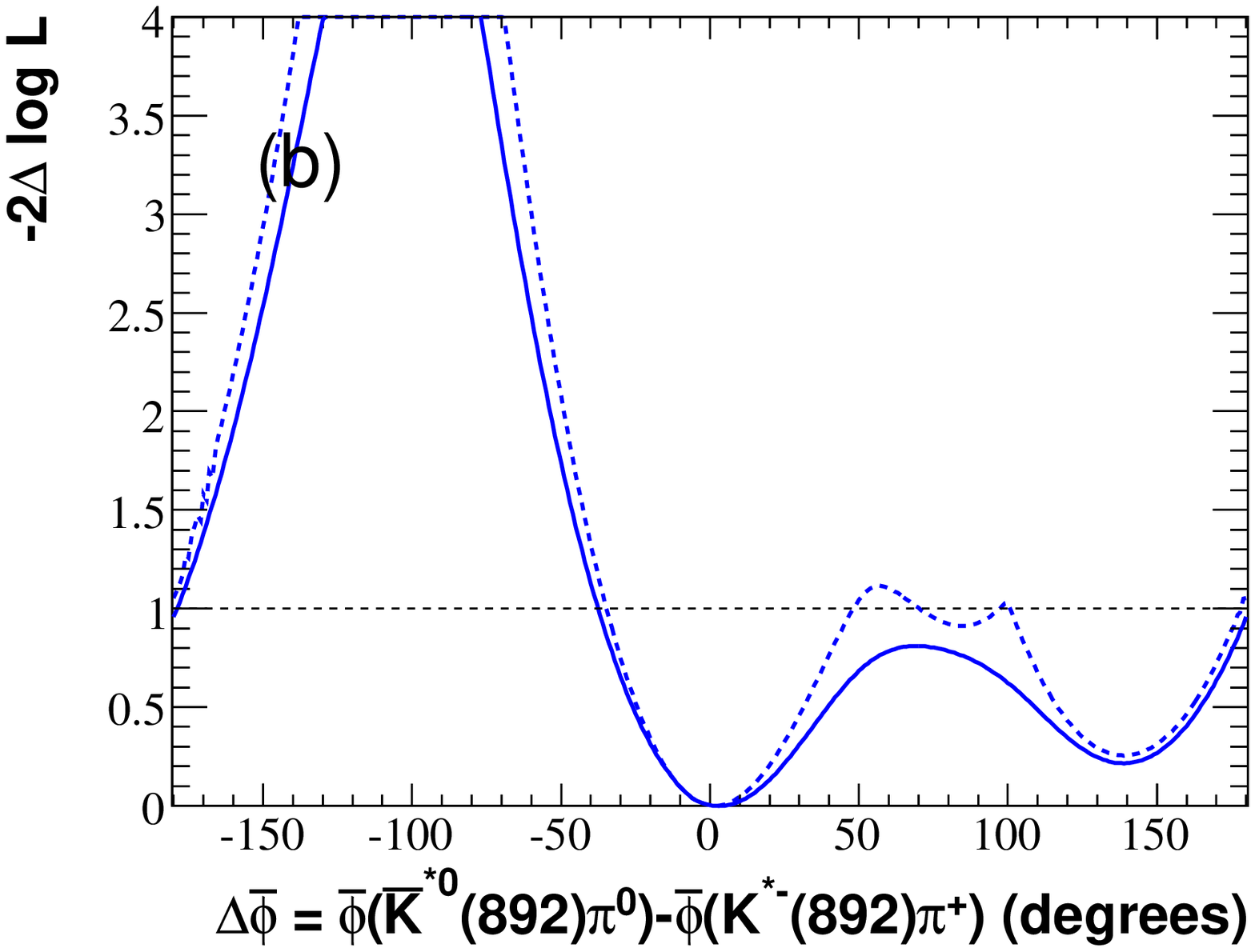,width=8.5cm} 
\epsfig{file=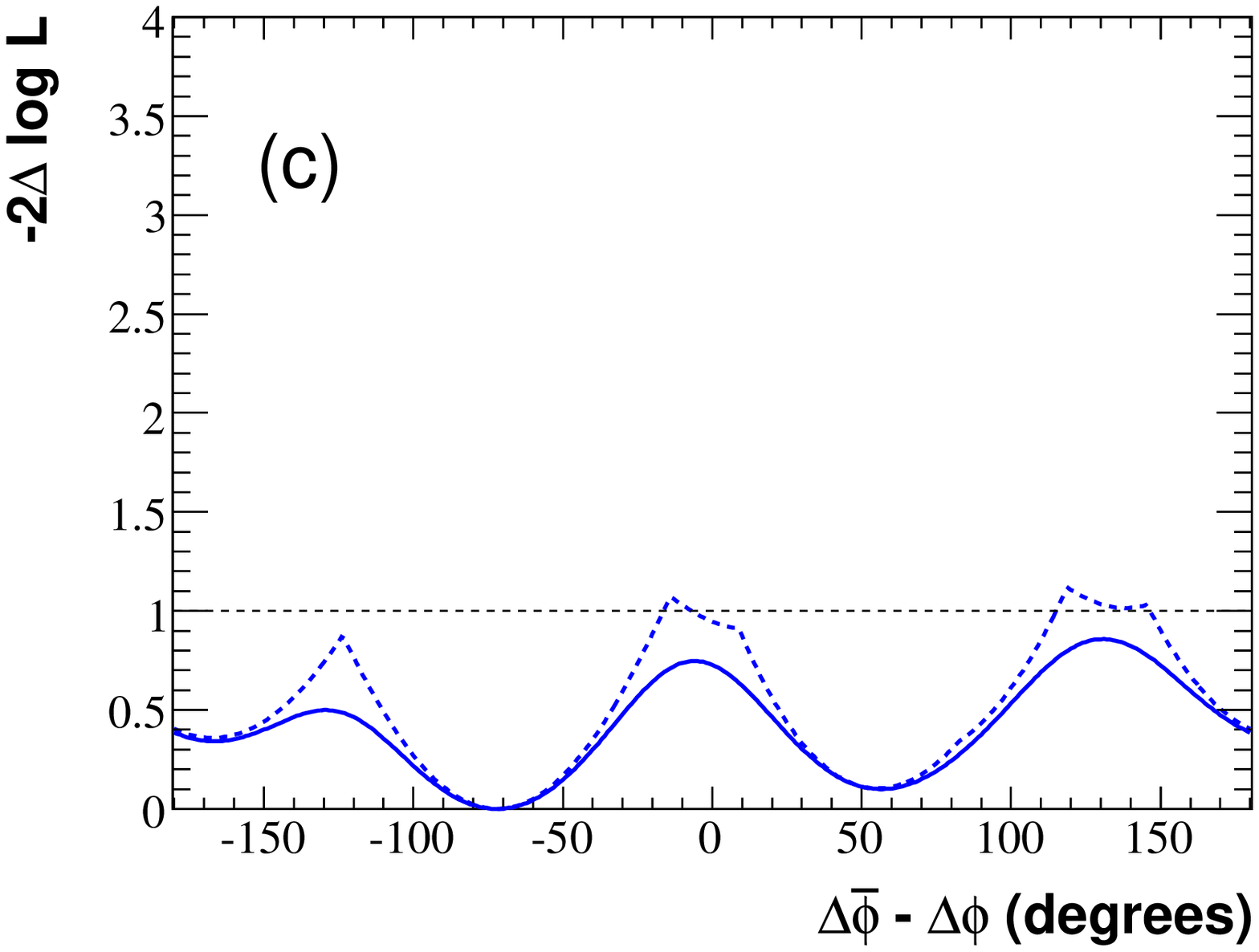,width=8.5cm} 
\caption{
\label{fig:PzerominusPplusphasediff} The phase difference between the $\Kstarz$ and the charged $\Kstar$ for the P-wave $\Kstar(892)$. 
The three diagrams are the NLL scans for the $\Bz$ (a) and $\Bzb$ (b) decays as well as their differences (c). 
The dashed line gives the statistical uncertainty, and the solid line,
the total uncertainty. The data do not indicate preferred angles,
except for the [-131, -75] degree range which is excluded for \Bzb at the two
standard deviation level.
The four fit solutions find their NLL minimum for distinct phases. The vertical scale stops at $2\Delta$(NLL)=4 slightly above 3.84 which is 
the 95\% confidence level. A horizontal dashed line at $2\Delta$(NLL)=1 shows the one standard deviation level.
}
\end{figure}
\begin{figure}[!h]
\epsfig{file=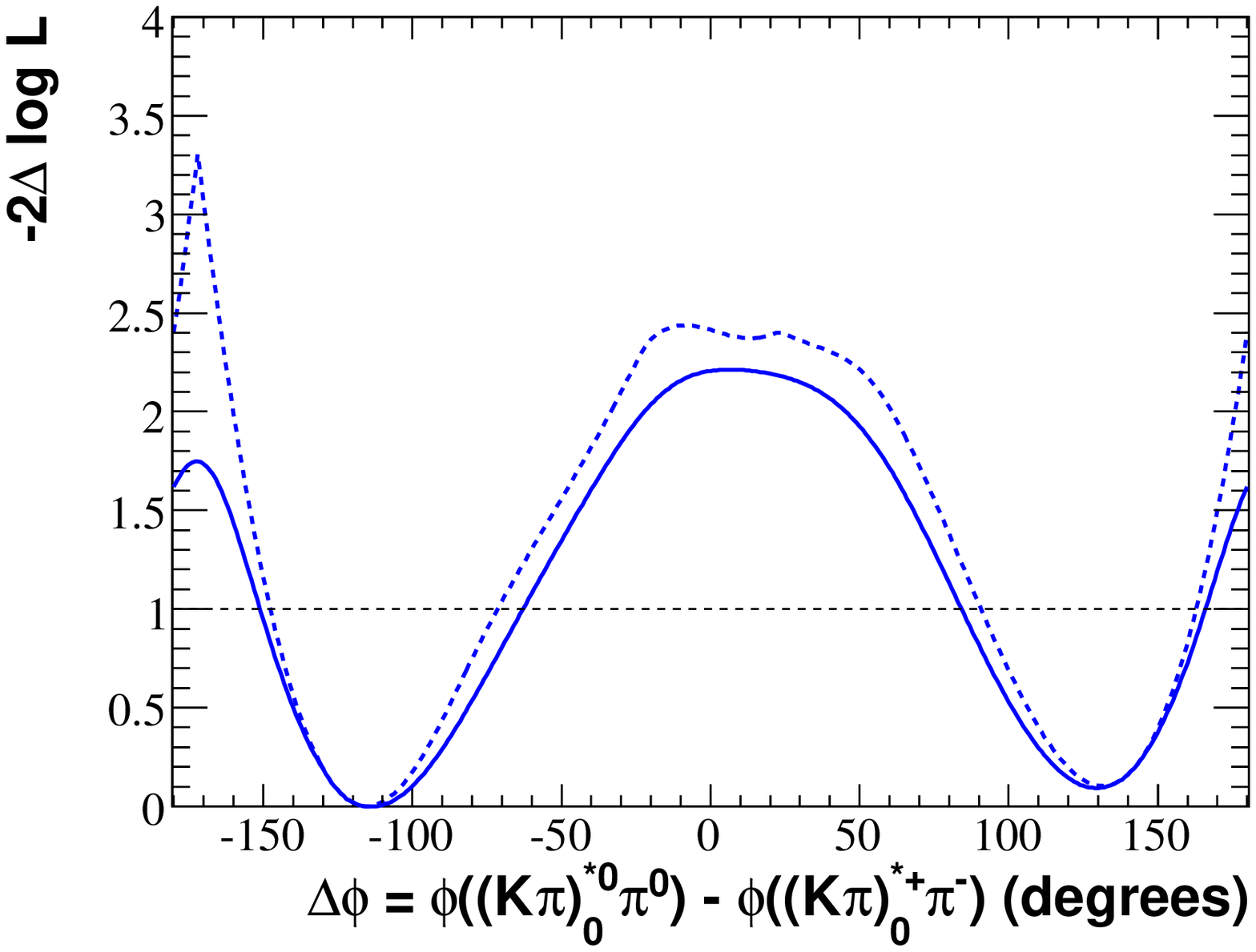,width=8.5cm} 
\epsfig{file=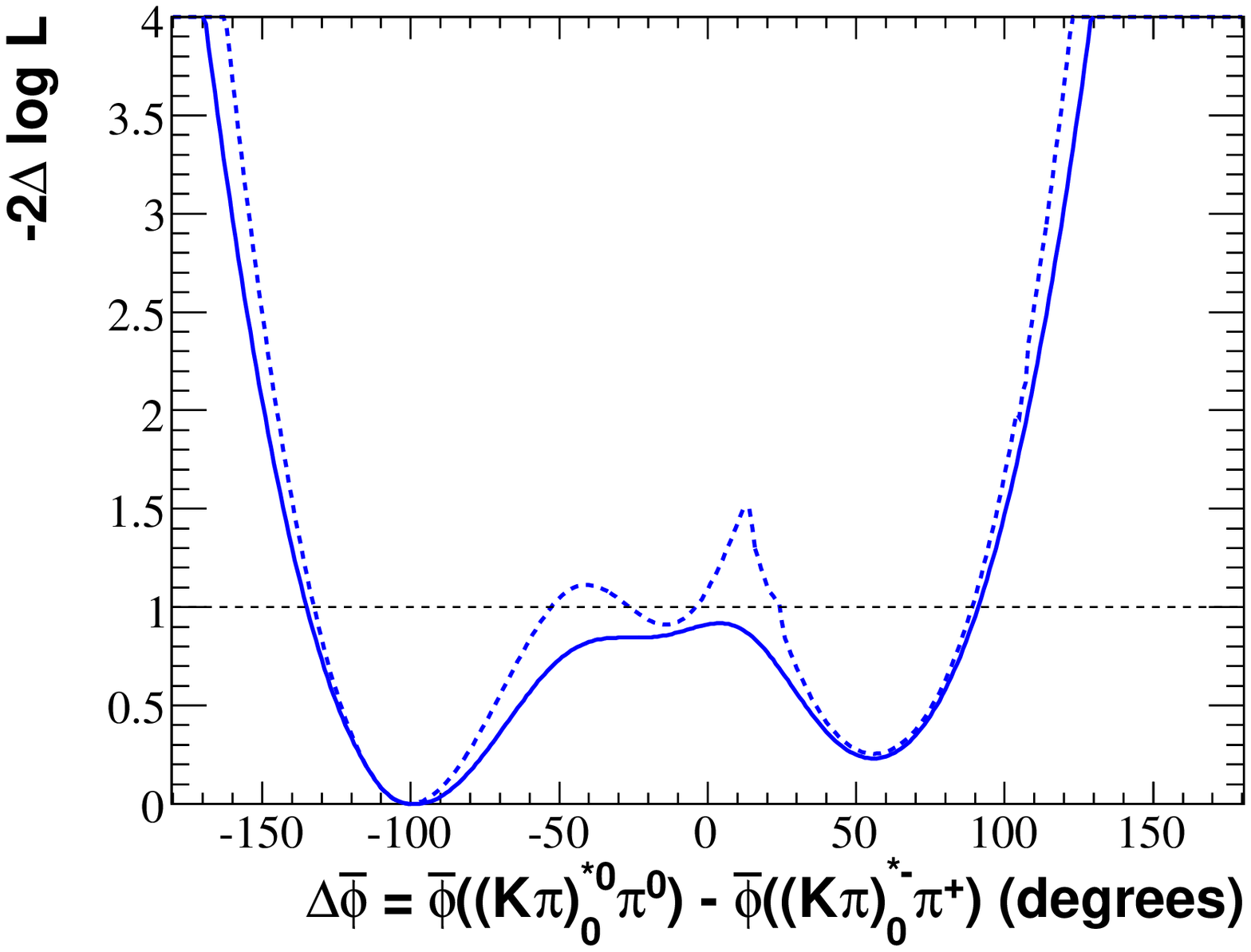,width=8.5cm}
\epsfig{file=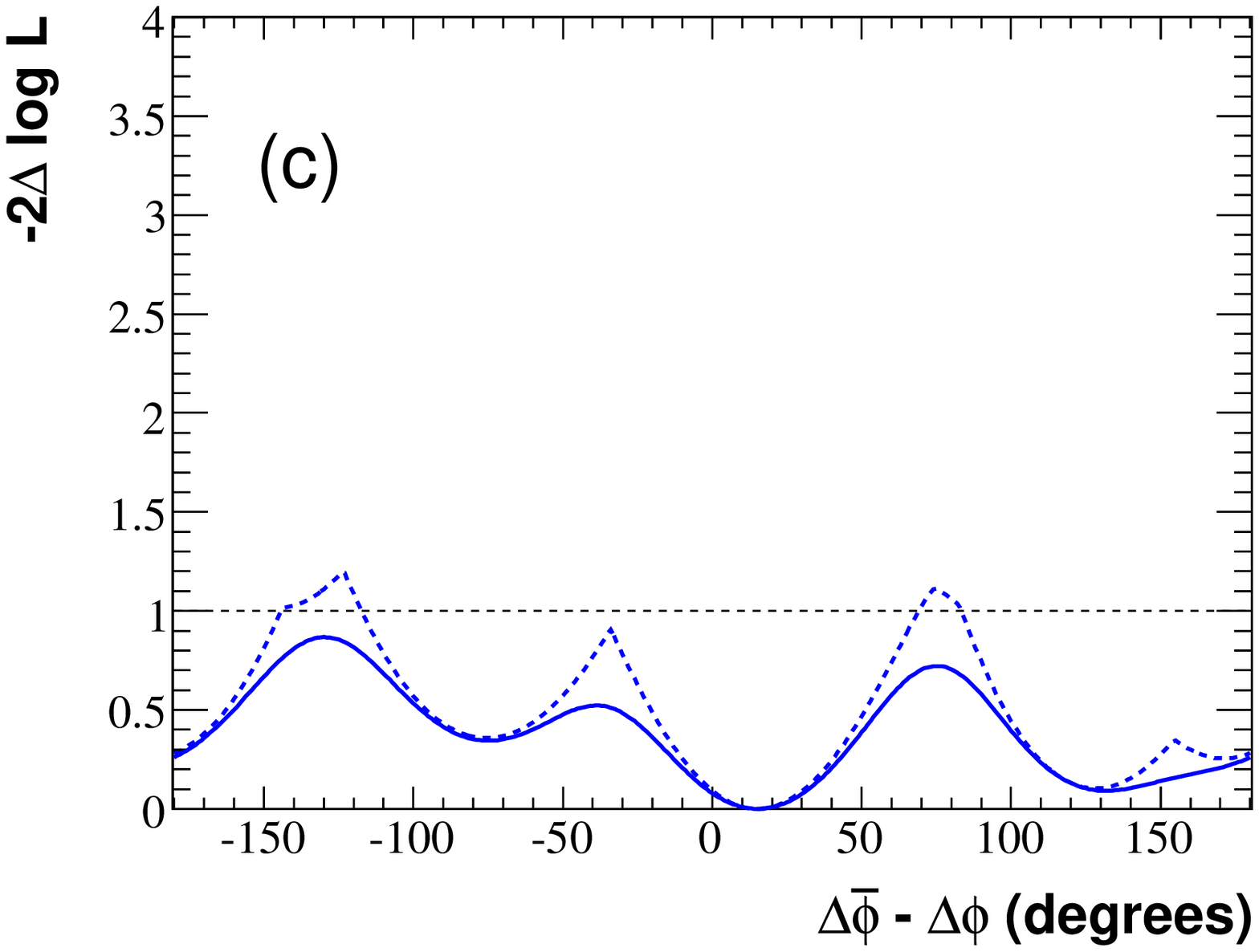,width=8.5cm} 
\caption{
\label{fig:SzerominusSplusphasediff} The phase difference between the $(K\pi)_0^{*0}$
and the $(K\pi)_0^{*\pm}$
S-waves. The three diagrams are the NLL scans for the $\Bz$ (a) and $\Bzb$ decays (b) as well as their differences (c). 
The lines are drawn as for~\figref{PzerominusPplusphasediff}. The data do not indicate preferred angles. 
The four fit solutions find their NLL minimum for distinct phases. 
}
\end{figure}
\begin{figure}[!h]
\epsfig{file=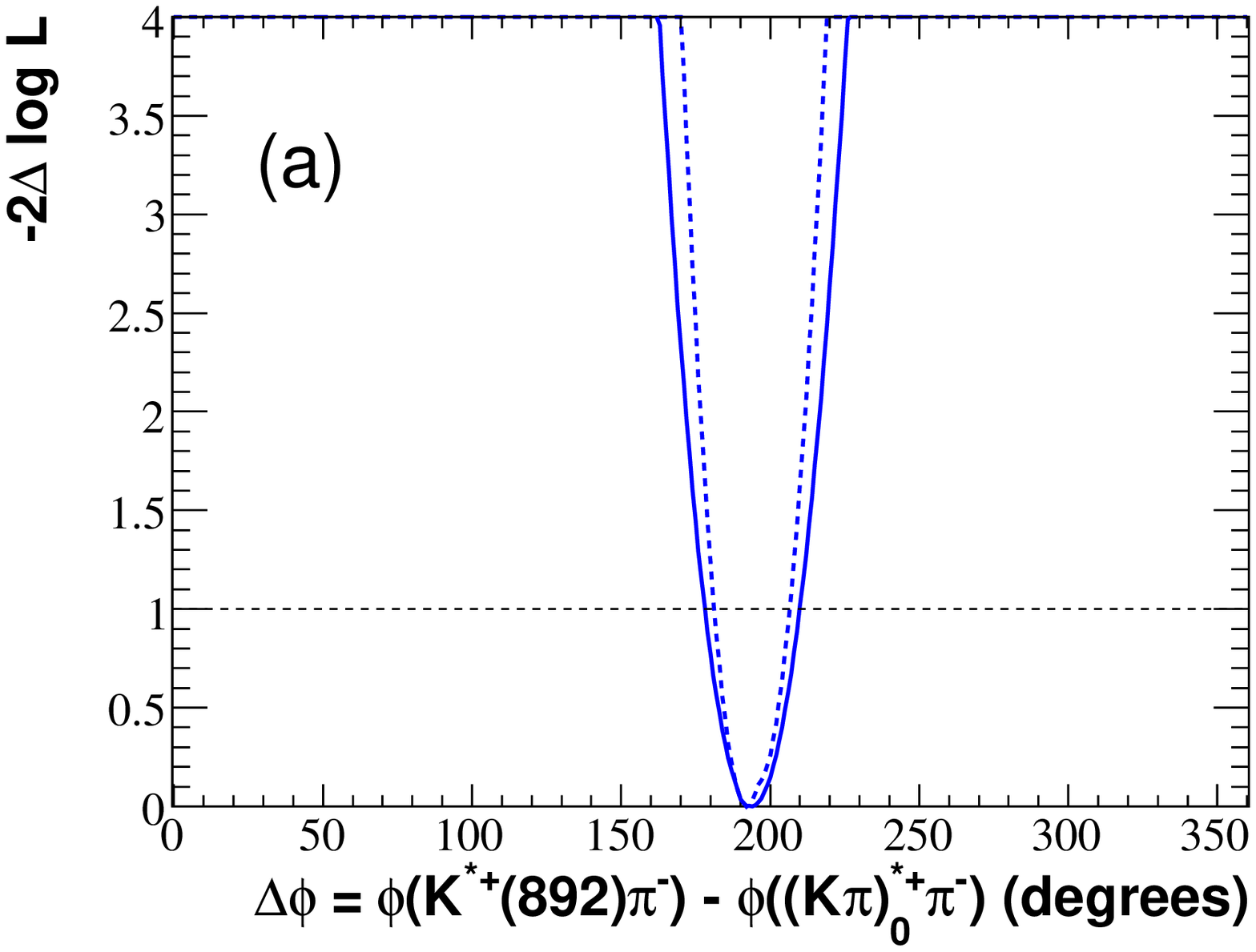,width=8.5cm} 
\epsfig{file=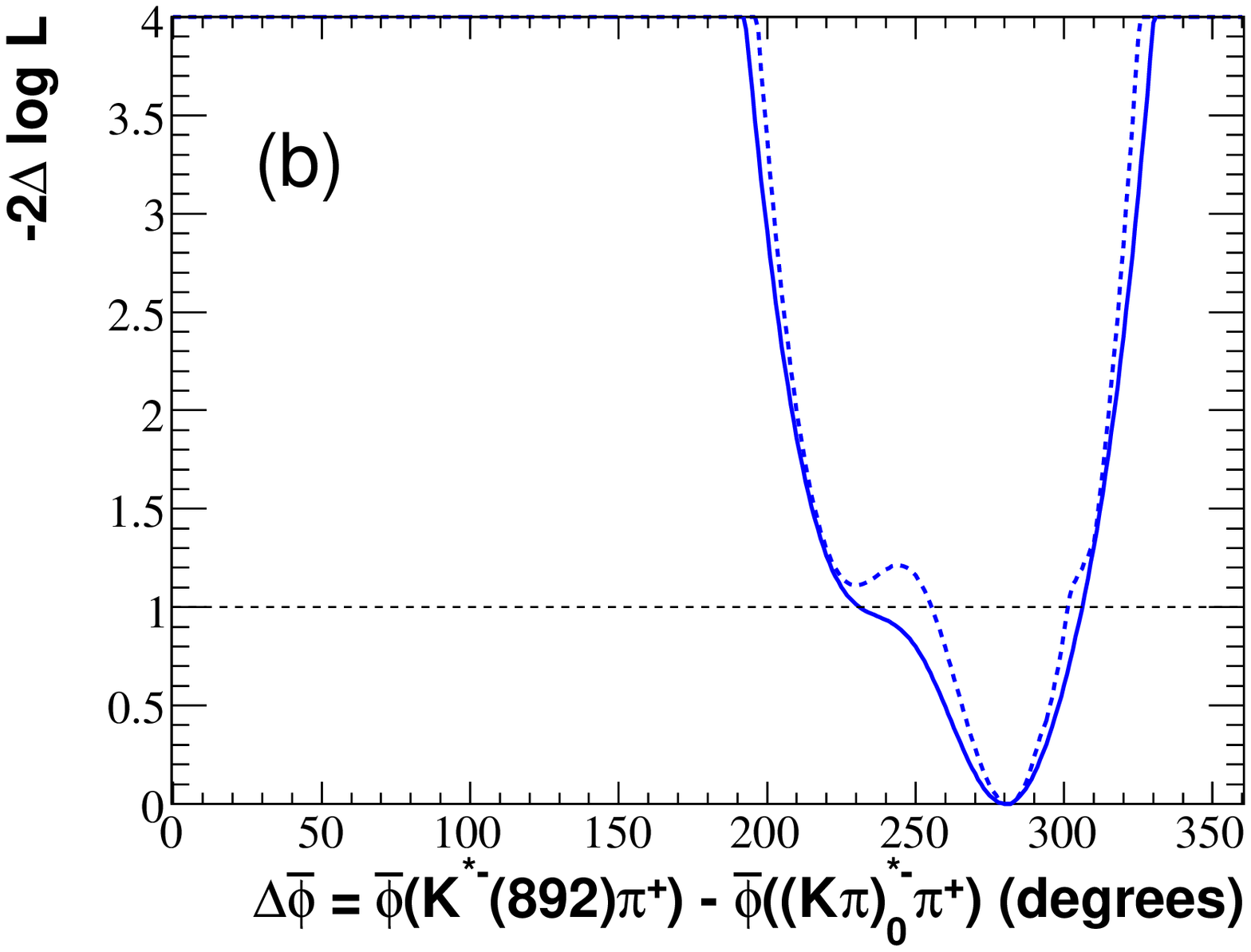,width=8.5cm}
\epsfig{file=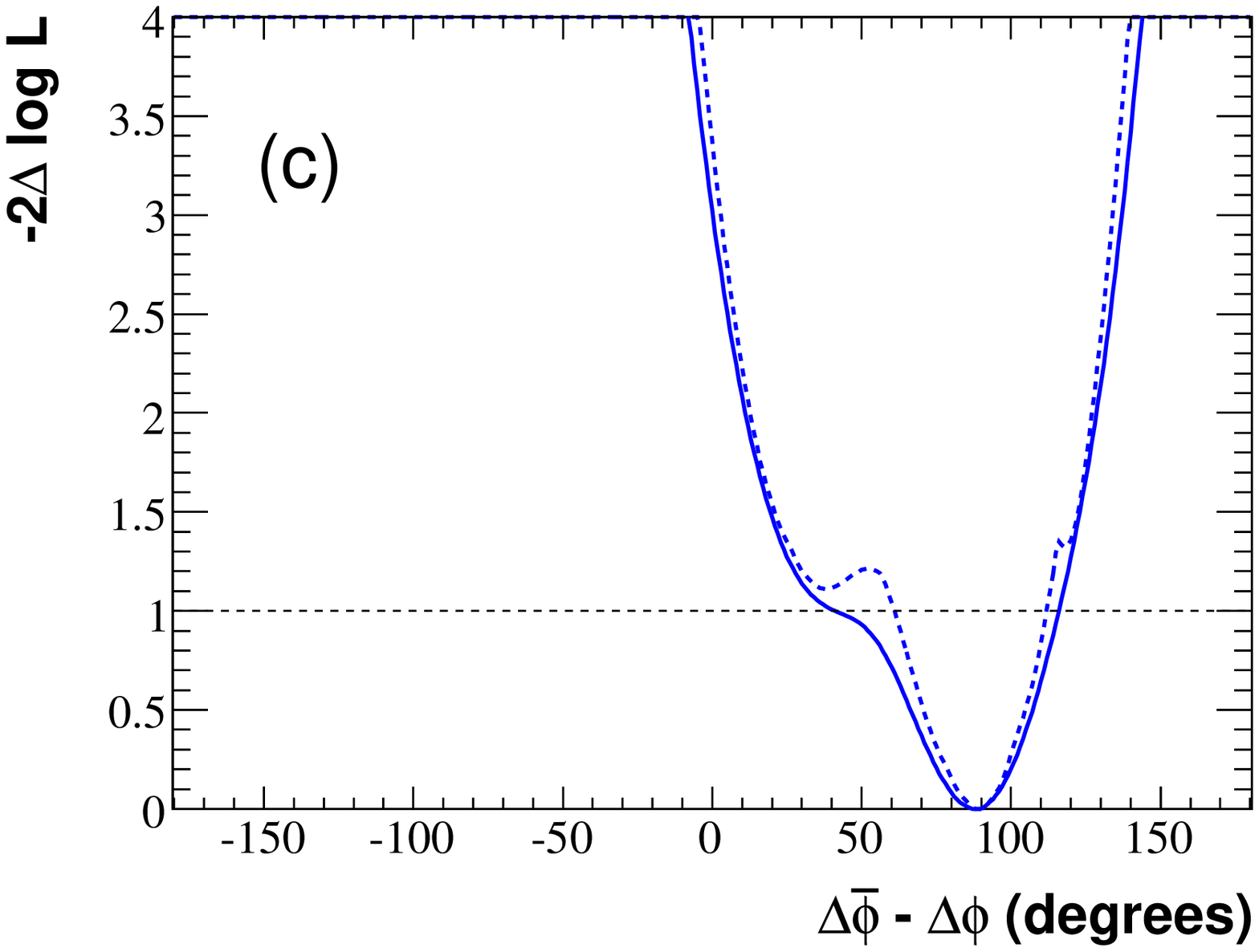,width=8.5cm} 
\caption{
\label{fig:PplusminusSplus} The phase difference between the charged $K\pi$ P~and S-waves. The three diagrams are the NLL scans for the $\Bz$ (a) 
and  $\Bzb$ (b) decays as well as their differences (c). 
The lines are drawn as for~\figref{PzerominusPplusphasediff}. The data provide significant constraints on these angles. 
The four fit solutions find their NLL minimum at approximately the same phase differences. 
}
\end{figure}
\begin{figure}[h]
\epsfig{file=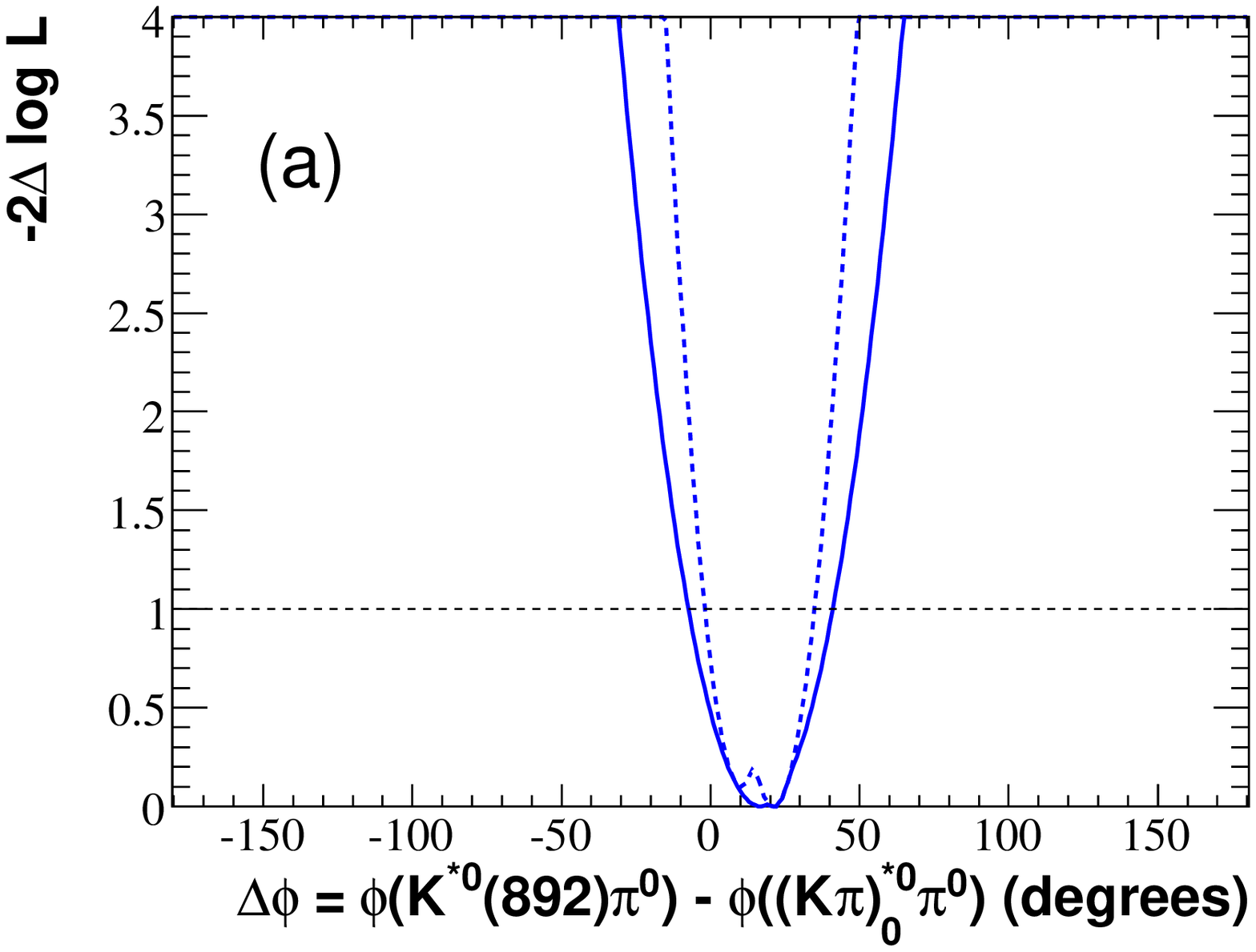,width=8.5cm} 
\epsfig{file=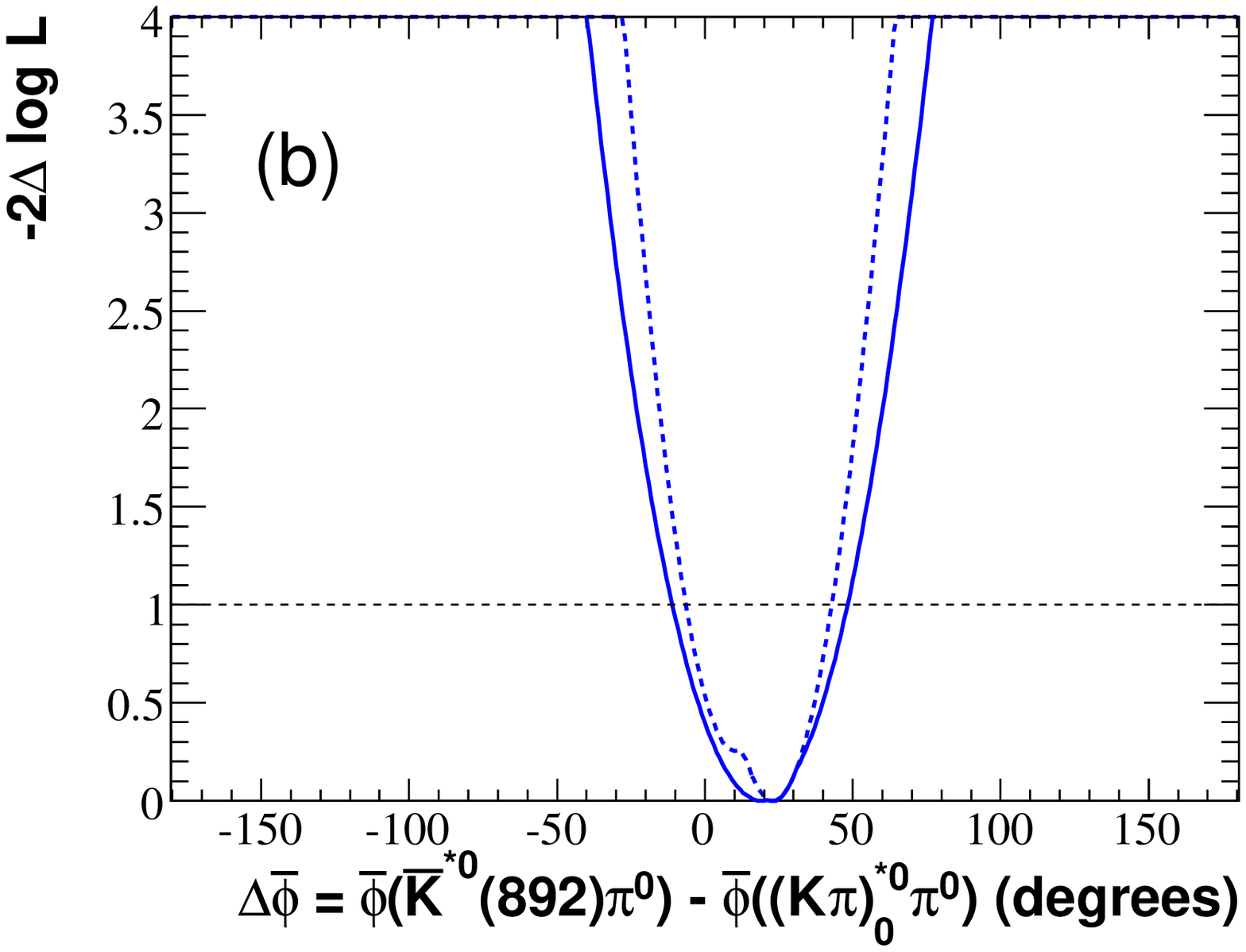,width=8.5cm}
\epsfig{file=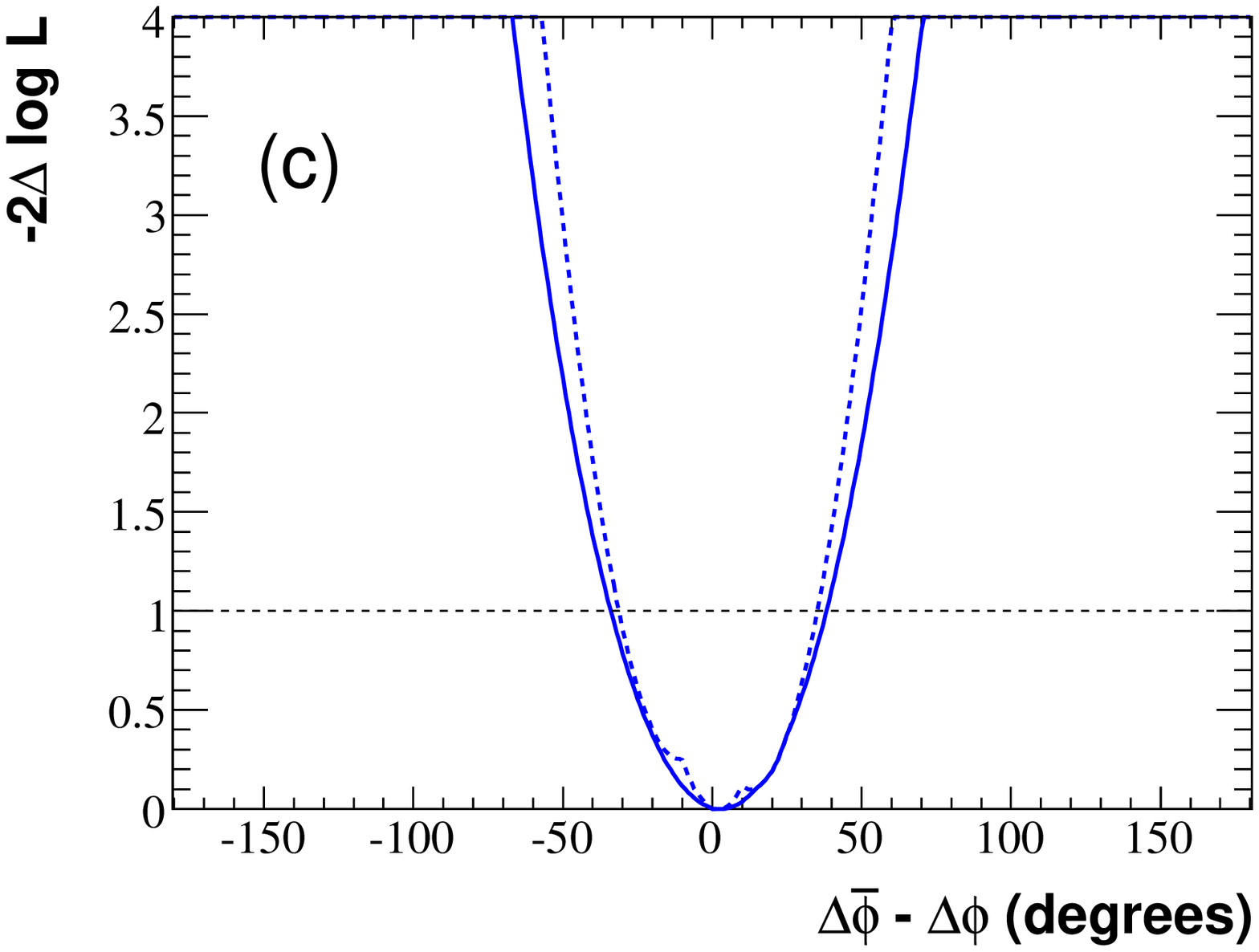,width=8.5cm} 
\caption{
\label{fig:PzerominusSzero} The phase difference between the neutral $K\pi$ P~and S-waves. 
The three diagrams are the NLL scans for the $\Bz$ (a) and  $\Bzb$ (b) decays as well as their differences (c). 
The lines are drawn as for~\figref{PzerominusPplusphasediff}. The data provide significant constraints on these angles. 
The four fit solutions find their NLL minimum at approximately the same phase differences. 
}
\end{figure}
\begin{figure}[h]
\epsfig{file=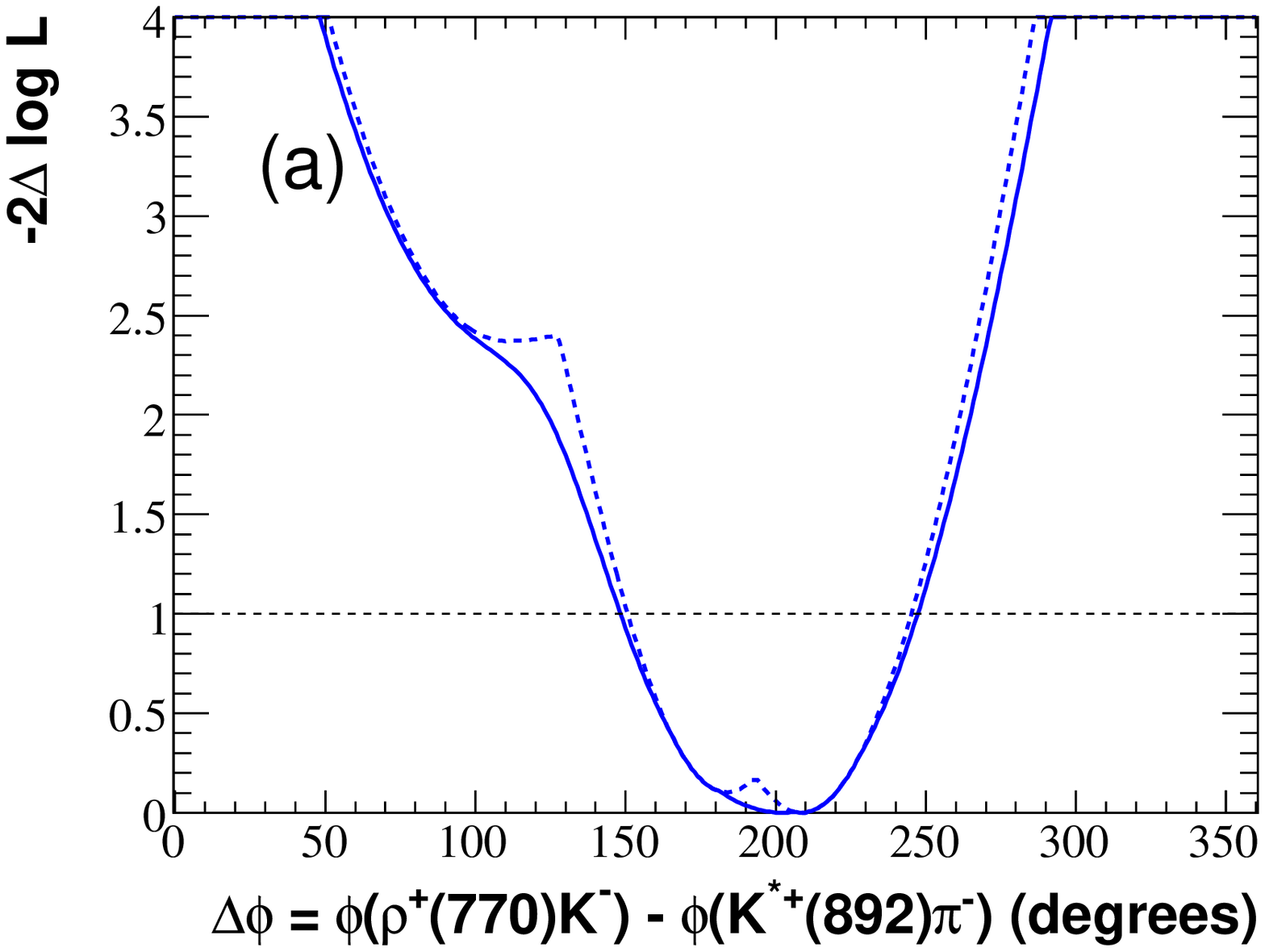,width=8.5cm} 
\epsfig{file=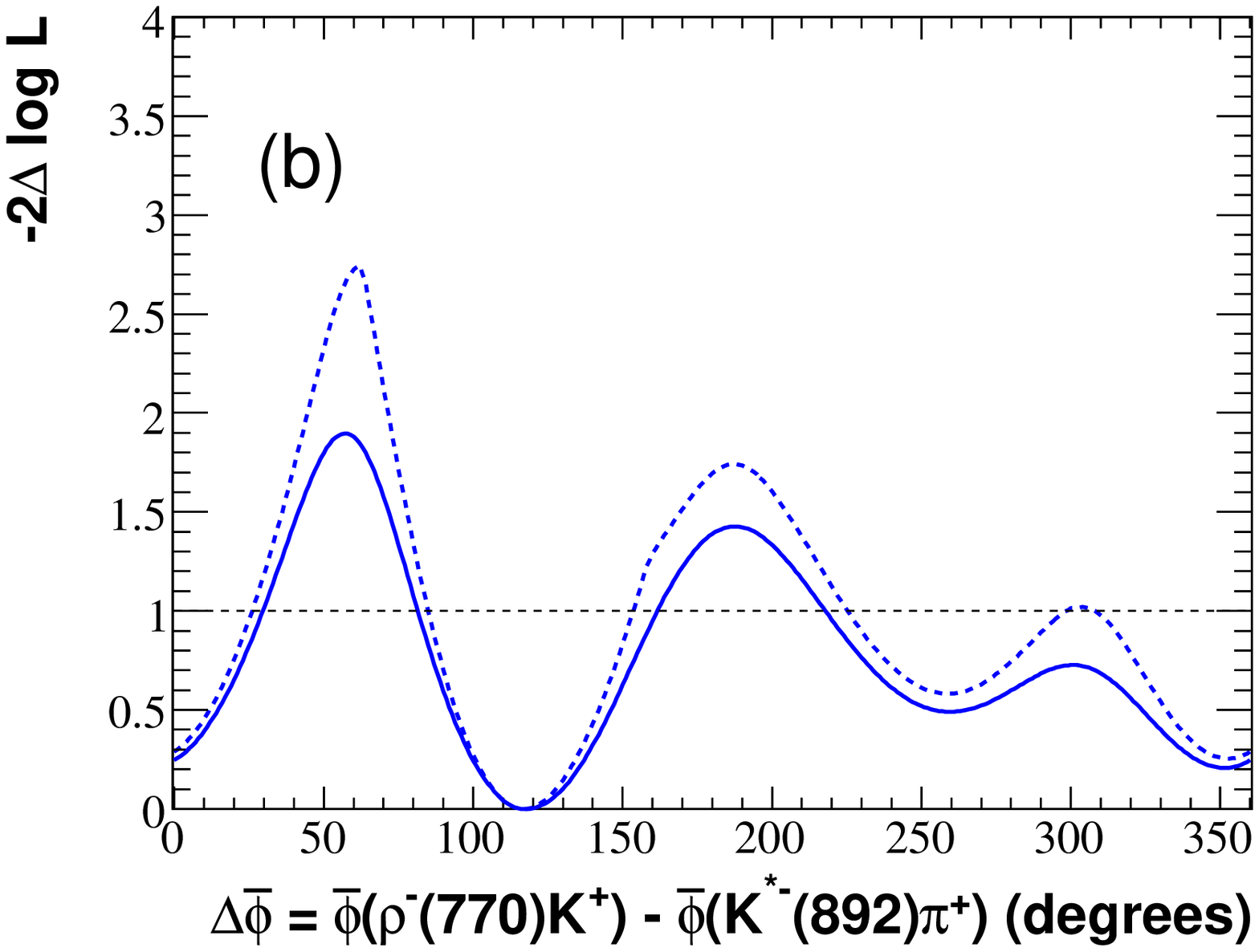,width=8.5cm}
\epsfig{file=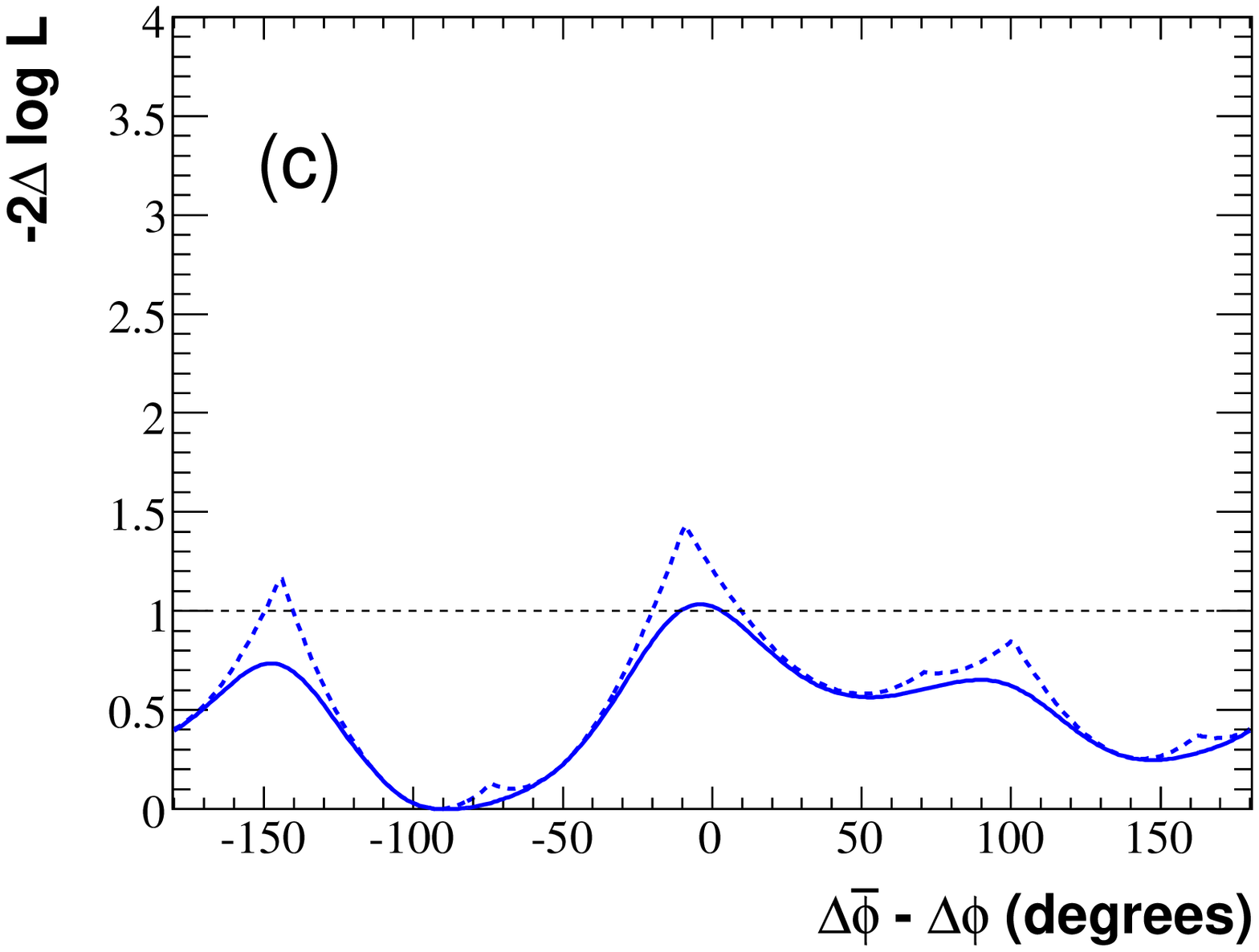,width=8.5cm} 
\caption{
\label{fig:RhoKPwave} The phase difference between the $\rho K$ and the charged $K\pi$ P-wave.
The three diagrams are the NLL scans for the $\Bz$ (a) and  $\Bzb$ (b) decays as well as their differences (c). 
The lines are drawn as for~\figref{PzerominusPplusphasediff}. The
vertical scale cuts off $\Delta\chi^2=4$, however it has been checked
that all phase differences are consistent with the data at the three standard
deviation level.
}
\end{figure}
\begin{figure}[p]
\epsfig{file=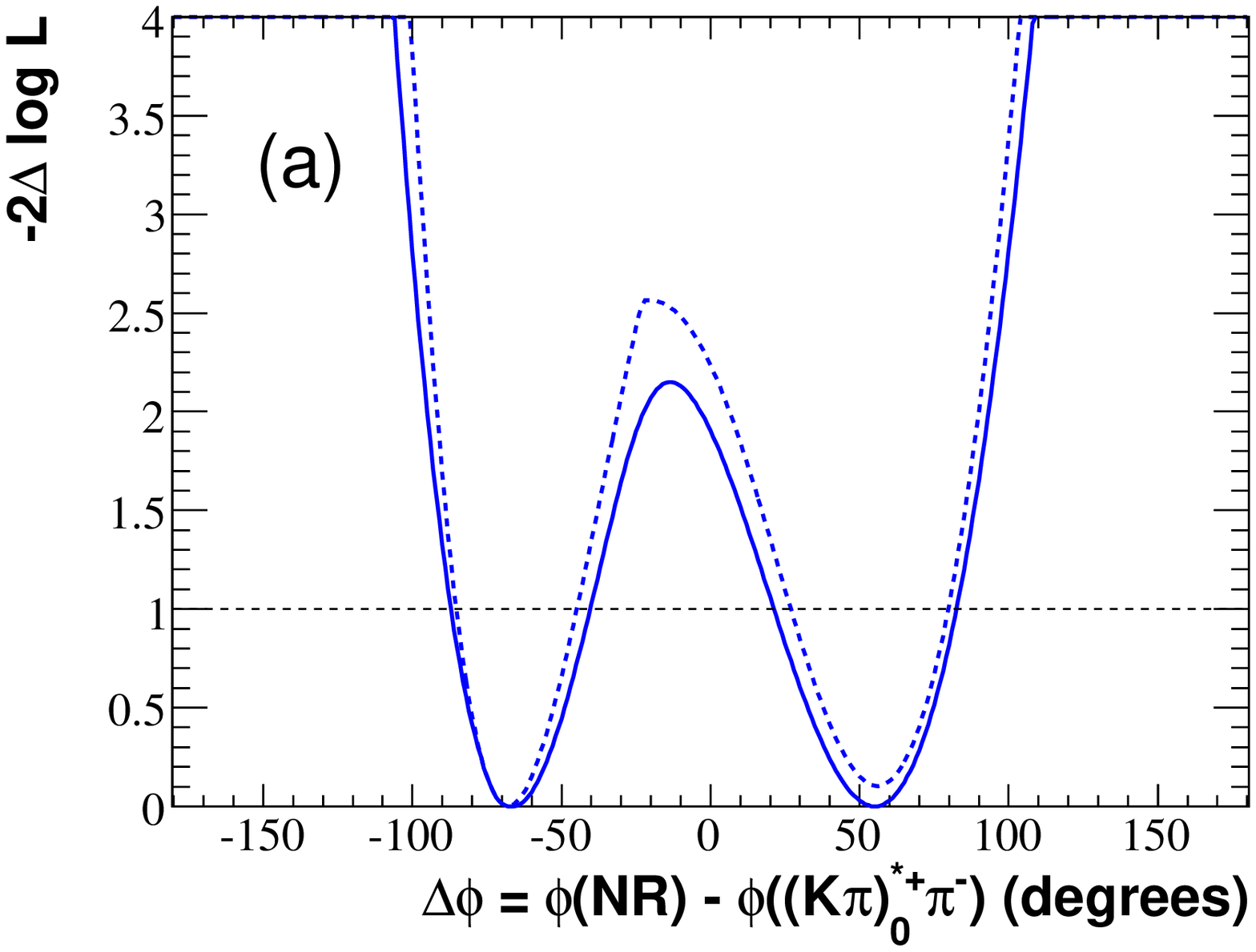,width=8.5cm} 
\epsfig{file=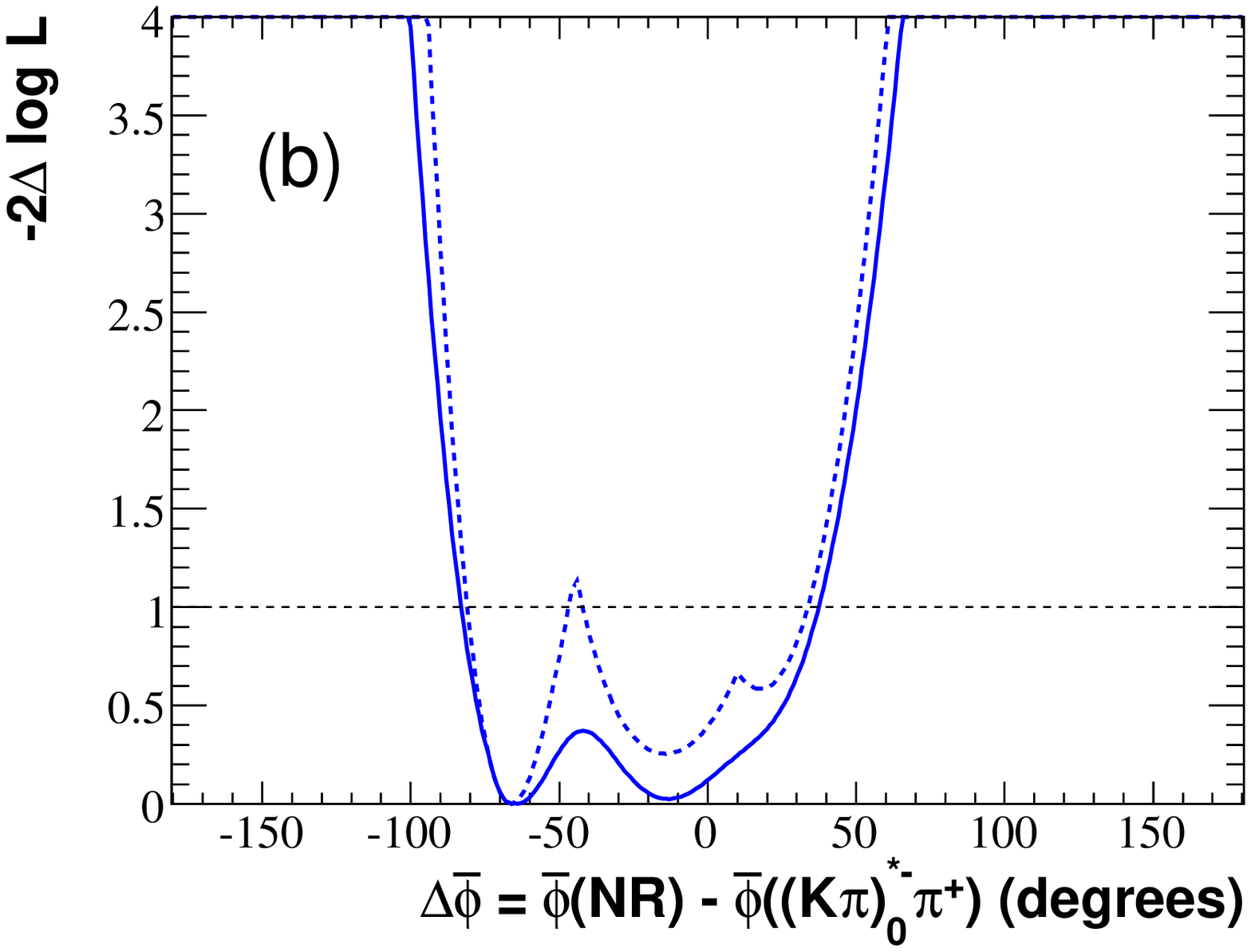,width=8.5cm}
\epsfig{file=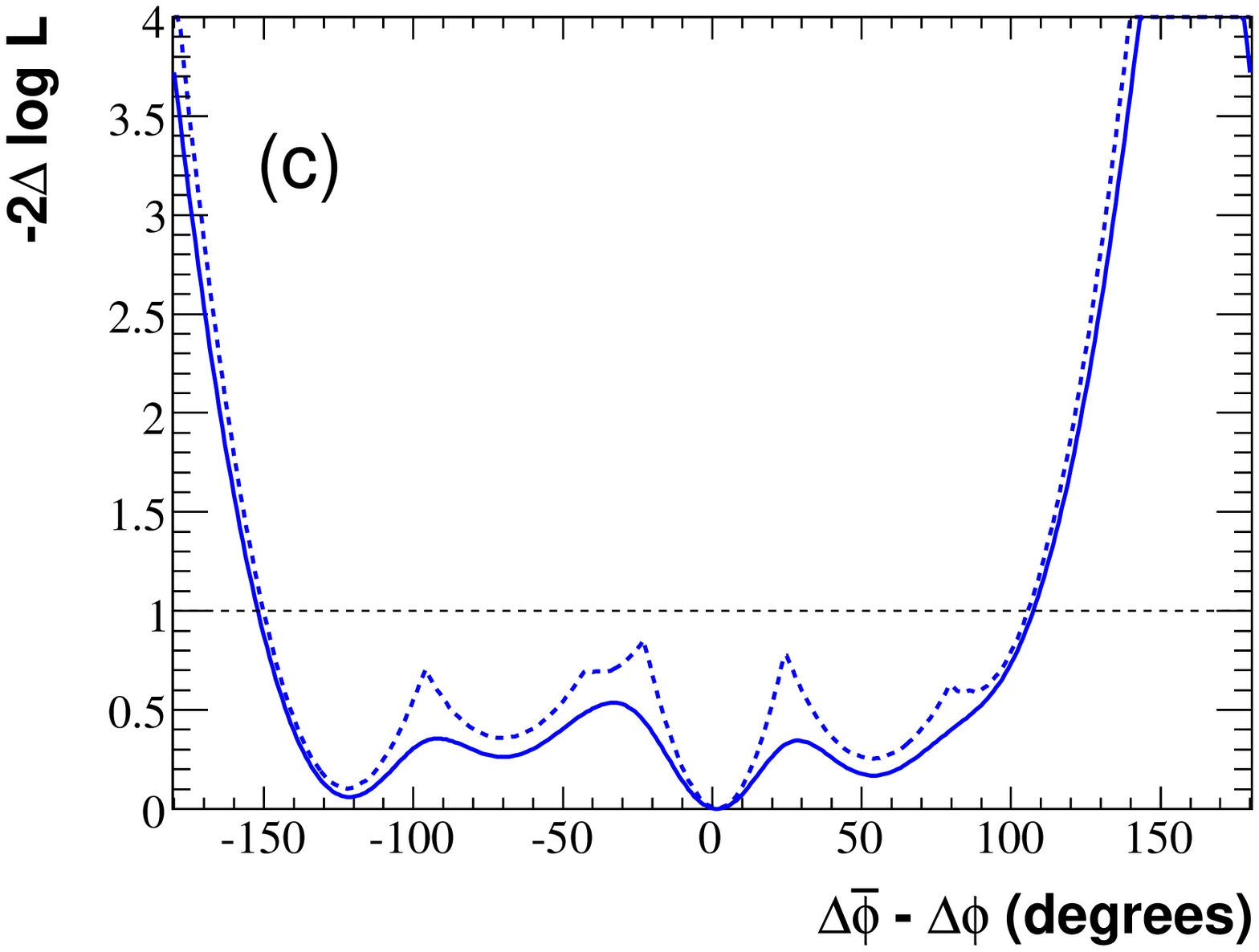,width=8.5cm} 
\caption{
\label{fig:NRminusSplus} The phase difference between the nonresonant $\Kp\pim\piz$ and the charged $K\pi$ S-wave. 
The three diagrams are the NLL scans for the $\Bz$ (a) and  $\Bzb$ (b) decays as well as their differences (c). 
The lines are drawn as for~\figref{PzerominusPplusphasediff}. Ranges
with widths of 140 (a) and 190 (b) degrees are excluded at the 95\% 
confidence level for the $\Bz$ and $\Bzb$ decays. No significant difference between the $\Bz$ and $\Bzb$ interference patterns is seen. 
}
\end{figure}
\begin{figure}[h]
\epsfig{file=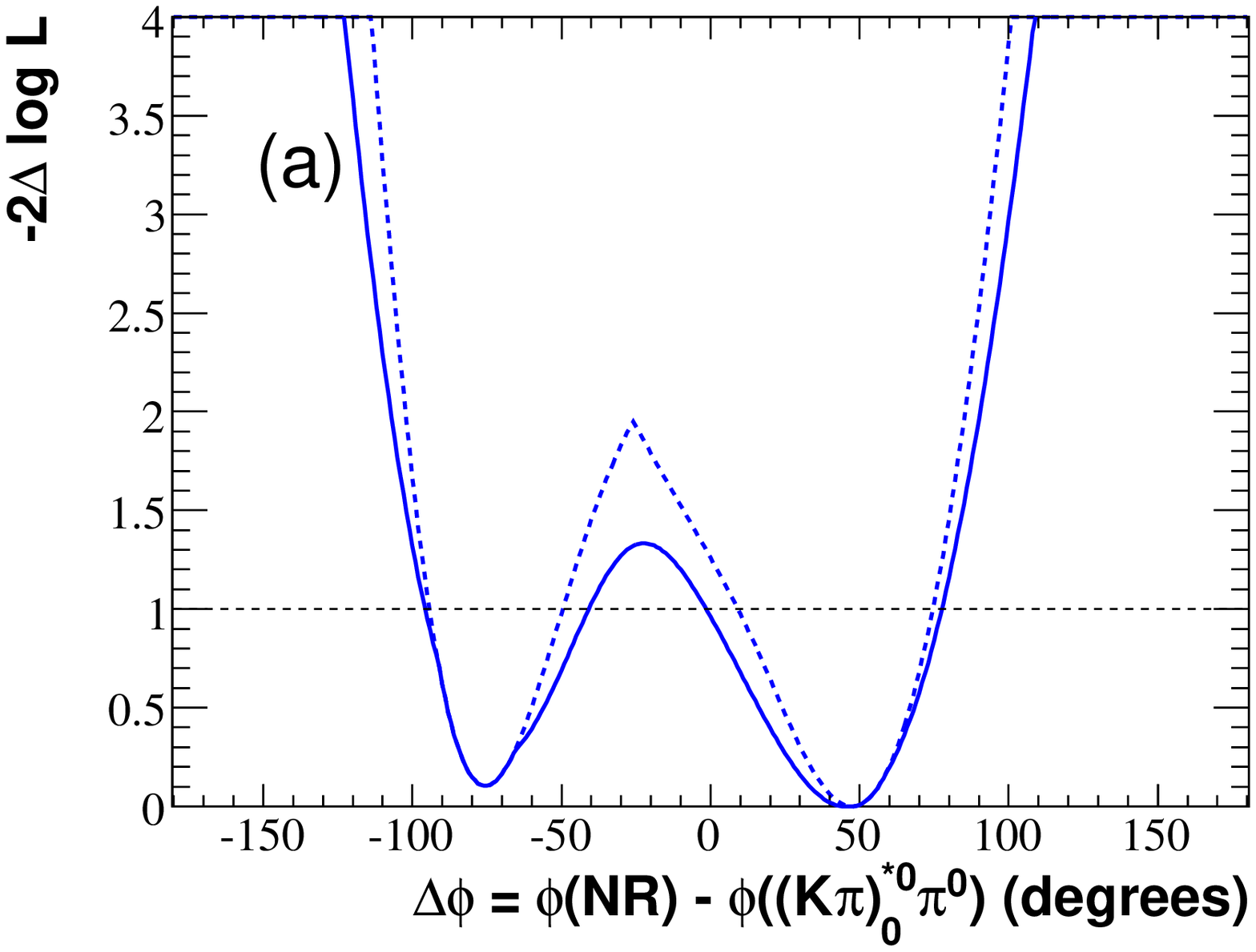,width=8.5cm} 
\epsfig{file=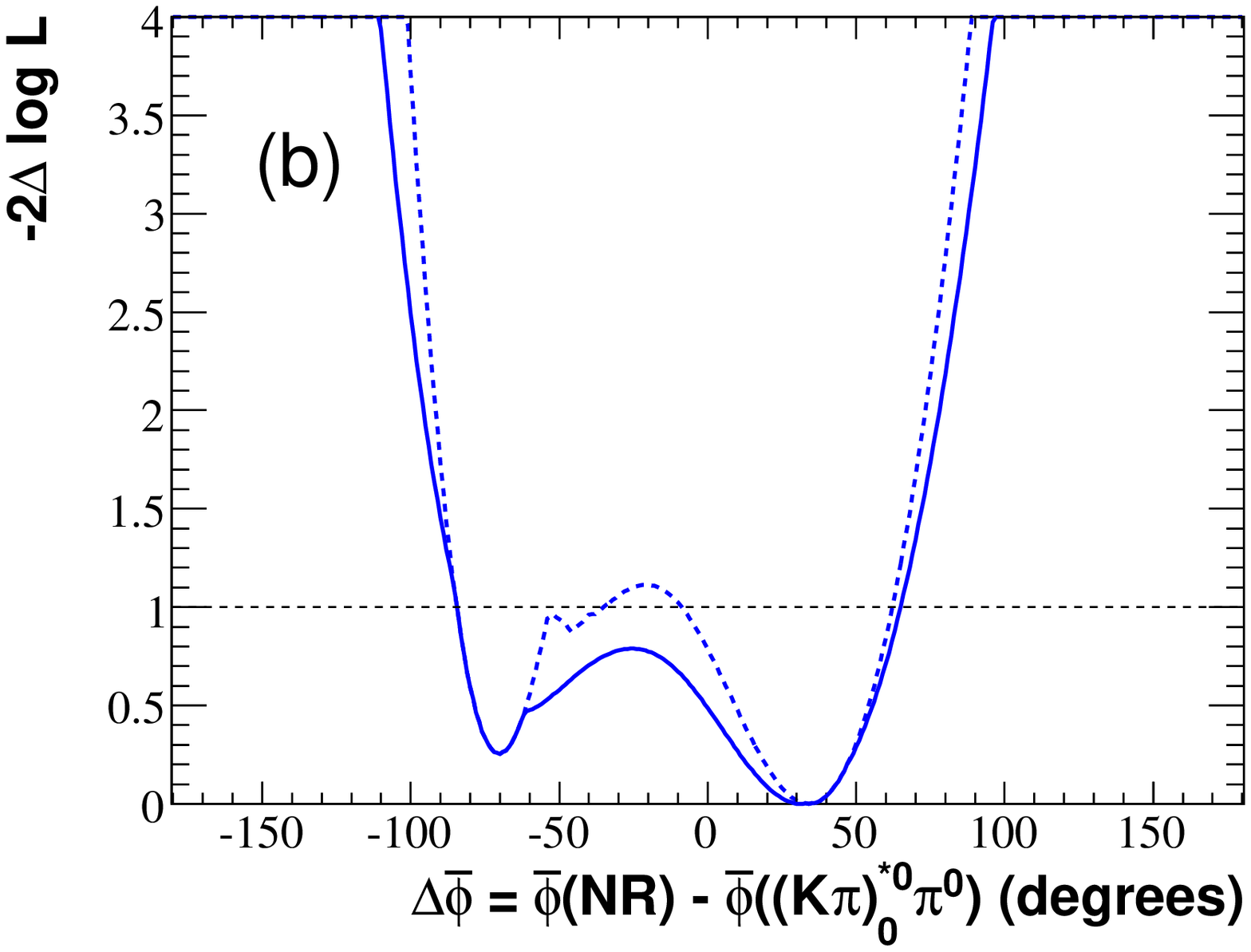,width=8.5cm}
\epsfig{file=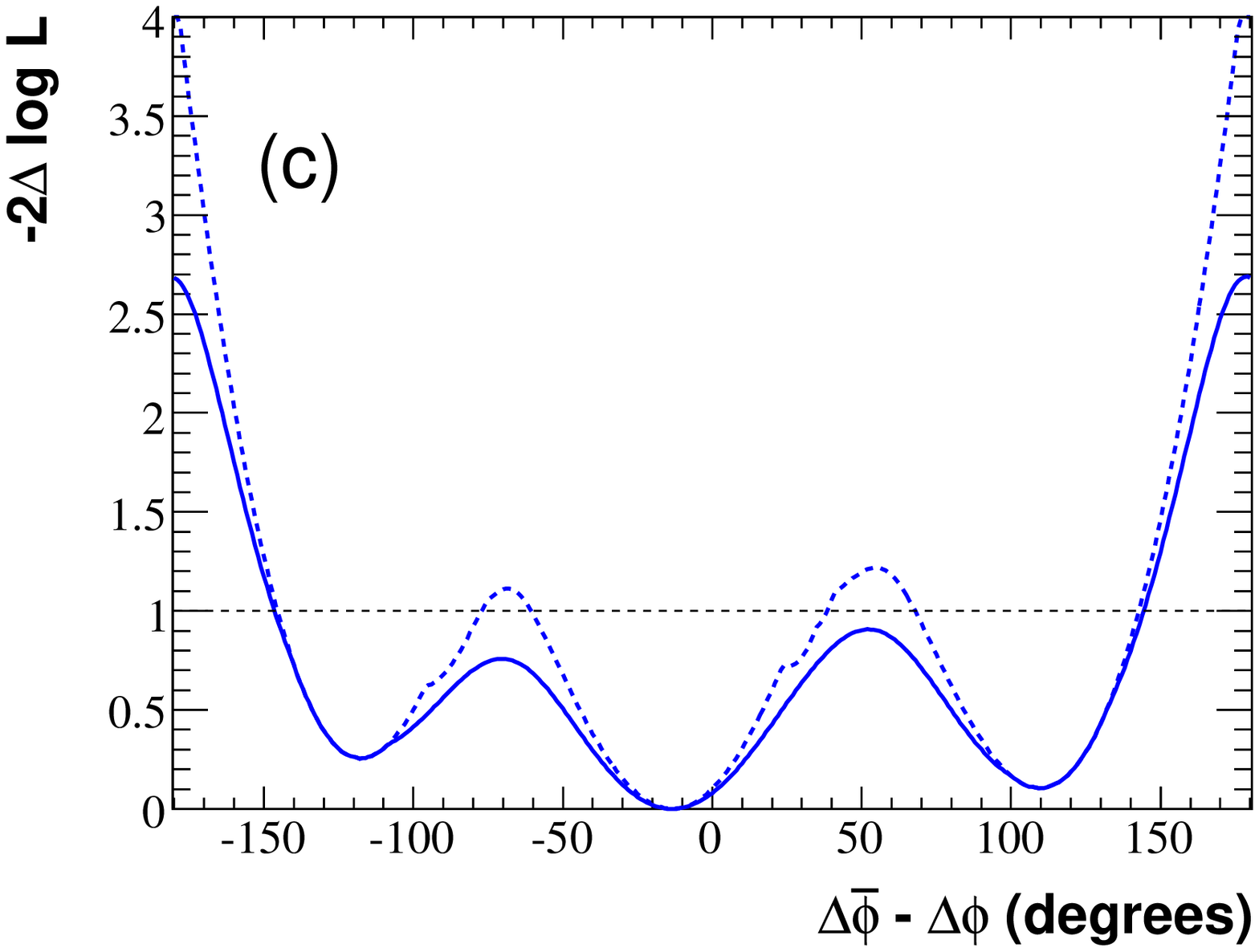,width=8.5cm} 
\caption{
\label{fig:NRminusSzero} The phase difference between the nonresonant $\Kp\pim\piz$ and the neutral $K\pi$ S-wave. 
The three diagrams are the NLL scans for the $\Bz$ (a) and  $\Bzb$ (b) decays as well as their differences (c). 
The lines are drawn as for~\figref{PzerominusPplusphasediff}. Ranges
with widths of 90 (a) and 110 (b) degrees are excluded at the 95\% 
confidence level for the $\Bz$ and $\Bzb$ decays. No significant difference between the $\Bz$ and $\Bzb$ interference patterns is seen.
}
\end{figure}
We also search for direct $CP$-violation in the interference between pairs of isobars ($R_i,\ R_j$) by comparing the interference patterns in 
the \Bz and \Bzb Dalitz plots. In the figures described in what follows, we display three NLL scans for in turn  $\Delta\phi_{ij}\equiv\phi_j-\phi_i$, 
the phase difference between the resonances in $\Bz$ decays, $\Delta\overline{\phi}_{ij}\equiv\overline{\phi_j}-\overline{\phi_i}$, 
the same for $\Bzb$ decays and $\delta\phi_{ij}\equiv\Delta\overline{\phi}_{ij}-\Delta\phi_{ij}$.
A marked minimum in a $\Delta\phi$ ($\Delta\overline{\phi}$) scan indicates a sizable interference in $\Bz$ ($\Bzb$) decays. 
Evidence for direct $CP$-violation would be seen if $\delta\phi_{min}\equiv\Delta\overline{\phi}_{min}-\Delta\phi_{min}$ were significantly away from zero. 
One standard deviation confidence intervals ($1\sigma~c.i.$) bounds are graphically seen as
the intersection points between the solid scan curves (which incorporate both statistical and systematic uncertainties) and the $\Delta\chi^2=1$ horizontal line. 
The results are collected in~\tabref{phasedifferences}. 
All isobar pairs for which the 95\% confidence intervals in $\delta\phi$ are non trivial (smaller than $\pm 180$ degrees) exhibit sizeable interference
patterns in \Bz and \Bzb decays. The $\Delta\chi^2$ value for $\delta\phi=0$
corresponds to the square of the direct $CP$-violation significance in standard deviation units. A scan of the $R_i$ and $R_j$ lineshapes
with more statistics could enable one to disentangle the strong phase motions and determine the weak phase. For the scans where interference is elusive, we quote 
in~\tabref{phasedifferences} the maximum $\Delta\chi^2$ over the $\pm 180$ degrees scanned 
range. For the $K\pi$ systems, we see no significant interference pattern between charged and neutral 
P-waves (\figref{PzerominusPplusphasediff}) at the 95\% confidence level;  similarly for the S-waves (\figref{SzerominusSplusphasediff}). Only a
small range ($\Delta\overline{\phi}$ between -131 and -75 degrees) is excluded at the two standard deviation level for the
\Bzb P-waves. These observations are not unexpected for
the $\Kstar(892)\pi$ final states since the \Kstar resonances are quite narrow and therefore have a small overlap in the Dalitz plot. Furthermore the coherent
sum of amplitudes that interfere might behave like the model in ~\cite{Ciuchini:2006kv,Gronau:2006qn} with a single weak phase (equal to $\gamma_{\rm CKM}$ 
in the absence of electroweak penguin diagrams). In such a scheme, one weak phase would be missing to enable direct $CP$-violation to take place.
In contrast, \figref{PplusminusSplus} and~\figref{PzerominusSzero} show that we measure, with uncertainties smaller than $\pm 36$~degrees, 
the phase differences between the $K\pi$ S- and P-waves of the same charge for the \Bz and \Bzb decays.
Moreover, the associated $CP$-observables $\delta\phi$ are measured to 35~degrees with negligible systematic uncertainties (less than 10 degrees). 
An interval of order 220 degrees is excluded at the 95\% confidence level. The charged and neutral S- and P-wave interference patterns thus provide 
sensitivity to two weak phases. \par
The scans of the phase differences between $\rho K$ and
$\Kstar(892)\pi$ show no evidence for interference at the three
standard deviation level as shown in~\figref{RhoKPwave}. 
Here again the overlap in phase space between the interfering resonances is small.  
This contrasts with what we observe in the scans of the phase differences between the nonresonant $\Kpm\pimp\piz$ and the S-waves in~\figref{NRminusSplus}
and~\figref{NRminusSzero}. We see that they are somewhat constrained. These observations are in agreement with the fact that the fit finds a 
sizable nonresonant $\Bz\to\Kp\pim\piz$ component that populates the Dalitz plot far from the boundary.\par
\begin{table}
\begin{center}
\caption{\label{tab:isospin-corrected-bf}
The branching fractions $\mathcal{B}$ of \B decays to quasi-two-body final states
assuming that all $\Kstar$ resonances are isospin-$1/2$ states. The
branching fractions of the $K^{*}_2(1430)$ and $K^{*}(1680)$ resonances to $K\pi$ from
reference~\cite{PDG2006} have been used. The upper limits
at 90\% confidence level, $\mathcal{UL}$ are based on statistical uncertainties only.}
\setlength{\tabcolsep}{1.2pc}
\begin{tabular}{ll}
\hline\hline
\B decay final state                                &\multicolumn{1}{c}{$\mathcal{B}$ ($10^{-6}$)}  \\\hline
$K^{*+}(892)\pi^{-}$                                & $ 12.6_{-1.6}^{+2.7}\pm0.9$       \\
$K^{*0}(892)\pi^{0}$                                & $\ 3.6\pm0.7\pm0.4$       \\
$(K\pi)^{*+}_0\pi^{-}$;~$(K\pi)^{*+}_0 \to \Kp\piz$ & $\ 9.4_{-1.3}^{+1.1}$$_{-1.1}^{+1.4}\pm1.8$  \\
$(K\pi)^{*0}_0\pi^{0}$;~$(K\pi)^{*0}_0 \to \Kp\pip$ & $\ 8.7_{-0.9}^{+1.1}$$_{-1.3}^{+1.8}\pm2.2$  \\
$\rho^{-}(770)K^{+}$        & $\ 8.0_{-1.3}^{+0.8}\pm0.6$       \\
N.R.                        & $\ 4.4\pm0.9\pm0.5$       \\ \hline
                            &                                                    \\
                            &\multicolumn{1}{c}{$\mathcal{UL}$ ($10^{-6}$)} \\
$\rho^-(1450)K^+$           & $\ 2.1$                                              \\
$\rho^-(1700)K^+$           & $\ 1.1$                                              \\
$K^{*0}_2(1430)\piz$        & $\ 4.0$                                              \\
$K^{*+}_2(1430)\pim$        & $16.2$                                              \\
$\Kstarz(1680)\piz$         & $\ 7.5$                                              \\
$\Kstarp(1680)\pim$         & $25.$                                              \\ 
                            &                                                    \\
\hline\hline
\end{tabular}
\end{center}
\end{table}

\section{SUMMARY}
\label{sec:Summary}
%In this paper 
We have measured the branching fraction and $CP$-asymmetry of the $\Bpm\to\Kpm\pimp\piz$ decay and compared the Dalitz plots of 
$\Bz\to\Kp\pim\piz$ and $\Bzb\to\Km\pip\piz$ using an isobar model. We have extracted the $CP$-averaged isobar branching fractions and $CP$-asymmetries
assuming no interference. We observe the $\Bz\to\Kstarz(892)\piz$ with 5.6 standard deviation  significance. 
We have looked at the interference patterns in the Dalitz plots and put significant constraints on phase differences between wide
intermediate states which have a sizable overlap in phase space. The phase shifts between S and P-waves in the charged and neutral $K\pi$ and $\Kbar\pi$
are constrained to within $\pm 70$ degrees or less at the 95\% confidence level. Weaker constraints are observed for the phase shifts between the 
$K(\Kbar)\pi$ and nonresonant components which extend widely over the Dalitz plots. The phase shift differences between $K\pi$ and $\Kbar\pi$ S and P-waves 
are measured and found to be consistent with no direct $CP$-violation within 2~standard deviations. Additionally we 
determine the branching fraction for the decay $\Bz\to\Dzb\piz$ with an accuracy comparable to that of the world average value of this quantity.  
\clearpage

\section{Acknowledgments}
\label{sec:acknowledgments}
We are grateful for the 
extraordinary contributions of our \pep2\ colleagues in
achieving the excellent luminosity and machine conditions
that have made this work possible.
The success of this project also relies critically on the 
expertise and dedication of the computing organizations that 
support \babar.
The collaborating institutions wish to thank 
SLAC for its support and the kind hospitality extended to them. 
This work is supported by the
US Department of Energy
and National Science Foundation, the
Natural Sciences and Engineering Research Council (Canada),
the Commissariat \`a l'Energie Atomique and
Institut National de Physique Nucl\'eaire et de Physique des Particules
(France), the
Bundesministerium f\"ur Bildung und Forschung and
Deutsche Forschungsgemeinschaft
(Germany), the
Istituto Nazionale di Fisica Nucleare (Italy),
the Foundation for Fundamental Research on Matter (The Netherlands),
the Research Council of Norway, the
Ministry of Education and Science of the Russian Federation, 
Ministerio de Educaci\'on y Ciencia (Spain), and the
Science and Technology Facilities Council (United Kingdom).
Individuals have received support from 
the Marie-Curie IEF program (European Union) and
the A. P. Sloan Foundation.

%\clearpage
\section*{APPENDIX}
\label{sec:appendix}
\begin{figure*}
\centerline{\epsfxsize9cm\epsffile{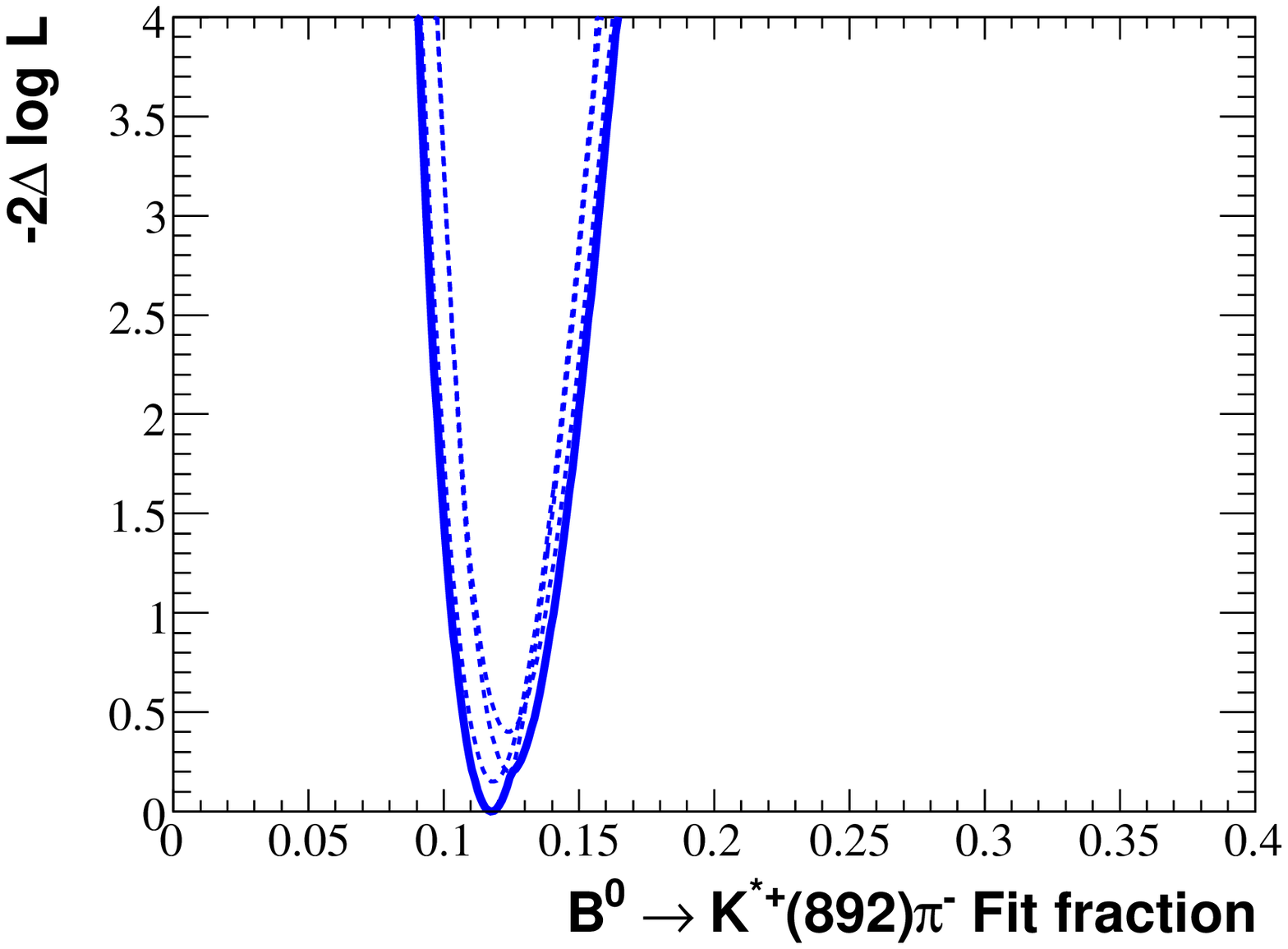} \epsfxsize9cm\epsffile{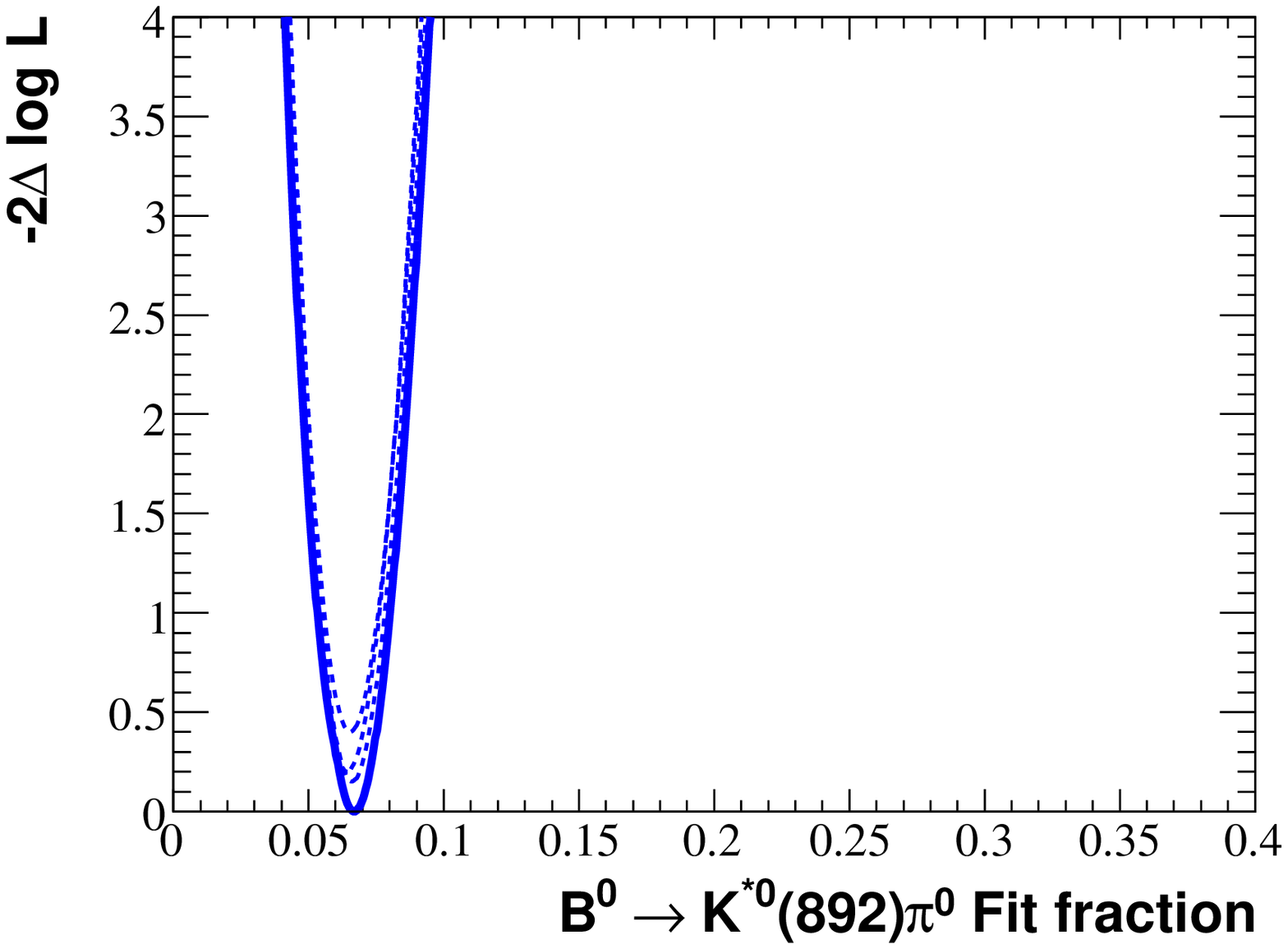} }
\centerline{\epsfxsize9cm\epsffile{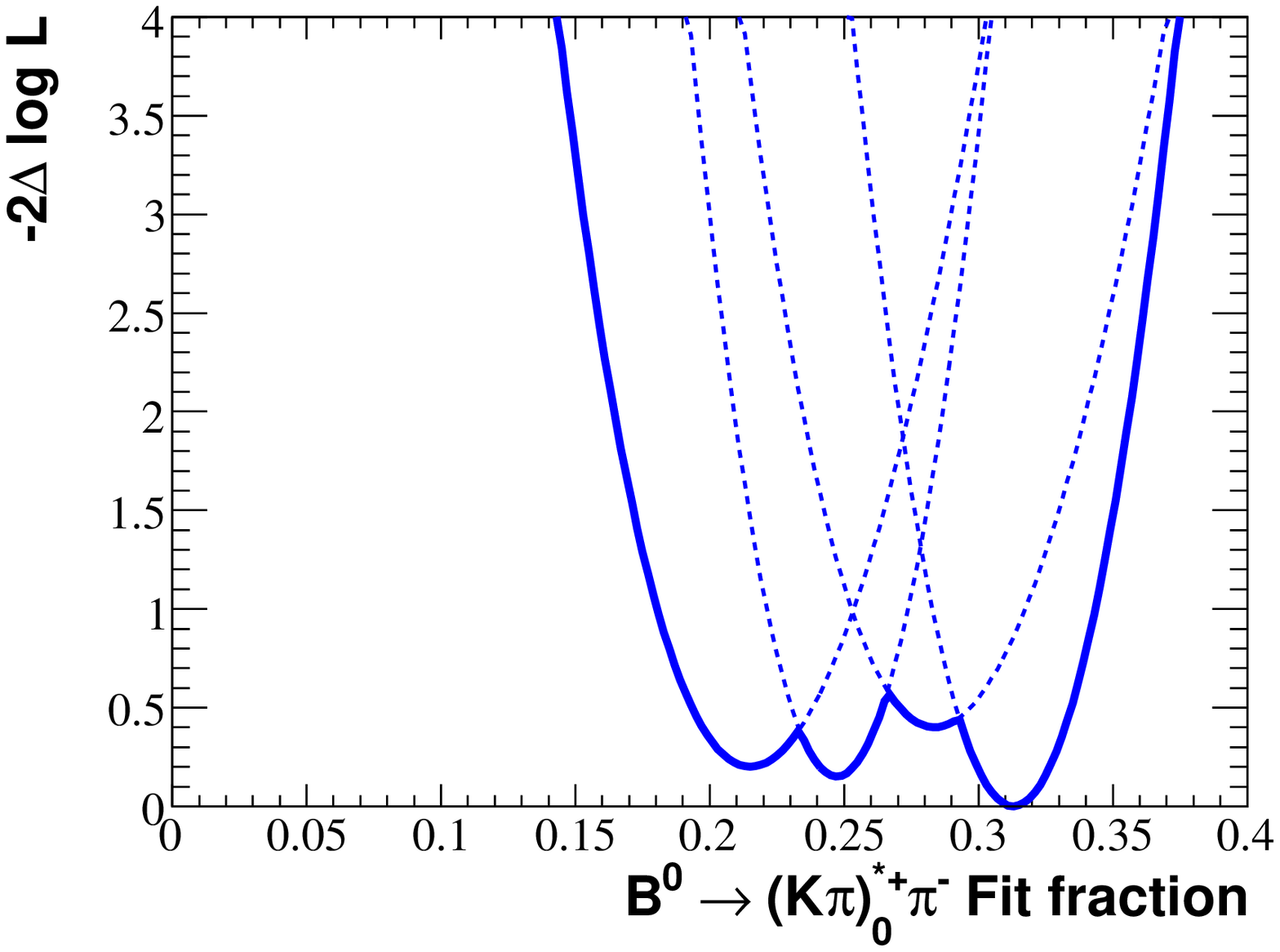} \epsfxsize9cm\epsffile{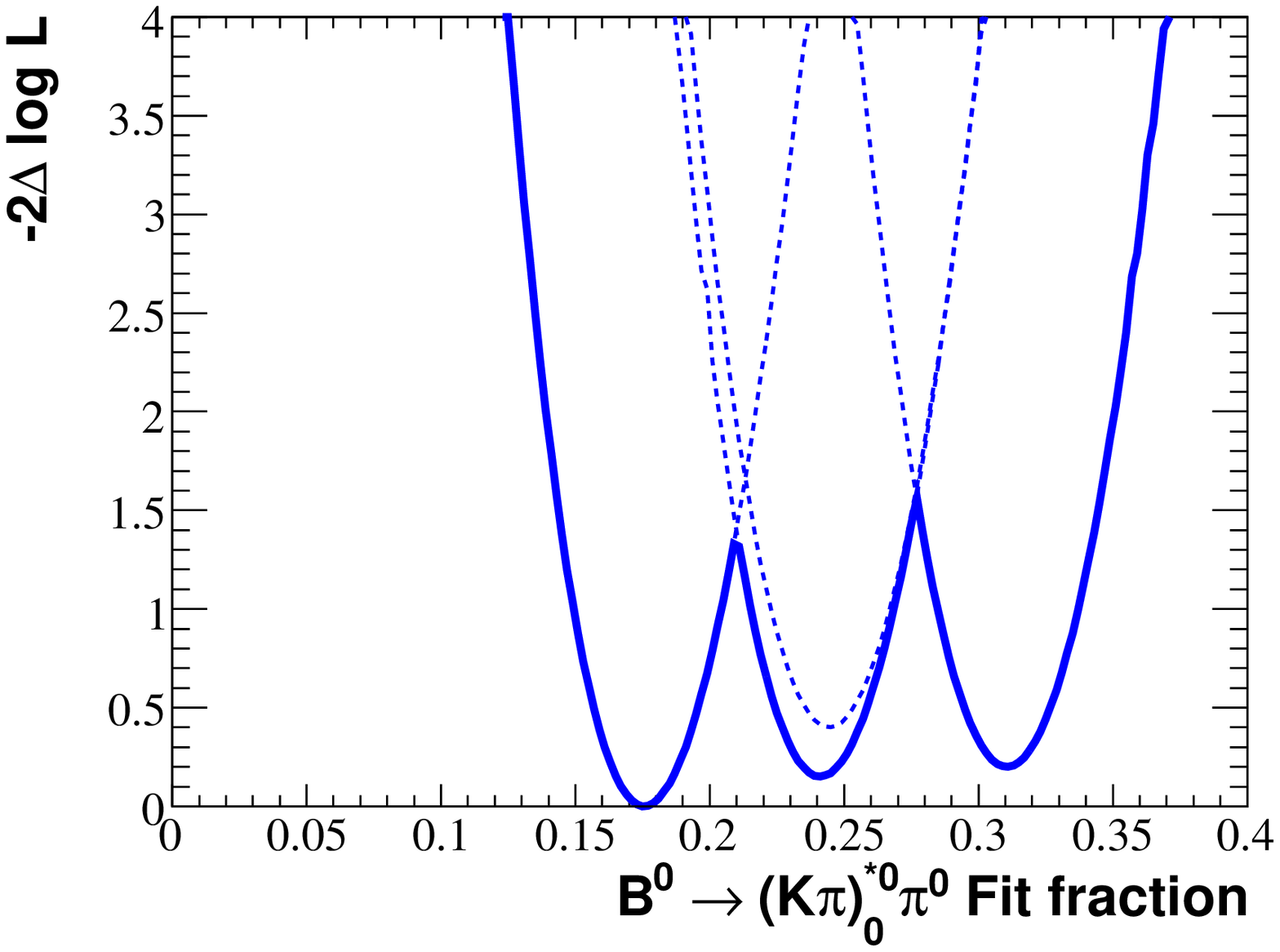} }
\centerline{\epsfxsize9cm\epsffile{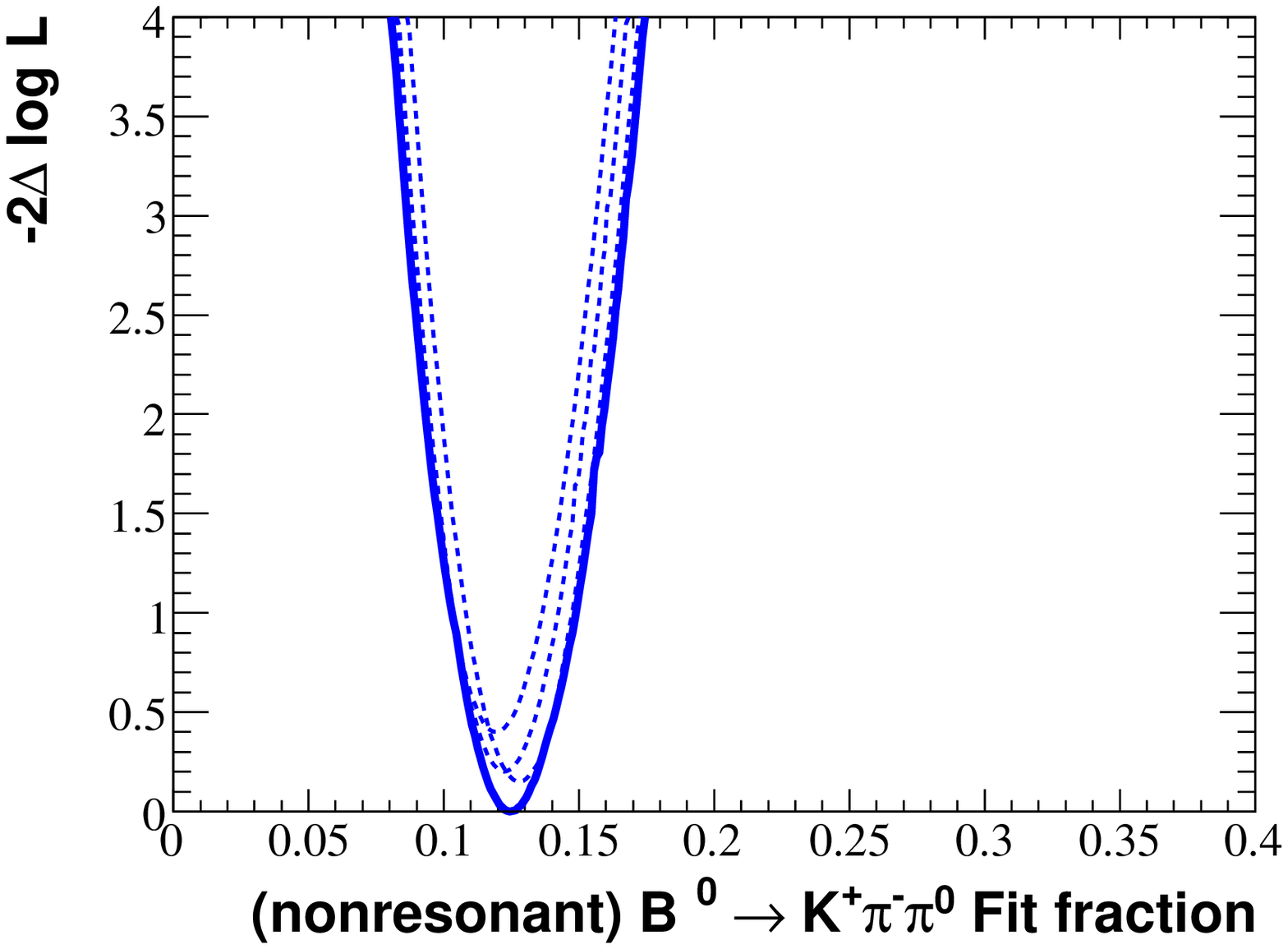} \epsfxsize9cm\epsffile{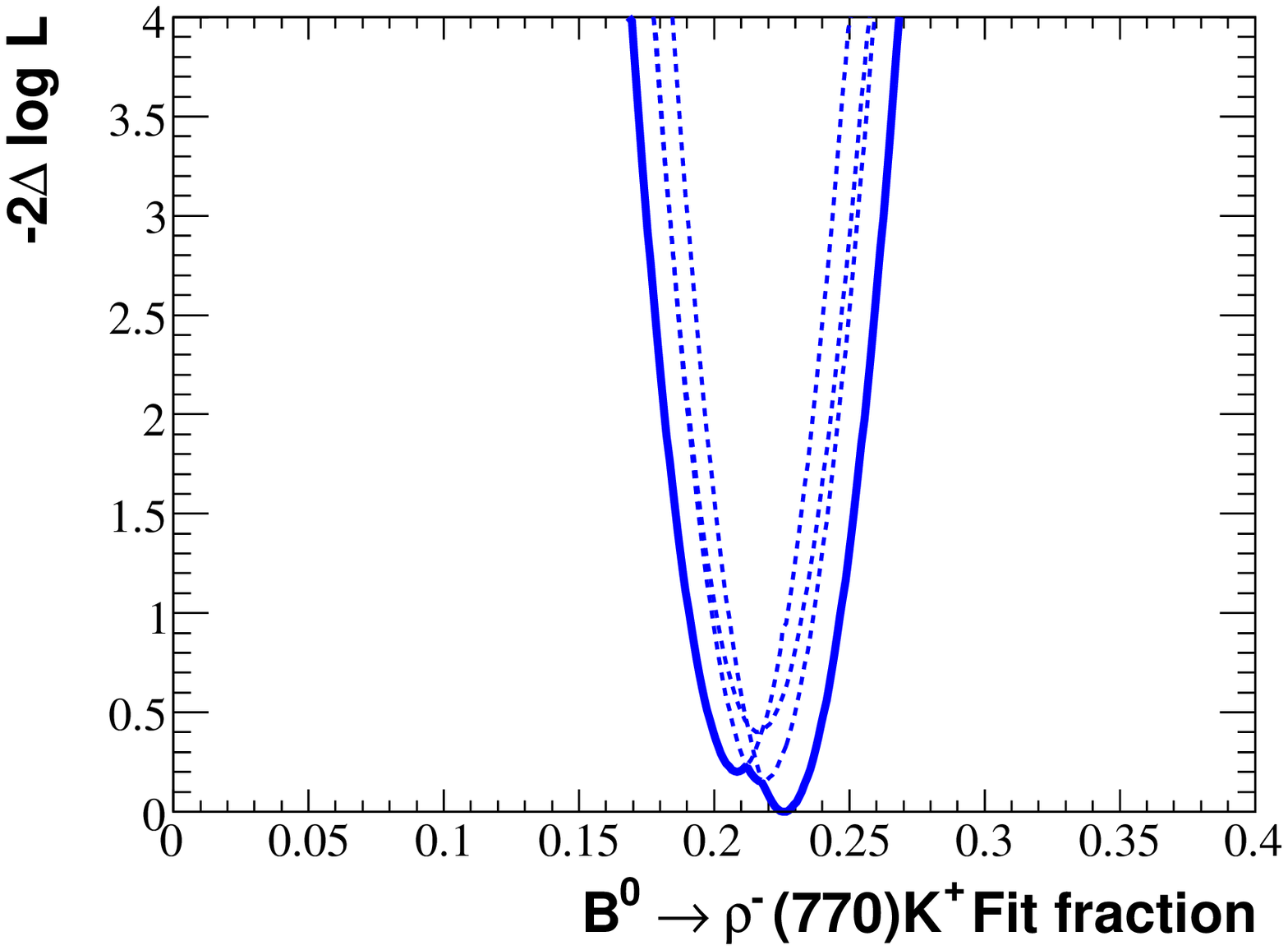} }
\caption{
\label{fig:neutralSwave-ff-scan} The NLL functions for the isobar fractions.  The NLL functions of each solution are shown with dashed lines.
The fitted values for the $\B\to(K\pi)^{*0}_0\pi$ in the four solutions are quite distinct.
The envelope curves (solid lines) are used to quote the physical results. }
\end{figure*}

\begin{figure*}
\centerline{\epsfxsize9cm\epsffile{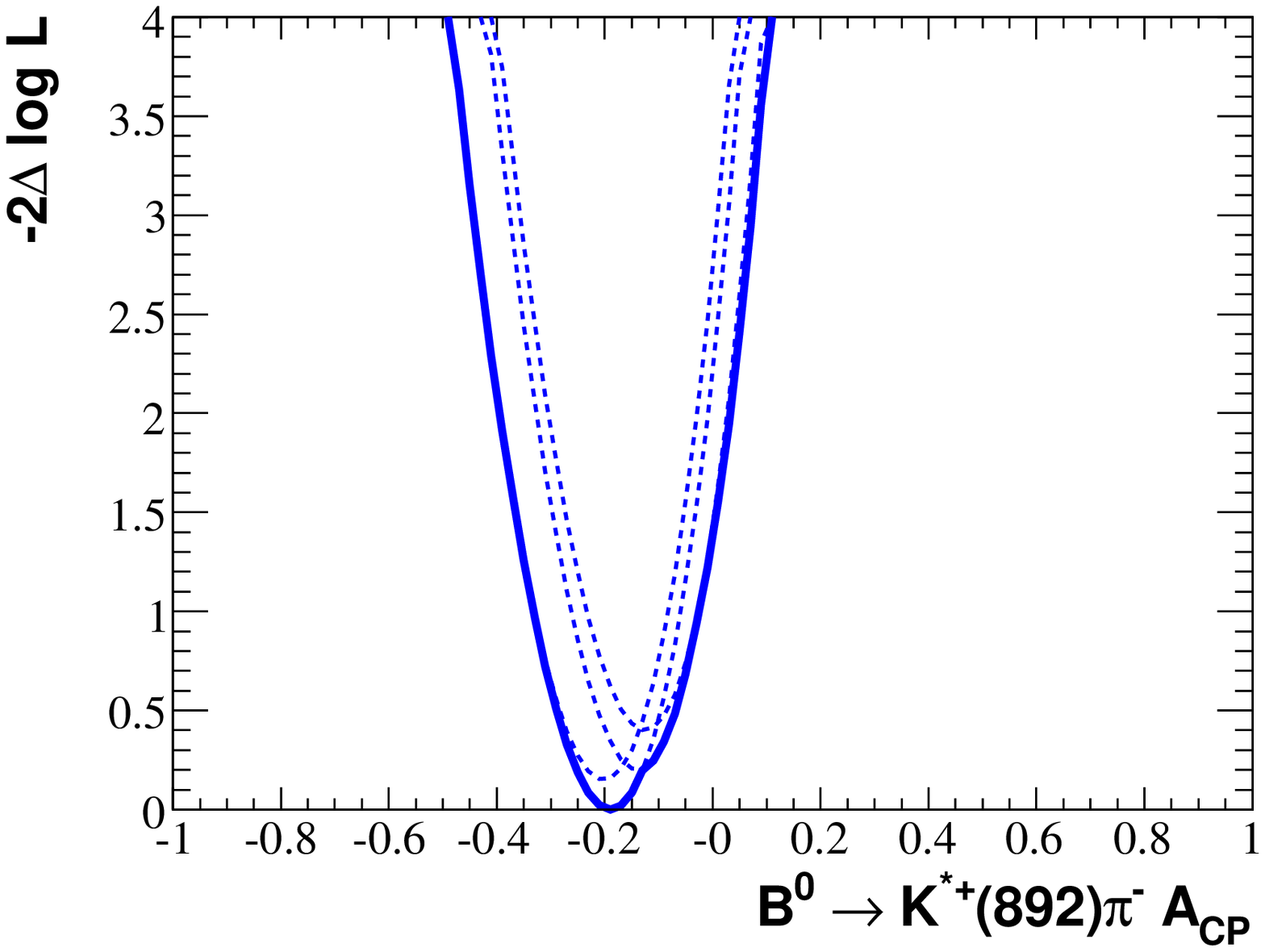} \epsfxsize9cm\epsffile{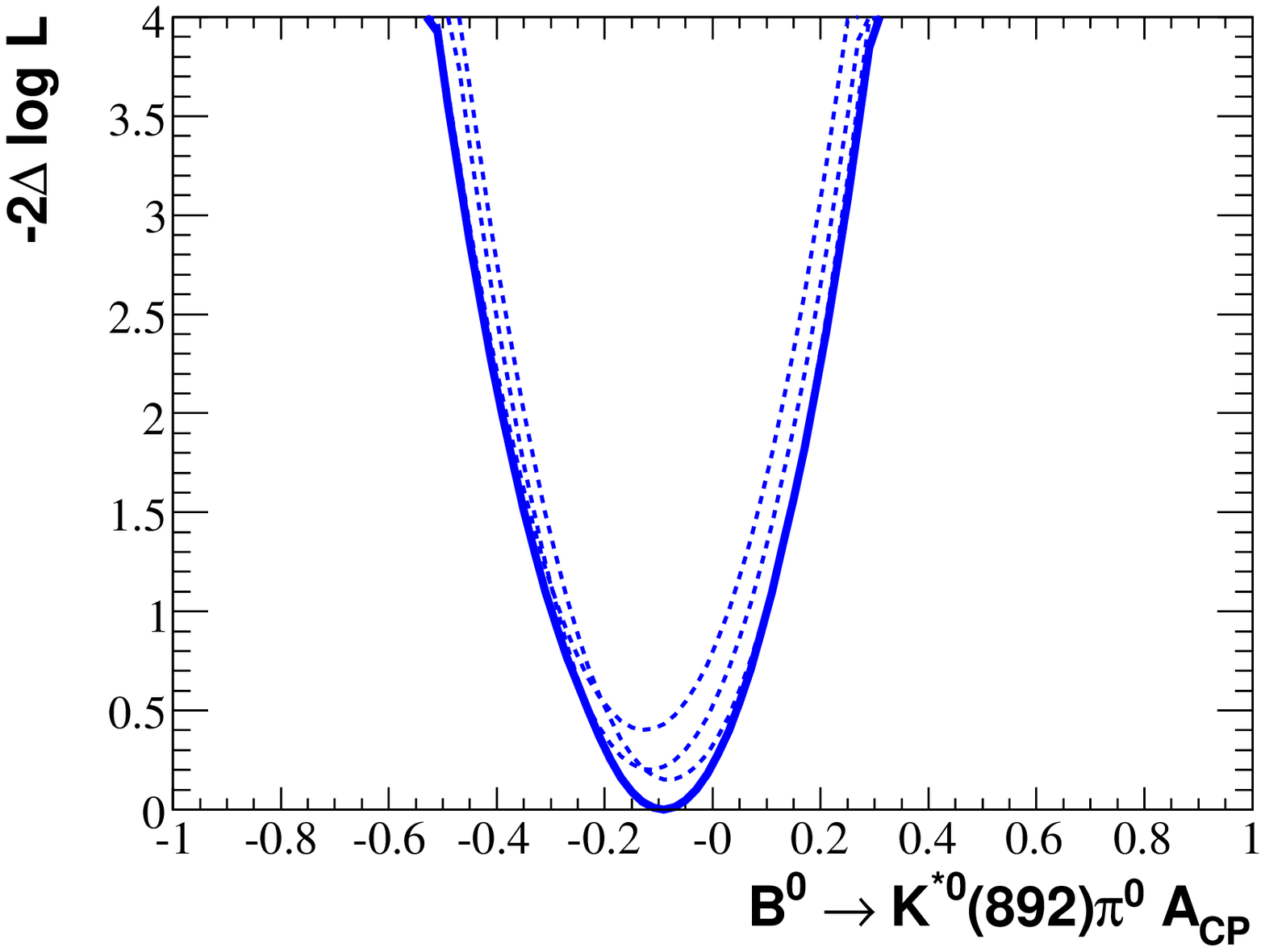} }
\centerline{\epsfxsize9cm\epsffile{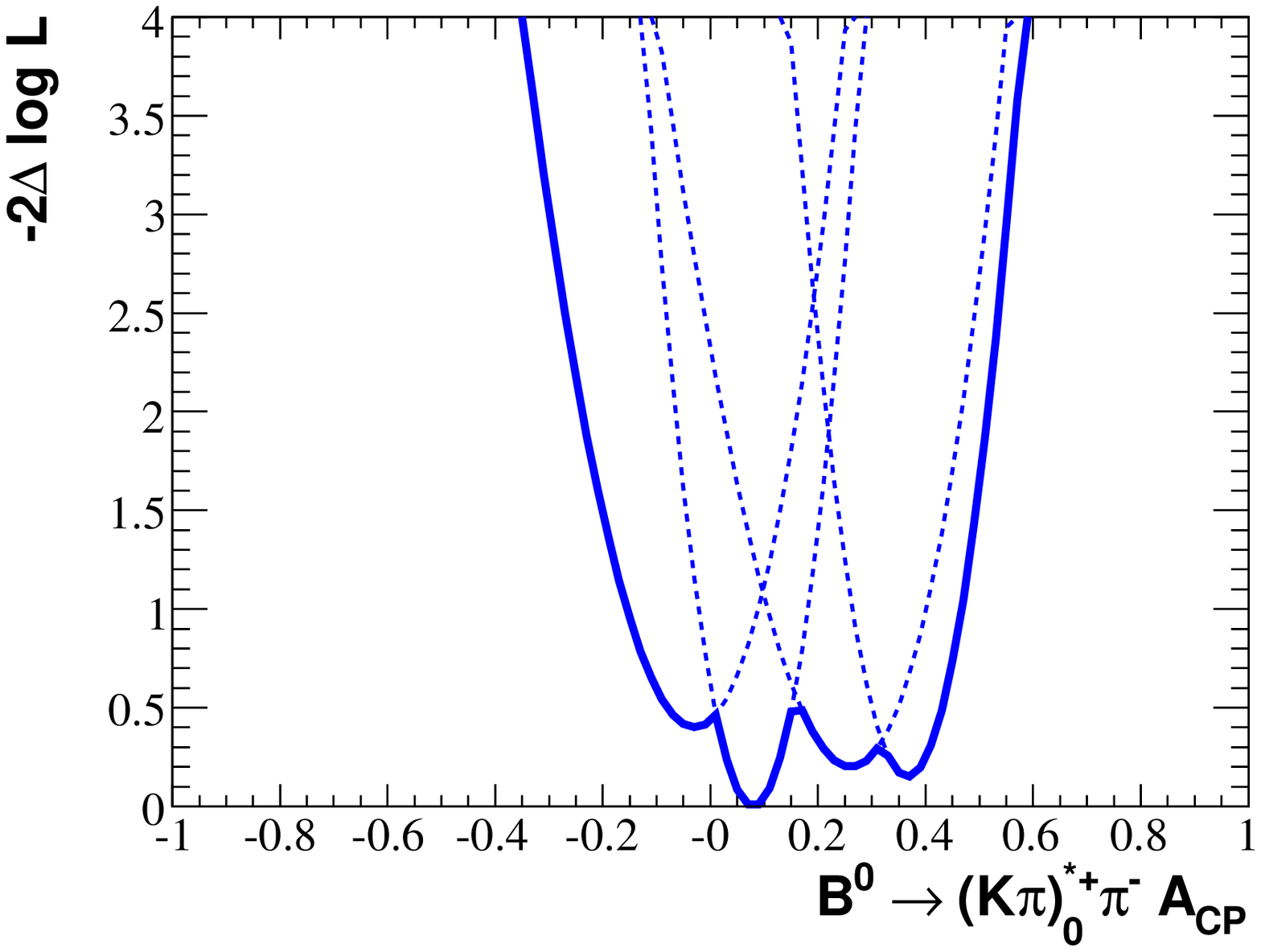} \epsfxsize9cm\epsffile{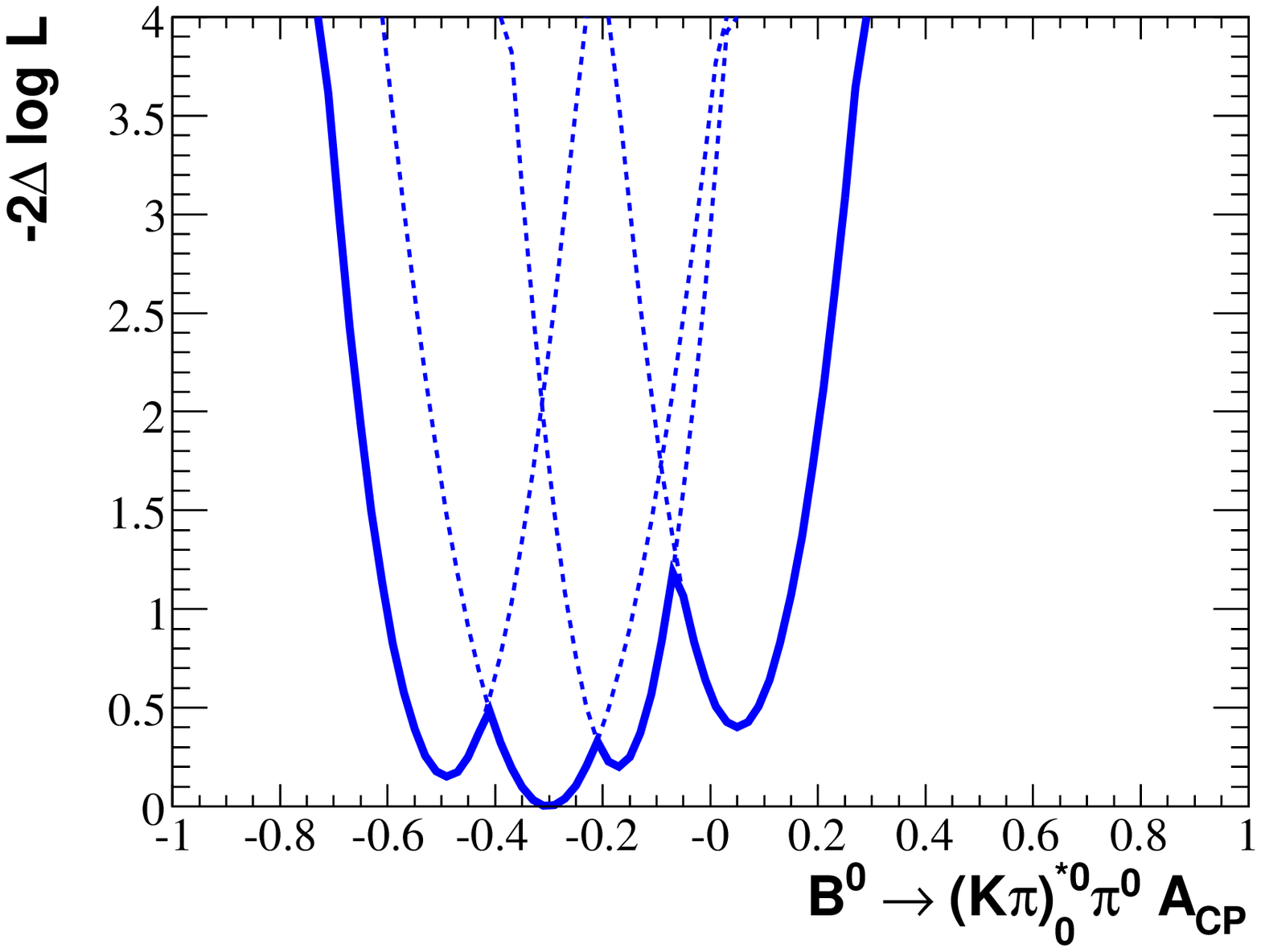} }
\centerline{\epsfxsize9cm\epsffile{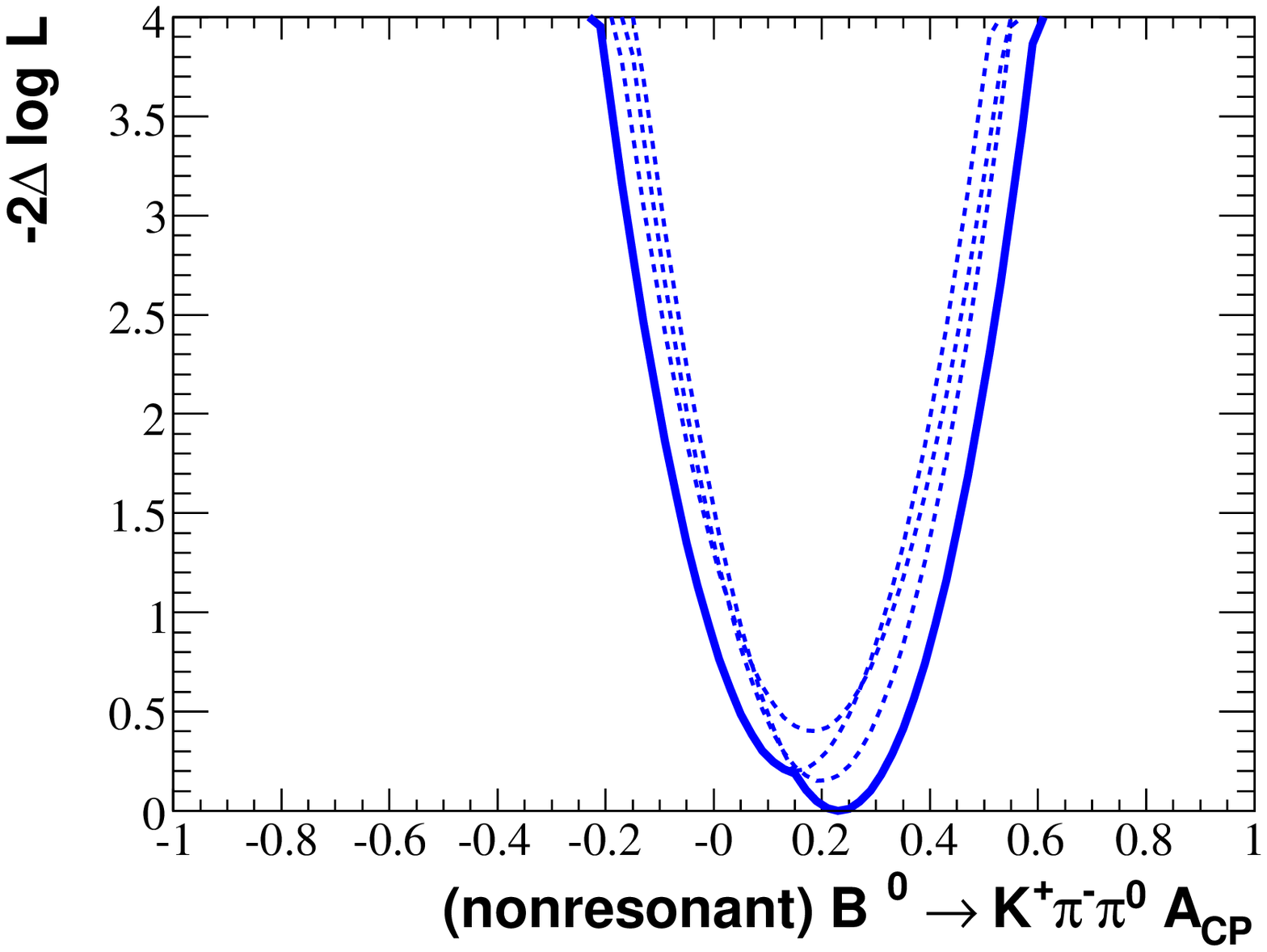} \epsfxsize9cm\epsffile{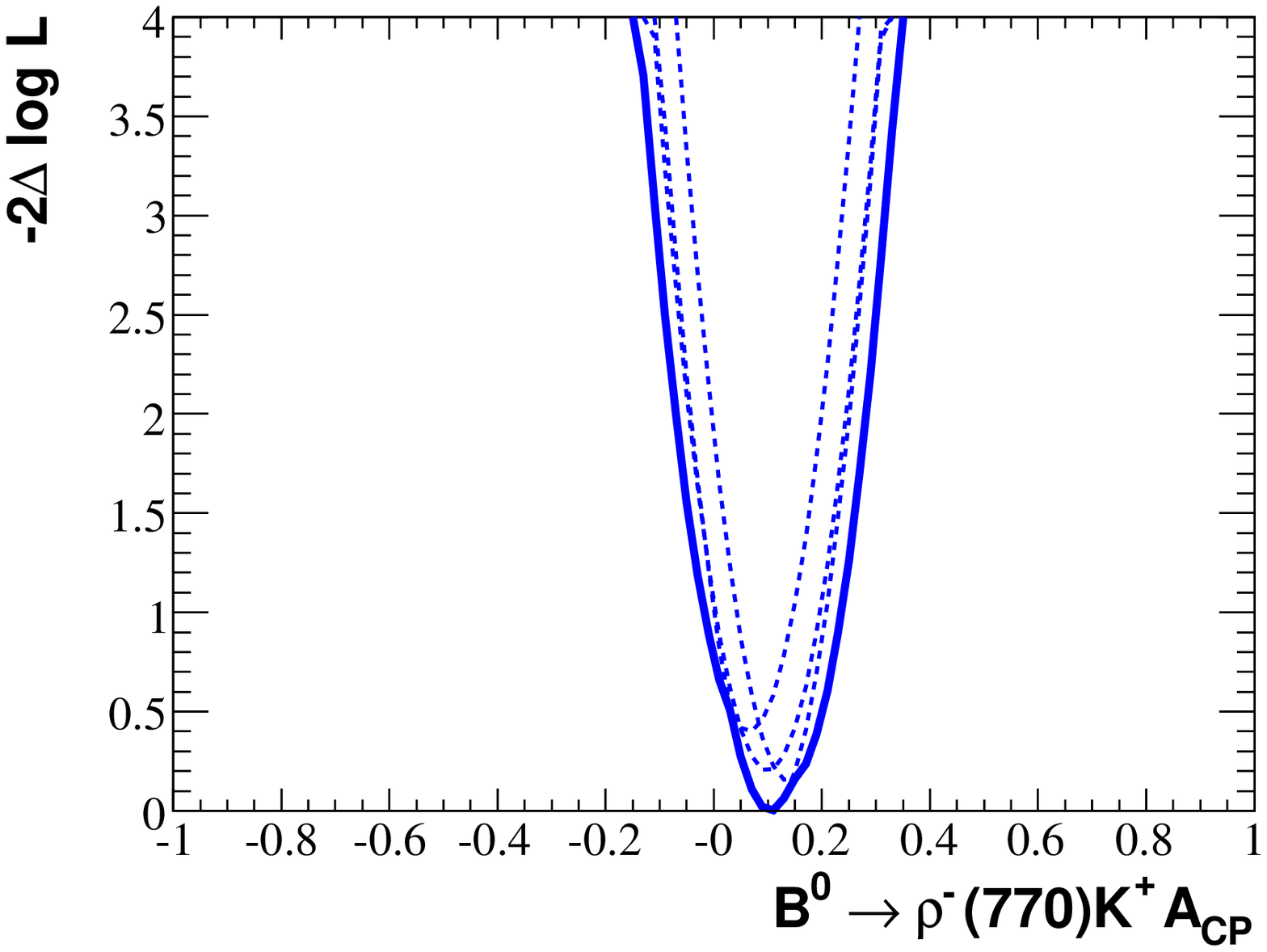} }
\caption{
\label{fig:neutralSwave-acp-scan} The NLL functions for the $CP$-asymmetries. The NLL functions of each solution are shown with dashed lines.
The fitted values for the $\B\to(K\pi)^{*0}_0\pi$ in the four solutions are quite distinct.
The envelope curves (solid lines) are used to quote the physical results.  }
\end{figure*}

The four solutions of the fit are displayed in~\tabref{fourfitsolutions}. The correlation coefficients of solution-I are given 
in~\tabtworef{correlationB-I}{correlationBbar-I}. 
As explained in~\secref{Results} the statistical uncertainty of each solution does not reflect the actual experimental 
uncertainty and should not be used. The procedure we have devised blends all four solutions and determines reliable statistical and systematic uncertainties. 
For illustration, we display the four NLL around their minima in~\figref{neutralSwave-ff-scan} for the isobar fractions 
and~\figref{neutralSwave-acp-scan} for the $CP$-asymmetries.
For the neutral S-wave final states, the spread due to the degeneracy of the fitted fractions and asymmetries is quite large.\par
\begin{figure*}[ht]
\begin{center}
\centerline{\epsfxsize9cm\epsffile{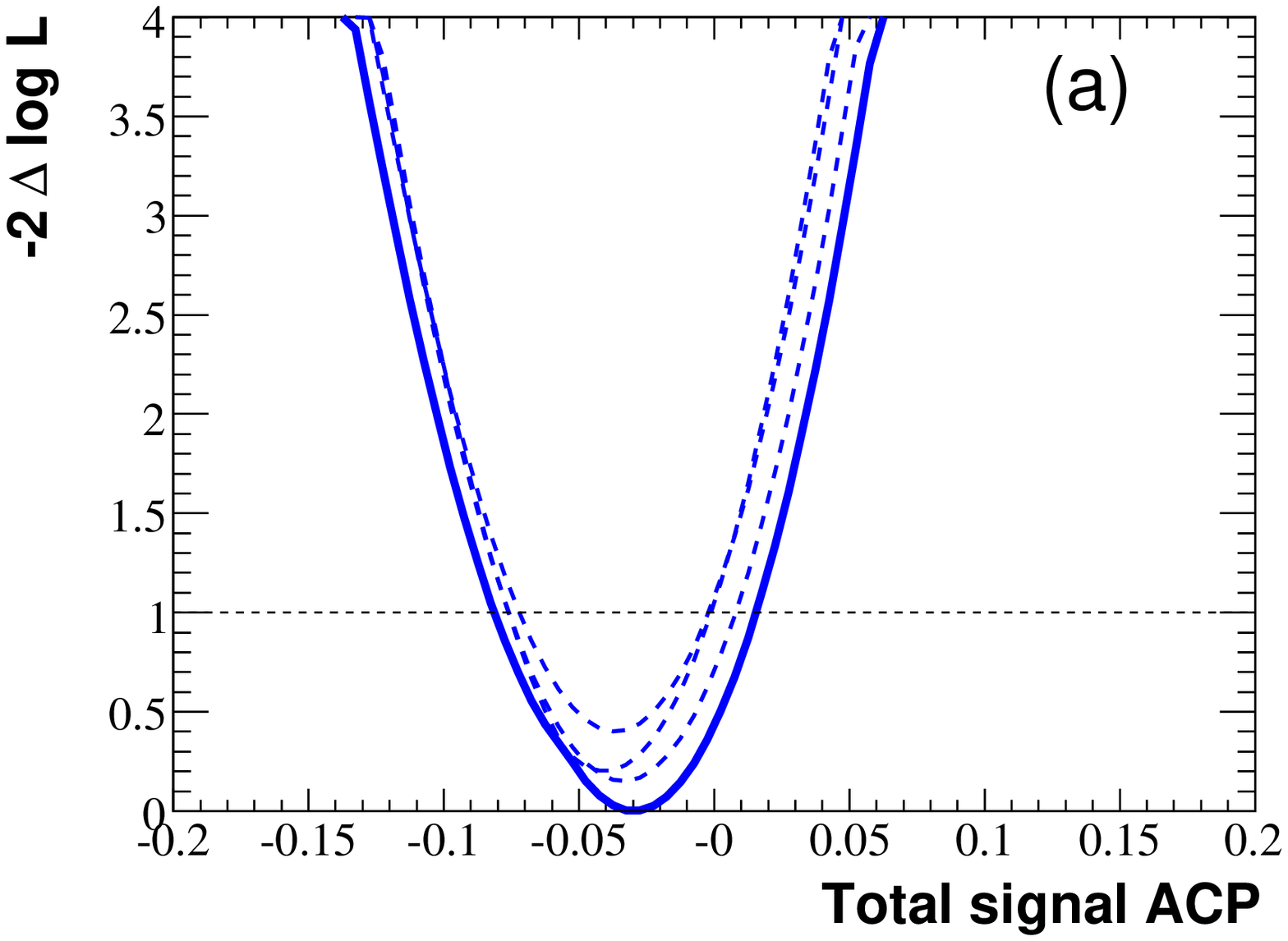} \epsfxsize9cm\epsffile{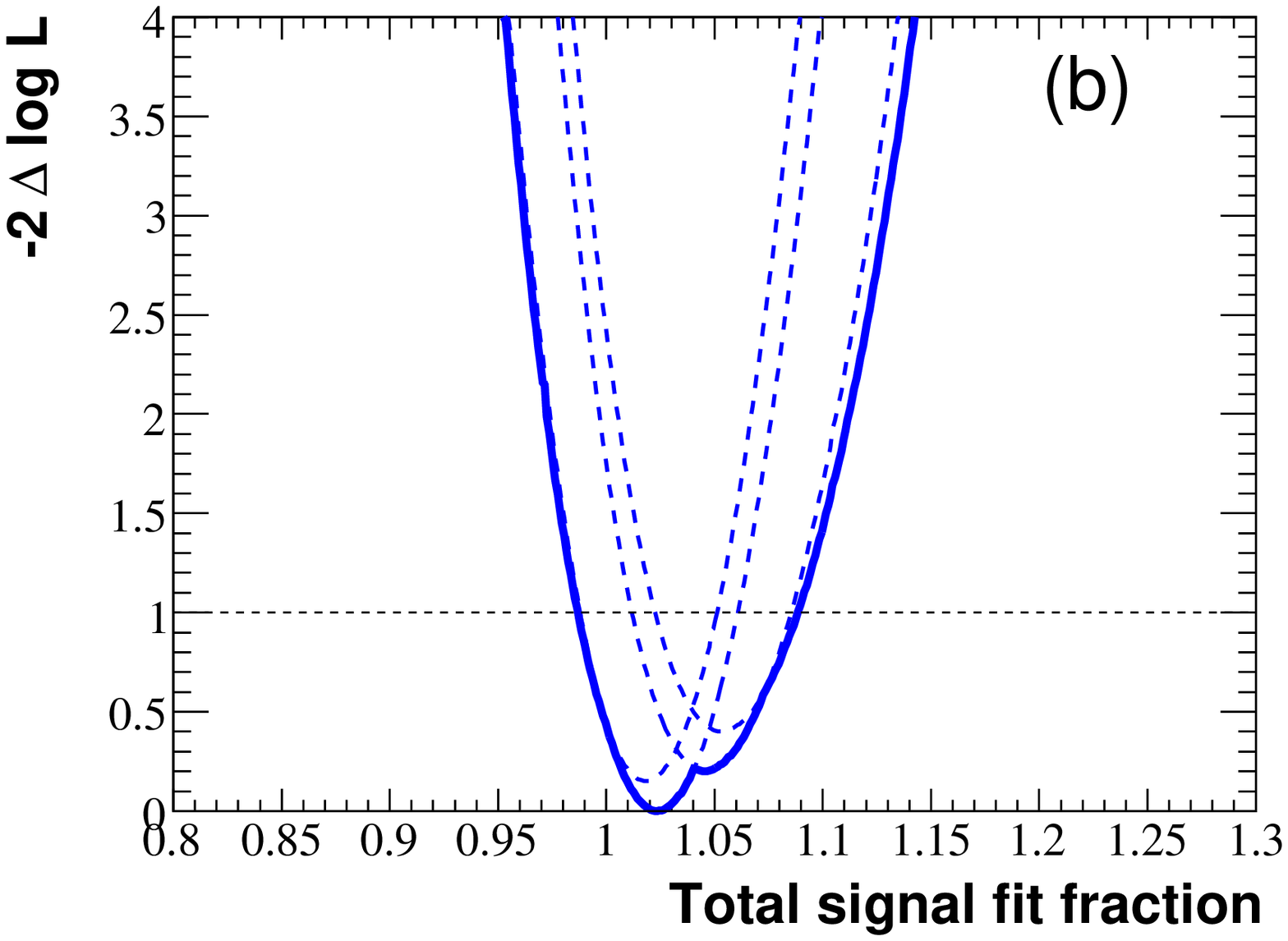} }
\caption{
\label{fig:global-ff-acp-scan} NLL scans for the  global $CP$-asymmetry (a) and the total isobar fraction (b).  
The scans of each solution are shown with dashed lines. The envelope curves (solid lines) are the scans that are used to quote the physical results. }
\end{center}
\end{figure*}

When the NLL are far from being parabolic at their minima, actual scans as described at the end of~\secref{TheFit} are performed to derive the results. 
\figref{global-ff-acp-scan} shows two examples of such scans for the sum of the isobar fit fractions (or the total fit fraction) and for the global 
$CP$-asymmetry $\mathcal{A}_{CP}$ [\equaref{global-bf-acp}].%\clearpage
\begin{table*}
\begin{center}
\caption{\label{tab:fourfitsolutions}
Results of the four solutions of the fit. The fractions are the $CP$-averaged isobar fractions ($FF_j$) defined with the $CP$-asymmetries $A_{CP}$
in~\secref{DecayAmplitudes}~[\equaref{PartialFractions}]. 
The phases $\phi$ for the $\Bz$ decays and $\overline{\phi}$ for the $\Bzb$ decays are measured relative to $\Bz(\Bzb)\to\Kstarpm\pimp$. 
The uncertainties are statistical only. They are underestimated because a parabolic approximation is made for the shape of the NLL close to minimum.}
\setlength{\tabcolsep}{1.2pc}
\begin{tabular}{cccccc}
\hline\hline
                          &                  & I                      & II                      & III                     & IV \\\hline
$NLL_{min}$               &                  & $-91079.6$             & $-91079.5$              & $-91079.4$              & $-91079.5$              \\\hline
$\Kstarp(892)\pim$        & Fraction~(\%)    & $11.75^{+1.80}_{-1.47}$& $11.81^{+1.80}_{-1.44}$ & $12.34^{+1.87}_{-1.46}$ & $12.48^{+1.78}_{-1.52}$ \\
                          & $A_{CP}$         & $-0.19^{+0.13}_{-0.14}$& $-0.20^{+0.13}_{-0.14}$ & $-0.12^{+0.13}_{-0.14}$ & $-0.14^{+0.13}_{-0.14}$ \\
                          & $\phi$~(deg.)    & $\ \ \ 0$~(fixed)      & $\ \ \ 0$~(fixed)       & $\ \ \ 0$~(fixed)       &$\ \ \ 0$~(fixed)        \\
                  & $\overline{\phi}$~(deg.) & $\ \ \ 0$~(fixed)      & $\ \ \ 0$~(fixed)       & $\ \ \ 0$~(fixed)       &$\ \ \ 0$~(fixed)        \\\hline
$\Kstarz(892)\piz$        & Fraction~(\%)    & $\ 6.72^{+1.29}_{-1.26}$&$\ 6.57^{+1.35}_{-1.19}$&$\ 6.52^{+1.36}_{-1.21}$ & $\ 6.47^{+1.29}_{-1.27}$\\
                          & $A_{CP}$         & $-0.09\pm0.19$         & $-0.08\pm0.19$          & $-0.12^{+0.19}_{-0.21}$ & $-0.12^{+0.21}_{-0.19}$ \\
                          & $\phi$~(deg.)    & $73.4\pm37.1$          & $306.6\pm37.8$          & $73.8\pm37.4$          & $305.8\pm37.9$           \\
                  & $\overline{\phi}$~(deg.) & $\ 1.5\pm38.8$         & $\ 1.0\pm38.7$          & $139.3\pm45.4$          & $140.5\pm45.4$          \\\hline
$(K\pi)_0^{*+}\pim$       & Fraction~(\%)    & $31.20^{+3.14}_{-2.91}$& $24.77^{+2.83}_{-2.86}$ & $28.40^{+4.36}_{-3.97}$ & $21.41^{+4.43}_{-3.75}$ \\
                          & $A_{CP}$         & $+0.07^{+0.11}_{-0.09}$& $+0.37\pm0.11$          & $-0.03^{+0.15}_{-0.16}$ & $+0.27^{+0.15}_{-0.18}$ \\
                          & $\phi$~(deg.)    & $167.8\pm10.8$         & $164.7\pm11.8$          & $168.6\pm10.8$          & $165.4\pm11.8$          \\
                  & $\overline{\phi}$~(deg.) & $\ 79.0\pm19.1$        & $\ 78.8\pm19.2$         & $\ 72.7\pm16.6$         & $\ 72.5\pm16.7$         \\\hline
$(K\pi)_0^{*0}\piz$       & Fraction~(\%)    & $17.56^{+2.87}_{-2.62}$& $24.12^{+2.96}_{-2.81}$ & $24.42^{+2.94}_{-2.77}$ & $31.00^{+3.02}_{-2.83}$ \\
                          & $A_{CP}$         & $-0.31^{+0.17}_{-0.15}$& $-0.49^{+0.13}_{-0.12}$ & $+0.05\pm0.12$          & $-0.17^{+0.10}_{-0.11}$ \\
                          & $\phi$~(deg.)    & $\ 52.3\pm36.9$        & $296.3\pm34.6$          & $\ 53.0\pm37.2$         & $295.8\pm34.6$          \\
                  & $\overline{\phi}$~(deg.) & $338.5\pm38.9$         & $337.9\pm38.8$          & $128.9\pm37.5$          & $130.0\pm37.5$          \\\hline
$\rho(770)^-\Kp$          & Fraction~(\%)    & $22.60^{+2.07}_{-2.08}$& $21.77^{+2.07}_{-2.03}$ & $21.64^{+2.10}_{-2.04}$ & $20.88^{+2.08}_{-2.03}$ \\
                          & $A_{CP}$         & $+0.10\pm0.10$         & $+0.14^{+0.10}_{-0.11}$ & $+0.06^{+0.10}_{-0.11}$ & $+0.10\pm0.11$          \\
                          & $\phi$~(deg.)    & $208.5\pm35.8$         & $183.8\pm33.5$          & $206.8\pm36.7$          & $181.4\pm33.7$          \\
                  & $\overline{\phi}$~(deg.) & $117\pm33.7$           & $115.9\pm33.6$          & $351.1\pm40.5$          & $351.4\pm39.8$          \\\hline
$NR$                      & Fraction~(\%)    & $12.51^{+2.22}_{-2.17}$& $12.78^{+2.28}_{-2.12}$ & $11.90^{+2.27}_{-2.05}$ & $12.24^{+2.22}_{-2.09}$ \\
                          & $A_{CP}$         & $+0.23^{+0.18}_{-0.19}$& $+0.19^{+0.19}_{-0.17}$ & $+0.18\pm0.19$          & $+0.15^{+0.18}_{-0.19}$ \\
                          & $\phi$~(deg.)    & $\ 99.9\pm22.9$        & $220.8\pm24.8$          & $100.0\pm22.8$          & $220.5\pm25.0$          \\
                  & $\overline{\phi}$~(deg.) & $\ 12.7\pm23.7$        & $\ 12.0\pm23.6$         & $\ 58.6\pm34.9$         & $\ 59.8\pm35.0$         \\\hline
                          &Total Fraction~(\%)&$102.4\pm3.6$            &$101.8^{+3.6}_{-3.4}$   &$105.3^{+4.6}_{-3.9}$    & $104.5^{+4.5}_{-3.7}$  \\
\hline\hline
\end{tabular}
\end{center}
\end{table*}
\begin{table*}[t]
\begin{center}
\caption{\label{tab:correlationB-I}
Matrix of the correlation coefficients between the fitted parameters for the \Bz\ Dalitz plot in solution-I.}
\begin{tabular}{crrrrrrrrrr}\hline\hline
Variable  	  &$  t_{(K\pi)_0^{*+}}    $&$  t_{(K\pi)_0^{*0}}   $&$  t_{\Kstarz(892)}  	  $&$  t_{NR}  	  $&$  t_{\rho(770)}  	  $&$  \phi_{(K\pi)_0^{*+}}  $&$  \phi_{(K\pi)_0^{*0}} $&$  \phi_{\Kstarz(892)}   $&$  \phi_{NR} 	  $&$  \phi_{\rho(770)} 	$\\\hline
$t_{(K\pi)_0^{*+}}  	  $&$  100.0 	  $&&&&&&&& \\
$t_{(K\pi)_0^{*0}}  	  $&$  -0.9  	  $&$  100.0 	  $&&&&&&&& \\
$t_{\Kstarz(892)}  	  $&$  7.3  	  $&$  -9.3  	  $&$  100.0 	  $&&&&&&&\\
$t_{NR}  	  $&$  40.6  	  $&$  -27.4  	  $&$  1.9  	  $&$  100.0 	  $&&&&&&\\
$t_{\rho(770)}  	  $&$  17.3  	  $&$  11.9  	  $&$  8.6  	  $&$  3.9  	  $&$  100.0 	  $&&&&&\\
$\phi_{(K\pi)_0^{*+}}  	  $&$  -9.1  	  $&$  0.8  	  $&$  0.2  	  $&$  7.0 	  $&$  1.3  	  $&$  100.0 	  $&&&&\\
$\phi_{(K\pi)_0^{*0}}  	  $&$  -31.7  	  $&$  53.0 	  $&$  -6.0 	  $&$  -15.0 	  $&$  -6.9  	  $&$  32.1  	  $&$  100.0 	  $&&&\\
$\phi_{\Kstarz(892)}  	  $&$  -31.0 	  $&$  50.1  	  $&$  -5.5  	  $&$  -14.2  	  $&$  -4.9  	  $&$  31.5  	  $&$  93.2  	  $&$  100.0 	$&&\\
$\phi_{NR}  	  $&$  -47.2  	  $&$  12.9  	  $&$  3.8  	  $&$  -19.5  	  $&$  -6.6  	  $&$  54.2  	  $&$  61.0 	  $&$  61.1  	  $&$  100.0 $&\\
$\phi_{\rho(770)}  	  $&$  -35.0 	  $&$  5.3  	  $&$  -7.5  	  $&$  -17.5  	  $&$  -20.4  	  $&$  31.9  	  $&$  44.1  	  $&$  37.4  	  $&$  52.4  	  $&$  100.0 	$\\
\hline\hline
\end{tabular}
\end{center}
\end{table*}
\begin{table*}
\begin{center}
\caption{\label{tab:correlationBbar-I}
Matrix of the correlation coefficients between the fitted parameters for the \Bzb\ Dalitz plot in solution-I.}
\begin{tabular}{crrrrrrrrrrr}\hline\hline
Variable  	  &$  \overline{t}_{(K\pi)_0^{*-}} 	  $&$  \overline{t}_{(\overline{K}\pi)_0^{*0}}   $&$  \overline{t}_{\Kstarm(892)}  	  $&$  \overline{t}_{\Kstarzb(892)}  	  $&$  \overline{t}_{NR}  	  $&$  \overline{t}_{\rho(770)}  	  $&$  \overline{\phi}_{(K\pi)_0^{*-}}  $&$  \overline{\phi}_{(\overline{K}\pi)_0^{*0}} $&$  \overline{\phi}_{\Kstarzb(892)}   $&$  \overline{\phi}_{NR} 	  $&$  \overline{\phi}_{\rho(770)} 	$\\\hline
$\overline{t}_{(K\pi)_0^{*-}}  	  $&$  100.0 	  $&&&&&&&&&&\\
$\overline{t}_{(\overline{K}\pi)_0^{*0}}  	  $&$  -1.4  	  $&$  100.0 	  $&&&&&&&&&\\
$\overline{t}_{\Kstarm(892)}  	  $&$  9.1  	  $&$  0.7  	  $&$  100.0 	  $&&&&&&&&\\
$\overline{t}_{\Kstarzb(892)}  	  $&$  7.0 	  $&$  -18.9  	  $&$  6.4  	  $&$  100.0 	  $&&&&&&&\\
$\overline{t}_{NR}  	  $&$  33.3  	  $&$  -18.4  	  $&$  6.2  	  $&$  0.9  	  $&$  100.0 	  $&&&&&&\\
$\overline{t}_{\rho(770)}  	  $&$  20.4  	  $&$  1.7  	  $&$  9.7  	  $&$  8.2  	  $&$  6.8  	  $&$  100.0 	  $&&&&&\\
$\overline{\phi}_{(K\pi)_0^{*-}}  	  $&$  5.2  	  $&$  -0.1  	  $&$  -9.7  	  $&$  -0.5  	  $&$  6.1  	  $&$  -3.3  	  $&$  100.0 	  $&&&&\\
$\overline{\phi}_{(\overline{K}\pi)_0^{*0}}  	  $&$  -27.7  	  $&$  59.1  	  $&$  -13.1  	  $&$  -10.2  	  $&$  -6.1  	  $&$  -13.9  	  $&$  42.2  	  $&$  100.0 	  $&&&\\
$\overline{\phi}_{\Kstarzb(892)}  	  $&$  -25.6  	  $&$  52.9  	  $&$  -11.7  	  $&$  -6.0 	  $&$  -6.6  	  $&$  -12.2  	  $&$  40.9  	  $&$  86.9  	  $&$  100.0 	  $&&\\
$\overline{\phi}_{NR}  	  $&$  -39.0 	  $&$  6.2  	  $&$  -16.2  	  $&$  -0.3  	  $&$  -7.5  	  $&$  -11.9  	  $&$  72.6  	  $&$  62.3  	  $&$  60.3  	  $&$  100.0 	  $&\\
$\overline{\phi}_{\rho(770)}  	  $&$  -38.1  	  $&$  6.2  	  $&$  -18.0 	  $&$  -10.3  	  $&$  -4.1  	  $&$  -11.8  	  $&$  50.0 	  $&$  53.8  	  $&$  47.3  	  $&$  70.0 	  $&$  100.0 	$\\
\hline\hline
\end{tabular}
\end{center}
\end{table*}


\clearpage
\begin{thebibliography}{99}

\bibitem{CKM}           N.~Cabibbo, 
                        Phys. Rev. Lett. {\bf 10}, 531 (1963);
                        M.~Kobayashi and T.~Maskawa, 
                        Prog. Theor. Phys. {\bf 49}, 652 (1973).

\bibitem{SnyderQuinn}   H.R. Quinn and A.E. Snyder, \jprd{48},2139-2144 (1993).

\bibitem{rhopibabar}    \babar\  Collaboration, B.~Aubert \ea, \jprd{76}, 012004 (2007). 

\bibitem{rhopibelle}    Belle Collaboration, A. Kusaka \ea,\jprl{98},221602 (2007). 

\bibitem{Ciuchini:2006kv}  M.~Ciuchini, M.~Pierini and L.~Silvestrini, \jprd{74}, 051301 (2006).

\bibitem{Gronau:2006qn} M.~Gronau, D.~Pirjol, A.~Soni and J.~Zupan, \jprd{75}, 014002 (2007).

\bibitem{Cmodeimplied}  Throughout the paper, whenever a mode is given, the charge conjugate is also implied.

\bibitem{josezhitang}   \babar\  Collaboration, B.~Aubert \ea, \verb!{arXiv:hep-ex/0408073}!.

\bibitem{Zhitang}       Zhitang Yu, PhD thesis, report SLAC-R-815.

\bibitem{cleo}          CLEO Collaboration, E. Eckhart \ea \jprl{89}, 251801 (2002)

\bibitem{belle}         Belle Collaboration, P. Chang \ea, \plb{599}, 148 (2004).

\bibitem{cleorhok}      CLEO Collaboration, C. P.  Jessop \ea, \jprl{85}, 2881-2885 (2000).

\bibitem{Feng}          \babar\  Collaboration, B.~Aubert \ea, \jprd{76},011103 (2007).

\bibitem{latham}        \babar\  Collaboration, B.~Aubert \ea, \jprd{72},072003 (2005); Erratum-ibid. \jprd{74}, 099903 (2006).

\bibitem{bellek+pi+pi-} Belle Collaboration, A Garmash \ea, \jprl{96}, 251803 (2006). 

\bibitem{babar-kspipi}  \babar\  Collaboration, B.~Aubert \ea, \jprd{73}, 031101 (2006). 

\bibitem{belle-kspipi}  Belle Collaboration, A Garmash \ea,, \jprd{75}, 012006 (2007).

\bibitem{Cheng:2007si}  H.~Y.~Cheng, C.~K.~Chua and A.~Soni, Phys.\ Rev.\  D {\bf 76}, 094006 (2007).

\bibitem{BlattWeissk}   J.~Blatt and V.~Weisskopf, {\em ``Theoretical 
                        Nuclear Physics''}, John Wiley \& Sons, New York,
                        1956.

\bibitem{Asner}         D.~Asner, \verb!{arXiv:hep-ex/0410014}!.

\bibitem{Zemach}        C. Zemach, Phys. Rev, {\bf 133}, B1201 (1964).

\bibitem{PDG2004}       Particle Data Group, S.~Eidelman \ea, 
                        Phys.\ Lett.\ {\bf B592}, 1 (2004).

\bibitem{LASS}          D.\ Aston \ea, Nucl. Phys. {\bf B296}, 493 (1988).

\bibitem{PDG2006}       Particle Data Group, W.-M. Yao \ea, Journal of Physics G{\bf 33}, 1 (2006).

\bibitem{rhoGS}         G.J.~Gounaris and J.J.~Sakurai,
                        Phys. Rev. Lett. {\bf 21}, 244 (1968).

\bibitem{Estabrooks}    P.\ Estabrooks, Phys. Rev. {\bf D19}, 2678 (1979).

\bibitem{Dunwoodie-SDP} W.M. Dunwoodie, private communication.

\bibitem{babarNim}      \babar\  Collaboration, 
                        B.~Aubert \ea, 
                        Nucl. Instrum. Methods {\bf A 479}, 1 (2002).

\bibitem{geant4}        GEANT4 Collaboration, S. Agostinelli \ea, \nima 506, 250 (2003).
 
\bibitem{NN}            P.~Gay, B.~Michel, J.~Proriol and O.~Deschamps, {\it ``Tagging Higgs Bosons in Hadronic LEP-2 Events with Neural Networks''}, 
                        in Pisa 1995, New computing techniques in physics research, 725 (1995).

\bibitem{XtalBall}      J. E. Gaiser \ea, Phys. Rev. {\bf D34}, 711 (1986).

\bibitem{cranmer}       K. S. Cranmer, Comput. Phys. Commun. 136, 198 (2001). 

\bibitem{Argus}         ARGUS Collaboration, (H. Albrecht \ea), 
                        Z. Phys. C {\bf 48}, 543 (1990).

\bibitem{rhopiq2bbabar} \babar\  Collaboration, B.~Aubert \ea,  \jprl{91}, 201802 (2003).

\bibitem{hfag}          Heavy Flavor Averaging Group (HFAG), E. Barberio \ea, arXiv:0704.3575v1 [hep-ex].
                      
\end{thebibliography}
\end{document}